\documentclass[twocolumn]{aastex63}
\received{?? ?, 2021}
\revised{?? ?, 2021}
\accepted{August 6, 2021}%\today}
\submitjournal{ApJ}

\usepackage{color}
\newcommand\propol{{\em propol}}% projected polytropes 
\newcommand\sersic{S\'ersic}

\shorttitle{Mass surface densities fitted with projected polytropes}
\shortauthors{S\'anchez Almeida et al.}
\begin{document}

\title{
  %Polytrope-based mass surface density profiles reproduce observed galaxies\\
  % Projected-Polytropes reproduce the mass surface density profiles observed in galaxies
  Physically motivated  fit to mass surface density profiles observed in galaxies
  }

\correspondingauthor{J. S\'anchez Almeida}
\email{jos@iac.es}

\author[0000-0003-1123-6003]{Jorge S\'anchez Almeida} \affil{Instituto de Astrof\'\i sica de Canarias, La Laguna, Tenerife, E-38200, Spain} \affil{Departamento de Astrof\'\i sica, Universidad de La Laguna}

\author[0000-0001-8647-2874]{Ignacio Trujillo} \affil{Instituto de Astrof\'\i sica de Canarias, La Laguna, Tenerife, E-38200, Spain} \affil{Departamento de Astrof\'\i sica, Universidad de La Laguna}

\author[0000-0001-5848-0770]{Angel R. Plastino} \affil{CeBio y Departamento de Ciencias B\'asicas, \\ Universidad Nacional del Noroeste de la Prov. de Buenos Aires, \\ UNNOBA, CONICET, Roque Saenz Pe\~na 456, Junin, Argentina}



\begin{abstract}
  % show, exhibit, capture, grasp , encapsulate
Polytropes have gained renewed interest because they account for several seemingly-disconnected observational properties of galaxies. Here we study if polytropes are also able to explain the stellar mass distribution within galaxies. We develop a code to fit surface density profiles using polytropes projected in the plane of the sky (\propol s).  \sersic\ profiles are known to be good proxies for the global shapes of galaxies and we find that, ignoring central cores, \propol s and \sersic\ profiles are indistinguishable within observational errors (within  5\,\% over 5~orders of magnitude in surface density). The range of  physically meaningful polytropes yields \sersic\ indexes between 0.4 and 6. The code has been systematically applied to $\sim 750$ galaxies with carefully measured mass density profiles and including all morphological types and stellar masses ($7 < \log [M_\star/{\rm M_\odot}] < 12$). The \propol\  fits are systematically better than \sersic\ profiles when $\log(M_\star/{\rm M}_\odot)\lesssim 9$ and systematically worst when $\log(M_\star/{\rm M}_\odot)\gtrsim 10$.  Although with large scatter, the observed  polytropic indexes increase with increasing mass and tend to cluster around $m=5$.  For the most massive galaxies,  \propol s are very good at reproducing their central parts, but they do not handle well cores and outskirts altogether. Polytropes are self-gravitating systems in thermal meta-equilibrium as defined by the Tsallis entropy. Thus, the above results are compatible with the principle of maximum Tsallis entropy dictating the internal structure in dwarf galaxies and in the central region of massive galaxies. 
%  \comment{239 words; max 250}  
\end{abstract}

\keywords{gravitation ---
  methods: data analysis ---
  galaxies:general ---
  galaxies: fundamental parameters ---
  galaxies: structure ---
  galaxies: statistics}


\section{Introduction} \label{sec:intro}

Galaxies are  self-gravitating structures that, among all possible equilibrium configurations, choose only  those consistent with a stellar mass surface density  profile approximately resembling a S\'ersic function \citep[e.g.,][]{1968adga.book.....S,1993MNRAS.265.1013C,2001MNRAS.326..869T,2003ApJ...594..186B,2005PASA...22..118G,2012ApJS..203...24V}.\footnote{The S\'ersic function includes exponential disks, observed in dwarf galaxies \cite[e.g.,][]{1994A&AS..106..451D}, and {\em de Vaucouleurs} $R^{1/4}$-profiles, characteristic of massive ellipticals \cite[e.g.,][]{1948AnAp...11..247D}.} Settling into this particular configuration could be due to either some fundamental physical process (as, e.g.,  thermodynamical equilibrium sets the velocities of the molecules in the air) or to the initial conditions that gave rise to the system \citep{2008gady.book.....B}. The mass distribution in galaxies is commonly explained as the outcome of cosmological initial conditions \citep{2014ApJ...790L..24C,2015ApJ...805L..16N,2017MNRAS.465L..84L,2020MNRAS.495.4994B}. The option of a  fundamental process like thermodynamic equilibrium determining the configuration is traditionally discredited because of two conceptually very different reasons. 

If the structure were set by thermodynamic equilibrium, following the principles of statistical physics, it should correspond to the most probable configuration of a self-gravitating system and, thus, it should result from maximizing its entropy. The use of the classical Boltzmann-Gibbs entropy to characterize  self-gravitating systems leads to a distribution with infinite mass and energy  \citep{2008gady.book.....B,2008arXiv0812.2610P}, disfavoring the thermodynamic equilibrium explanation. However, this difficulty of the theory has been overcome as follows. In the standard Boltzmann-Gibbs approach, the long-range forces that govern self-gravitating systems are not taken into account. Systems with long-range interactions admit long-lasting meta-stable states described by a maximum entropy formalism based on Tsallis ($S_q$) non-additive entropies \citep[][]{1988JSP....52..479T,2009insm.book.....T,2005PhyA..356..419C}. Observational evidence for the $S_q$ statistics has been found in connection with various astrophysical problems \citep{2013SSRv..175..183L,2013ApJ...777...20S}. In particular, the maximization under suitable constraints of the Tsallis entropy of a Newtonian self-gravitating N-body system leads to a polytropic distributions \citep{1993PhLA..174..384P,2005PhyA..350..303L}, which can have finite mass and a shape resembling the dark matter (DM) distribution found in numerical simulations of galaxy formation \citep[][]{2004MNRAS.349.1039N,2009PhyA..388.2321C,2013MNRAS.428.2805A,2021MNRAS.504.2832S}. %2005ApJ...624L..85M}. % take out Merrit if needed this is what is needed
In addition to having an identifiable physical origin, the polytropic shape has lately gained practical importance because of its association with real self-gravitating astrophysical objects.  The mass density profiles in the centers of dwarf galaxies are well reproduced by polytropes without any tuning or degree of freedom \citep{2020A&A...642L..14S}. The same type of profile also explains the stellar surface density profiles observed in globular clusters  (Trujillo et al. 2021, in preparation). Polytropic DM haloes fit the velocity dispersion observed in many early-type galaxies  \citep{2010MNRAS.405...77S}.

The second argument against thermodynamic equilibrium explaining galaxy shapes relies on the nature of DM. In the current cosmological model, DM provides most of the gravitational pull needed for the ordinary matter to colapse forming visible galaxies. The simplest form of DM is collision-less and so unable to reach thermodynamical equilibrium within the Hubble time \citep{2008gady.book.....B,2003MNRAS.338...14P,2019MNRAS.488.3663L}, which poses a potential problem for any distribution arising from thermodynamic equilibrium. However, this assumption on the nature of the DM causes some of the so-called small-scale problems of the cold DM (CDM) model \citep[e.g.,][]{2015PNAS..11212249W,2017ARA&A..55..343B,2017Galax...5...17D}, in particular, the {\em cusp--core problem}. Simulated CDM haloes have {\em cusps} in their central mass distribution \citep[e.g.,][]{1997ApJ...490..493N,2020Natur.585...39W} which disagree with the central plateau or {\em core} often observed in galaxies \citep[e.g.,][]{2015AJ....149..180O}. The various physical processes invoked to turn {\em cusps} into {\em cores}, in essence, produce the thermalization of the overall gravitational potential  within a finite timescale. Among others, the proposals have been:  feedback of the baryons on the DM particles through gravitational forces \citep[][]{2010Natur.463..203G,2014MNRAS.437..415D,2020MNRAS.499.2912F},  scattering with massive gas clumps \citep[][]{2013ApJ...775L..35E,2019MNRAS.489.5919S}, forcing by a central bar \citep{1971ApJ...168..343H}, or assuming an artificially large DM collision cross section that shortens the two-body collision timescales below the Hubble time \citep{2000PhRvL..84.3760S,2001ApJ...547..574D,2015MNRAS.453...29E}.
When the DM particles of a numerical simulation are allowed to interact through any of these mechanisms, it leads to a gravitational potential conforming with polytropes \citep[][]{2021MNRAS.504.2832S}, as expected if the maximum Tsallis entropy were dictating the stationary state.

Thus, the existing arguments against thermodynamic equilibrium setting galaxy shapes are questionable. Here, we contribute to the discussion studying if polytropes can reproduce the stellar mass distribution in galaxies. Previous works along this direction are restricted to only a few galaxies and yielded inconclusive results \citep[e.g.,][]{2005PhyA..356..419C,2021A&A...647A..29N}.
The question is addressed in various complementary ways.    
The paper starts off by formally  introducing polytropes and their projection in the plane of the sky (Sect.~\ref{sec:poly}).
Then, Sect.~\ref{sec:code} presents our {\tt python} code to fit surface density profiles using the projection of a polytrope in the plane of the sky (hereinafter denoted as \propol ). 
As we explain above, \sersic\ profiles are commonly used to model surface density profiles. We use the fitting code to address whether \sersic\ profiles and \propol s are related. We find that the range of observed \sersic\ shapes seem to emerge from the range of physically possible polytropes (Sect.~\ref{sec:sersic}). The correspondence holds even when stars do not strictly follow the total mass distribution.
In Sect.~\ref{sec:galaxies}, the \propol\  fitting code is systematically applied to a large number of galaxies ($\sim 750$) with carefully measured mass density profiles, including all morphological types, and spanning five orders of magnitude in stellar mass
($7 < \log [M_\star/{\rm M_\odot}] < 12$, where $M_\star/{\rm M}_\odot$ stands for the total mass in stars in solar mass units; \citeauthor{2020MNRAS.493...87T}~\citeyear{2020MNRAS.493...87T}).
The overall good fits provided by \propol s worsen for massive galaxies. Section~\ref{sec:monsters} is devoted to understand why through the examination of a few galaxies particularly well observed  \citep{2017A&A...603A..38S}.  
The main results are highlighted and discussed in Sect.~\ref{sec:conclusions}.
Various technical details are spelled out in Apps.~\ref{app:a} -- \ref{app:scatter}.
The fitting code is publicly available at %\atpublication{[to be completed at publication stage]}.
{\tt https://github.com/jorgesanchezalmeida/polytropic-fits}

%%%%
\section{Polytropes and their projection in the plane of the sky}\label{sec:poly}

A polytrope  of index $m$ is defined as the spherically-symmetric self-gravitating structure resulting from the solution of the Lane-Emden equation for the (normalized) gravitational potential $\psi$  \citep{1967aits.book.....C,2008gady.book.....B},
\begin{equation}
  \frac{1}{s^2}\frac{d}{ds}\Big(s^2\frac{d\psi}{ds}\Big)=
  \cases{
    -3\psi^m & $\psi > 0$,\cr
    0 & $\psi \le  0$.\cr
  }
\label{eq:lane_emden}
\end{equation}
 The symbol $s$ stands for the scaled radial distance in the 3D space and the mass volume density is recovered from $\psi$ as
 \begin{equation}
   \rho(r) = \rho(0)\,\psi(s)^m,
   \label{eq:densityle}
 \end{equation}
\begin{equation}
  r = b\, s,
   \label{eq:radius}
 \end{equation}
 where $r$ stands for the physical radial distance and $\rho(0)$ and $b$ are two arbitrary constants. Equation~(\ref{eq:lane_emden}) is solved under the initial conditions $\psi(0)=1$ and
 $d\psi(0)/ds =0$.\footnote{Equation~(\ref{eq:lane_emden}) also admits solutions with $d\psi(0)/ds \not= 0$, but those are discarded because they have infinite central density and total mass \citep[e.g.,][]{2008gady.book.....B}.}

 A number of properties are relevant for the present work. Polytropes with $m < 5$ have finite size, i.e., $\rho(r)=0$ for $r$ larger than a given radius (which depends on $m$).There is a preferred range of polytropic indexes,
\begin{equation}
  3/2 <  m \le 5,
  \label{eq:nlimits}
\end{equation}
for reasons detailed by \citet{2008gady.book.....B} \citep[see also][]{1993PhLA..174..384P}. Polytropes with $m > 3/2$ are stable according to the Doremus-Felix-Baumann theorem and Antonov second law for the stability of stellar systems. Stellar polytropes with $m \le 3/2$ are unrealistic because, at the cut-off point associated with particles of vanishing energy, the polytropic phase-space distribution has a singularity. Finally, polytropes with $m > 5$ are unphysical because they have infinite mass and energy.
The limit $m\rightarrow\infty$ correspond to the isothermal sphere, which is the solution corresponding to the thermodynamic equilibrium defined by the Boltzmann-Gibbs entropy \citep[][]{2001MNRAS.328..839H,2008gady.book.....B}. Since the isothermal sphere corresponds to  $m> 5$, it has infinite mass and energy. In general, the polytropes have to be evaluated by numerical integration of Eq.~(\ref{eq:lane_emden}), with two exceptions  \citep[e.g.,][]{2008gady.book.....B} corresponding to  $m=1$, with
\begin{equation}
  \psi(s)=
  \cases{
    \frac{\sin(\sqrt{3}\,s)}{\sqrt{3}\,s} & $s < \pi/\sqrt{3}$,\cr
    0 & {\rm elsewhere},\cr
    }
  \label{eq:poly1}
\end{equation}
and $m=5$
(\citeauthor{schuter84}~\citeyear{schuter84} sphere or \citeauthor{1911MNRAS..71..460P}~\citeyear{1911MNRAS..71..460P} model),
\begin{equation}
  \psi(s) = \frac{1}{\sqrt{1+s^2}},
  \label{eq:poly5}
  \end{equation}
expressions employed  in Sect.~\ref{sec:code} to test our numerical codes. 

 Note that all polytropes have cores, in the sense that $d\ln\rho/d\ln r\rightarrow 0$ when $r\rightarrow 0$. This result follows from the initial condition $d\psi(0)/ds=0$ and Eqs.~(\ref{eq:lane_emden}) -- (\ref{eq:radius}). Moreover, the cores of all polytropes look the same after suitable normalization \citep{2021MNRAS.504.2832S},
 \begin{equation}
   \frac{\rho(r)}{\rho_\alpha} \simeq
   \Big[1+\frac{\alpha}{2m}\Big(1-\frac{r^2}{r_\alpha^2}\Big)\Big]^m
   \simeq 1+\frac{\alpha}{2}\Big(1-\frac{r^2}{r_\alpha^2}\Big),
   \label{eq:central2}
 \end{equation}
 with $r_\alpha$ defined as
 \begin{equation}
   \frac{d\ln\rho}{d\ln r}(r_\alpha)=-\alpha,
   \label{eq:logder}
\end{equation}
so that $\rho_\alpha=\rho(r_\alpha)$. Equation~(\ref{eq:central2}) holds for $r<r_\alpha$ and $\alpha \ll  2m$.

We will compare the predictions from polytropic mass distributions with observed mass surface densities. Thus, the predicted volume densities have to be projected in the plane of the sky to make the comparison.  The surface density $\Sigma(R)$ corresponding to the volume density $\rho(r)$ is given by its Abel transform,
\begin{equation}
  \Sigma(R) = 2\,\int_R^\infty \frac{r\,\rho(r)\,dr}{\sqrt{r^2-R^2}},
  \label{eq:abeldirect0}
\end{equation}
with $R$ the projected distance from the center \citep[e.g.,][]{2008gady.book.....B}. Then, $\Sigma(R)$ can be expressed in terms of the normalized Abel transform
\begin{equation}
  f(x,m) = 2\,\int_x^\infty \frac{s\,\psi^m(s)\,ds}{\sqrt{s^2-x^2}},
  \label{eq:abeldirect}
\end{equation}
as
\begin{equation}
  \Sigma(R) = a\,f(R/b,m),
  \label{eq:needlabel}
\end{equation}
where  $x=R/b$  is the normalized projected distance to the center, and $a=\rho(0)\times b$ has units of surface density. The projected polytrope  $\Sigma(R)$ (\propol ) inherits its shape from $f$ but depends on two parameters  encoding the central surface density $a$ and the absolute radial distance $b$. The expression~(\ref{eq:needlabel}) will be used to fit observed surface density profiles, with $a$, $b$, and $m$ the free parameters to be obtained from the fit. The \propol\ shape $f$ can be evaluated analytically when $m=5$ (Eq.~[\ref{eq:poly5}]), and it corresponds to
\begin{equation}
  f(x,5) = \frac{4/3}{(1+x^2)^2},
  \label{eq:ppoly5}
\end{equation}
with its half-mass radius at $x=1$.  
\begin{figure}
  \centering
    \includegraphics[width=0.9\linewidth]{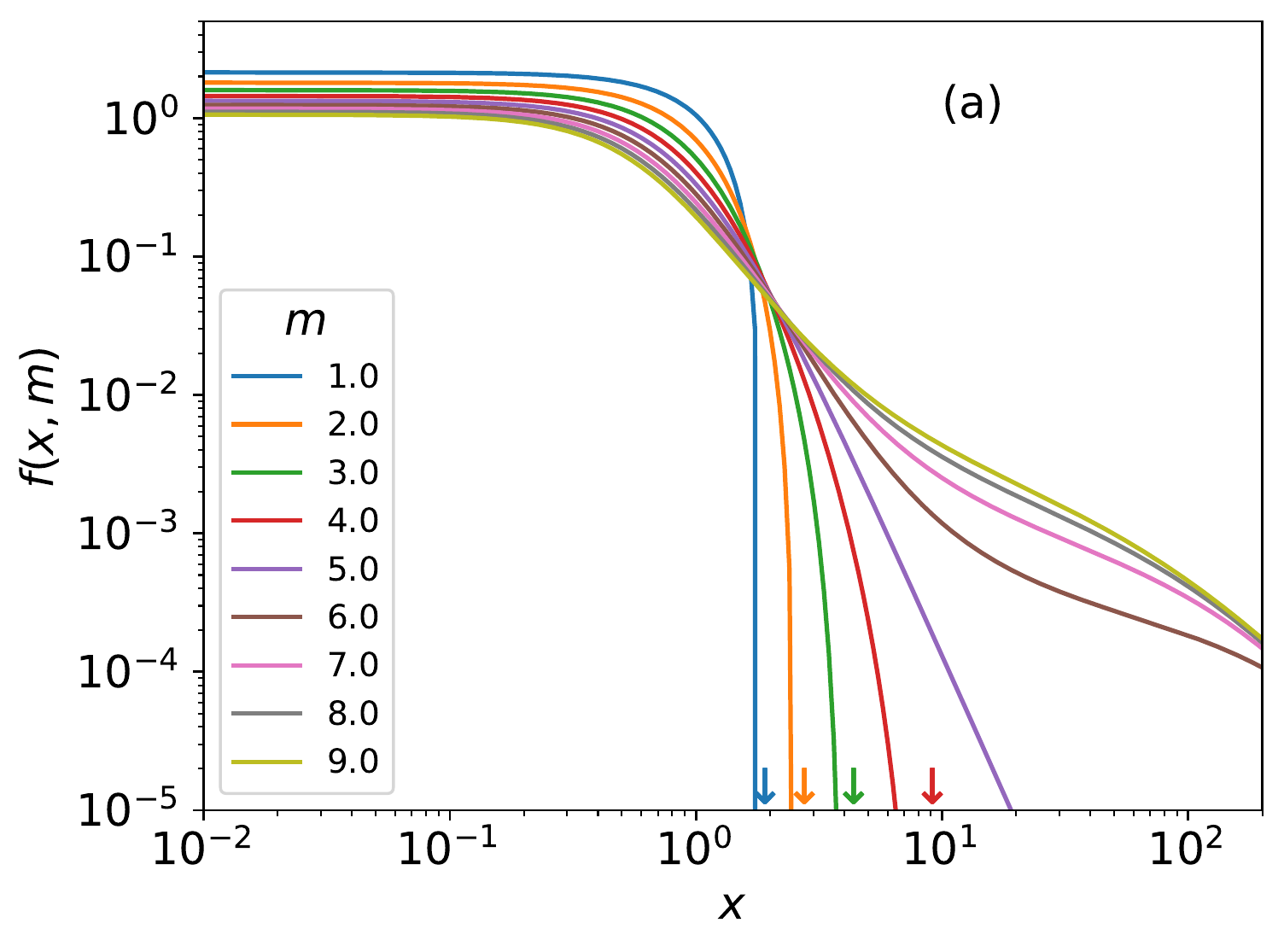}\\
    \includegraphics[width=0.9\linewidth]{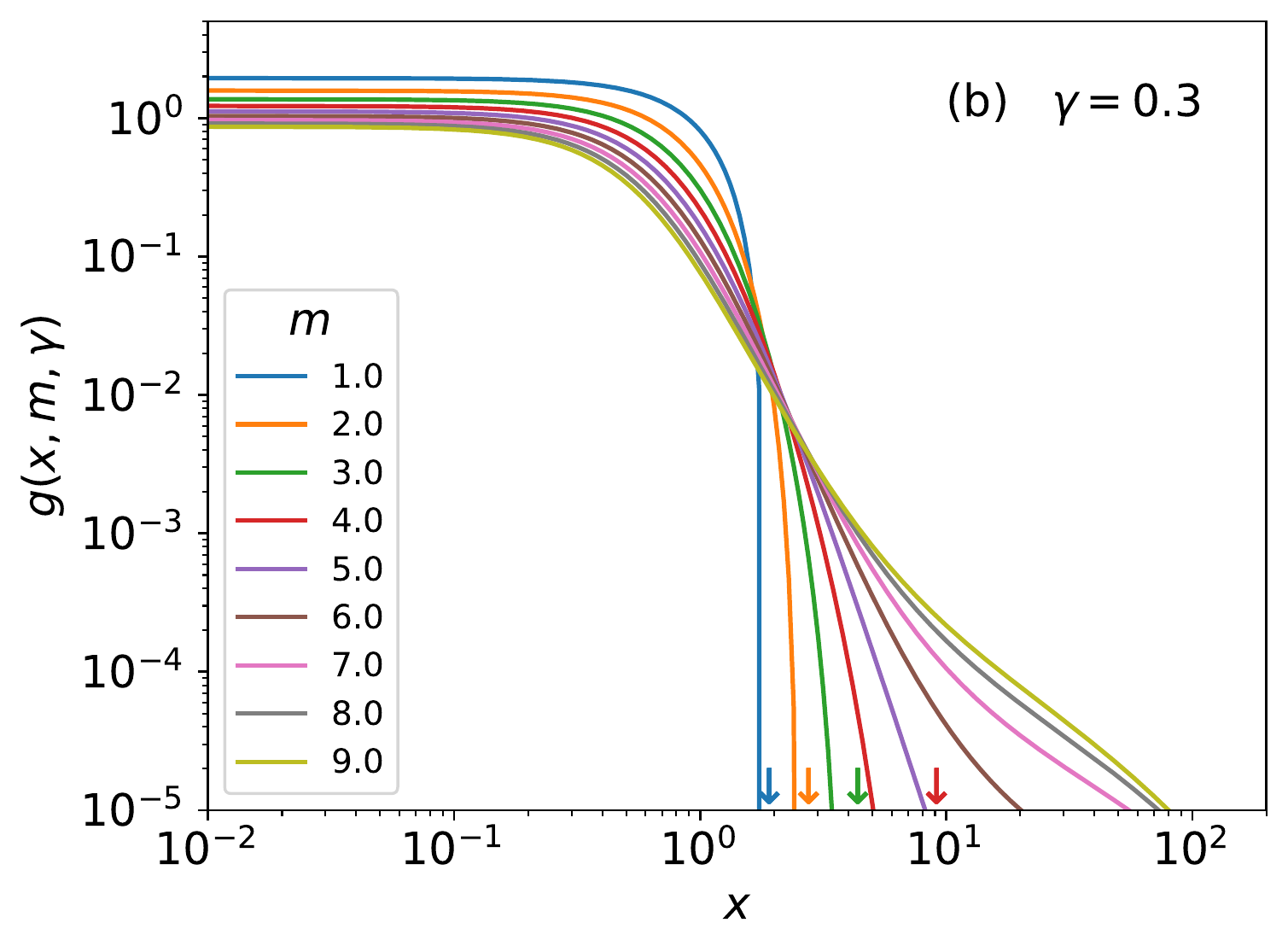}
  \caption{Shapes of \propol s as function of the normalized radial distance. The color code indicates polytropic index.  (a)~Shapes of plain projected polytropes (Eq.~[\ref{eq:abeldirect}]). The arrow heads point out the size of the \propol , which is finite only for $m<5$. (b)~Shape of \propol s when the ratio of total to stellar mass is assumed to vary according to Eq.~(\ref{eq:gamma}). The exponent describing the variation, $\gamma$, is taken to be 0.3.}
  \label{fig:propols}
\end{figure}
\begin{figure}
  \includegraphics[width=0.9\linewidth]{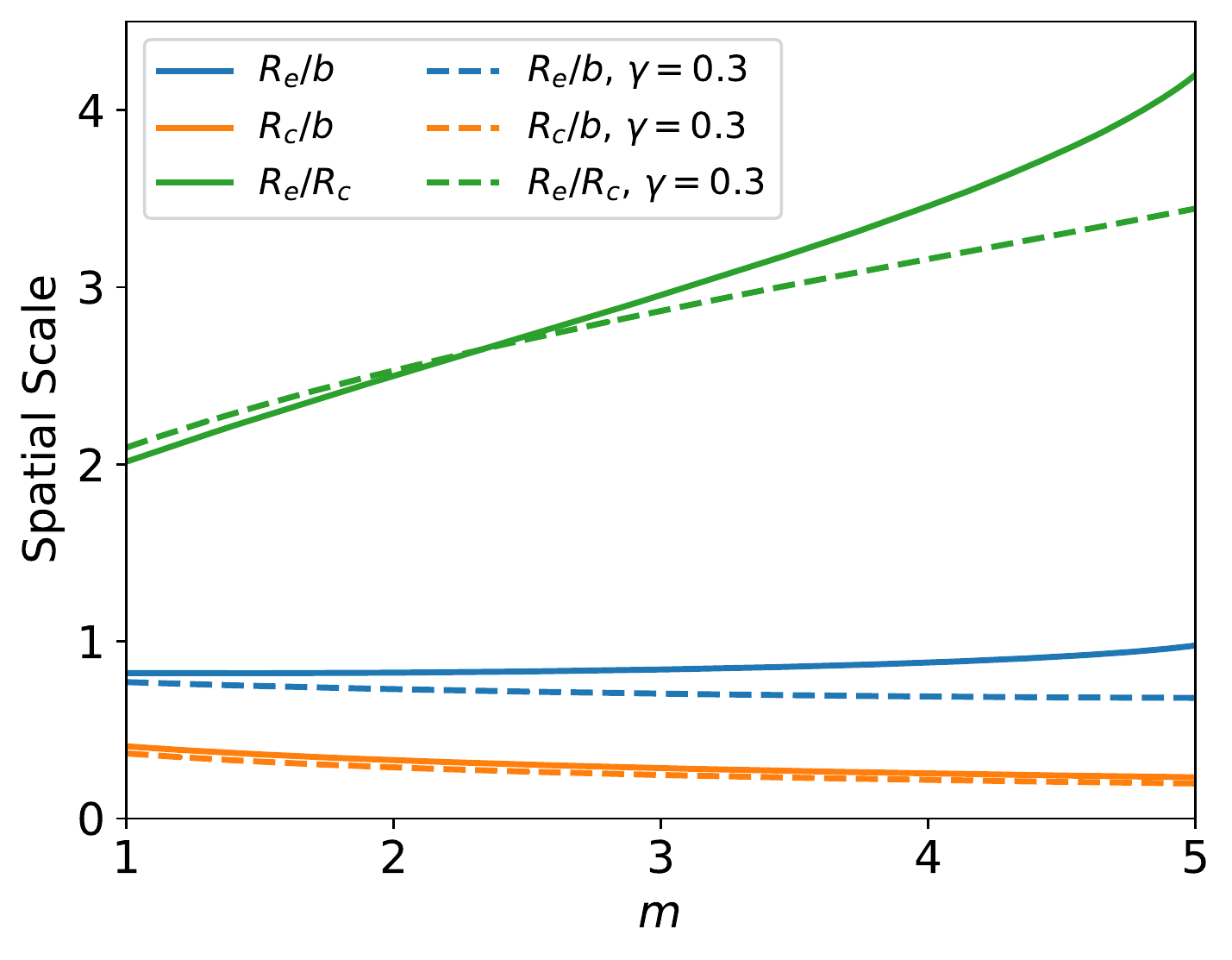}
  \caption{Dependence  on the index $m$ of the effective radius $R_e$ and core radius $R_c$  characterizing \propol s. Given $m$, both $R_e$ and $R_c$ scale linerly with $b$. {\em Propol}s with $m > 5$ are excluded because they have infinity mass and so undefined $R_e$. The solid lines show plain \propol s whereas the dashed lines correspond to varying stellar to total mass ratio with  $\gamma=0.3$ (Eq.~[\ref{eq:gamma}]).}
  \label{fig:propol_re}
\end{figure}

Comparing \propol s with observed stellar surface densities assumes the stellar mass  to scale with the total mass responsible for the overall gravitational potential. The stellar mass results from replacing  $\rho$ in  Eq.~(\ref{eq:abeldirect0}) with the stellar density $\rho_\star$. If $\rho_\star\propto \rho$ then the shape $\Sigma(R)$ in Eq.~(\ref{eq:needlabel}) is the same for $\rho$ and $\rho_\star$.  However, $\rho_\star$ is not proportional to $\rho$ in real galaxies. According to the current cosmological model, most of the gravitation is provided by DM, which is more spread-out than the centrally concentrated stars responsible for the visible light. This change can be easily  accommodated within the above formalism. Assuming a moderate change in the stellar to total mass ratio, then (App.~\ref{app:a})
\begin{equation}
  \rho_\star (r)\simeq\rho_\star(r')\, \Big[\frac{\rho(r)}{\rho(r')}\Big]^{1+\gamma},
\label{eq:gamma}
  \end{equation}
with
\begin{displaymath}
\gamma =  \frac{d\ln (\rho_\star/\rho)}{d\ln\rho},
\end{displaymath}
and $r$ and $r'$ are any two radii. In galaxies, both $\rho_\star/\rho$ and $\rho$ drop outward \cite[e.g.,][]{2013ApJ...764L..31K,2015NatPh..11..245I} leading to $\gamma > 0$. An educated guess (App.~\ref{app:a}) renders $\gamma\sim 0.3$, which corresponds to a change of one order of magnitude in $\rho_\star/\rho$ when $\rho$ changes by three orders of magnitude.  Thus, considering the difference between stellar mass and total mass is just a matter of replacing $f(x,m)$ in Eq.~(\ref{eq:needlabel}) with $g(x,m,\gamma)$,
\begin{equation}
  g(x,m,\gamma) = 2\,\int_x^\infty \frac{s\,[\psi(s)]^{(1+\gamma)m}\,ds}{\sqrt{s^2-x^2}},
  \label{eq:abeldirect2}
\end{equation}

In order to illustrate the range of shapes described by the \propol s, Fig.~\ref{fig:propols} shows $f$ and $g$ for $m$ from 1 to 9 when $\gamma =0$ (a) and  $\gamma =0.3$ (b). As it happens with the polytropes, all \propol s have a {\em core} (a central plateau) and their size is finite when $m<5$ (see Fig.~\ref{fig:propols}a).  As expected, the profiles drop with projected radius much faster when   $\gamma \not= 0$ (cf. panels a and b in Fig.~\ref{fig:propols}). 
The characteristic length scales corresponding to theses polytropes are shown in Fig.~\ref{fig:propol_re}. We include the effective radius $R_e$ (i.e., the half-mass radius of the \propol) as well as the core radius $R_c$, defined as the radius where $f$ (or $g$) drops to 90\,\% of its maximum value. Given $m$, both $R_e$ and $R_c$ scale linearly with $b$.  Indexes $m > 5$ are not shown since these \propol s have infinite mass and their $R_e$ is undefined (actually, it is infinite). Note that  $R_e$ is smaller but of the order of $b$, and that the ratio $R_e/R_c$ spans from 2 to 4 for $m$ from 1 to 5. The dashed lines in Fig.~\ref{fig:propol_re} correspond to varying stellar to total mass ratio with $\gamma=0.3$.

%%%%%
\section{Code to fit projected polytropes}\label{sec:code}

Fitting mass surface densities with \propol s requires evaluating $f(x,m)$ (or $g[x,m,\gamma]$) many times. Since there is no general analytic expression for the polytropes, neither is there for $f$ (or $g$).  In order to speed up the fitting procedure, the evaluation of $f$ was carried out by interpolating on a precomputed grid $f(x_i,m_i)$. We pre-compute several grids suited to the particular problem to be treated. We include grids covering different ranges of $m$, different sampling intervals, and different values of $\gamma$.  Their main characteristics are listed in Table~\ref{tab:grids}. The calculation of the grids and the assessment of uncertainty are described in App.~\ref{app:b}. From these tests, we know that the \propol\ shape, $f$, is evaluated with a relative precision better than $10^{-6}$, which suffices for most practical applications since other ubiquitous sources of error (from interpolation to observational uncertainties) are  larger.
\begin{deluxetable*}{lcccc}
\tablenum{1}
\tablecaption{Pre-computed grid for \propol\  fitting\label{tab:grids}}
\tablewidth{0pt}
\tablehead{
\colhead{Name} & \colhead{Range} & \colhead{Sampling}&\colhead{$\gamma$}&\colhead{Section}\\
& \colhead{$m$} & \colhead{$\Delta m$}&& \\
(1) & (2) & (3) & (4) & (5)
}
%\decimalcolnumbers
\startdata
      {\bf grid0} & 1--10 & 0.01 ($2 \le m \le 5$) &0 &  --- \\
      && 0.1 (elsewhere) &&\\
      {\bf grid1} & 2-- 5 & 0.01 &0 &---\\
      {\bf grid2} & 1 -- 10 & 0.01 ($3 \le m \le 6$) &0 & \ref{sec:sersic} \& \ref{sec:fit_propol_galax}\\
      && 0.1 (elsewhere) &&\\
      {\bf grid2g1} & & &0.1 &\ref{sec:fit_propol_galax}\\
      {\bf grid2g3} & & &0.3 &\ref{sec:sersic}\\ 
     % {\bf grid2g1} & 1 -- 10 & 0.01 ($3 \le m \le 6$) &0.1 &\\ && 0.1 (elsewhere) &&\\
     %{\bf grid2g3} & 1 -- 10 & 0.01 ($3 \le m \le 6$) &0.3 &\\ && 0.1 (elsewhere) &&\\
      {\bf grid3} & 1 -- 100 & 0.1  ($1 \le m \le 7$) &0 & \ref{sec:sersic} \& \ref{sec:fit_propol_galax}\\
      && 1 (elsewhere) && \\
      {\bf grid3g1} & & &0.1 & ---\\
      {\bf grid3g3} & & &0.3 & ---\\
      %{\bf grid3g1} & 1 -- 100 & 0.1  ($1 \le m \le 7$) &0.1 & \\ && 1 (elsewhere) && \\
     %{\bf grid3g3} & 1 -- 100 & 0.1  ($1 \le m \le 7$) &0.3 & \\ && 1 (elsewhere) && \\
      {\bf grid5} & 1 -- 1000 & 0.1  ($1 \le m \le 7$) &0 & \ref{sec:fit_propol_galax} \\
      && 1 ($7 \le m \le 100$) && \\
      && 10 ($100 \le m \le 1000$) && \\
      {\bf grid5g3} & & &0.3 & ---\\
\enddata
\tablecomments{The \propol\ $g(x,m,\gamma)$ is defined in Eq.~(\ref{eq:abeldirect2}). Note that $f(x,m)=g(x,m,0)$. 
The sampling in  $x$ is the same in all grids, and it goes from 0.01 to 200 equi-spaced in $\log x$ steps of 0.02. (1) Grid name. (2) range of $m$ covered by the grid. (3) Sampling interval in $m$. (4) Exponent of the relation stellar to total mass; see Eq.~(\ref{eq:gamma}). (5) Section of the paper where the grid is employed. 
}
\end{deluxetable*}

We use a least squares approach to carry out the fits. The final least squares algorithm minimizes the merit function,
\begin{equation}
  \chi^2=\sum_i\,\big[y_i-\log\,a-\log f(R_i/b,m)\big]^2,
  \label{eq:meritf}
  \end{equation}
  where $y_i$ is the log surface density observed at $R = R_i$, and the subscript $i$ varies from 1 to the number of observed radii in the profile. The minimization of the merit function is carried out using the classical Levenberg-Marquardt minimization algorithm \citep{1986nras.book.....P} as implemented in {\tt python} ({\tt scipy.least\_squares}). The free parameters to be determined are $m$, $\log a$, and $b$, and their formal errors are estimated from the diagonal of the covariance matrix. While fitting, the \propol\  are computed by 2D interpolation on the selected grid (Table~\ref{tab:grids}). We use a linear 2D interpolation as provided by {\tt interp2d} in {\tt python} ({\tt scipy.interpolate}).

%%% 
\section{Comparison of projected polytropes and S\'ersic profiles}\label{sec:sersic}
As we point out in Sect.~\ref{sec:intro}, the mass within galaxies drops with radial distance following a law  approximately given by \sersic\ functions \citep{1968adga.book.....S,1993MNRAS.265.1013C,2005PASA...22..118G},
\begin{equation}
  \Sigma(R)= \Sigma(0)\,\exp\large[-c_n\,(R/R_e)^{1/n}\large],
  \label{eq:sersicdef}
\end{equation}
with $\Sigma(R)$ the mass surface density at a distance $R$ from the center,  $R_e$ the radius enclosing half of the mass, and $c_n$ a constant which only depends on the parameter $n$, called \sersic\ index. The \sersic\ index $n$ controls the shape of the profile, and has been observed to vary mostly from 0.5 to 6 \citep{2003ApJ...594..186B,2012ApJS..203...24V}, going from disk-like galaxies to elliptical galaxies, and from low-mass dwarfs to the brightest galaxy in a cluster. As we also explain in Sect.~\ref{sec:intro},  the range of shapes includes exponential disks \citep[$n=1$;][]{1994A&AS..106..451D} and the {\em de Vaucouleurs} $R^{1/4}$-law characteristic of massive ellipticals \citep[$n=4$;][]{1948AnAp...11..247D}.

The question arises as to whether the empirical \sersic\ profiles and the theoretical \propol s are equivalent. All \propol s have cores, inherited from their parent polytrope (see Fig.~\ref{fig:propols}). \sersic\ profiles also present cores, but of different nature. Except when $n\lesssim 1$, they are too small to be comparable to the cores of the \propol s. This mismatch in the cores can be shown analytically: the logarithmic derivative of the surface density given by a \sersic\ function (Eq.~[\ref{eq:sersicdef}]) can be approximated as \citep[e.g.,][]{2005PASA...22..118G},
\begin{equation}
  \frac{d\ln\Sigma(R)}{d\ln R} \simeq -(2-0.33/n)\cdot (R/R_e)^{1/n}, 
\end{equation}
which tends to zero when $R\rightarrow 0$, but very slowly when $n\gtrsim 1$ because of the $1/n$ exponent.  Defining the core radius as the radius where the logarithmic derivative of the profile is small enough   \citep[0.3 is often used; e.g.,][]{2015AJ....149..180O},
then the resulting core radius  $R_{0.3}$, implicitly defined as
\begin{equation}
\frac{d\ln\Sigma(R_{0.3})}{d\ln R} \simeq  -0.3, 
\end{equation}
turns out to be comparable with the effective radius for $n=0.5$ ($R_{0.3}/R_e\simeq 0.5$) but becomes tiny  for $n=5$ ($\simeq 10^{-4}$).   % see page 141 of my handwritten notes
Therefore, except for $n\lesssim 1$, the \propol s cannot reproduce the \sersic\ profiles for $R \lesssim R_e$.

\begin{figure}
  \centering 
\includegraphics[width=0.9\linewidth]{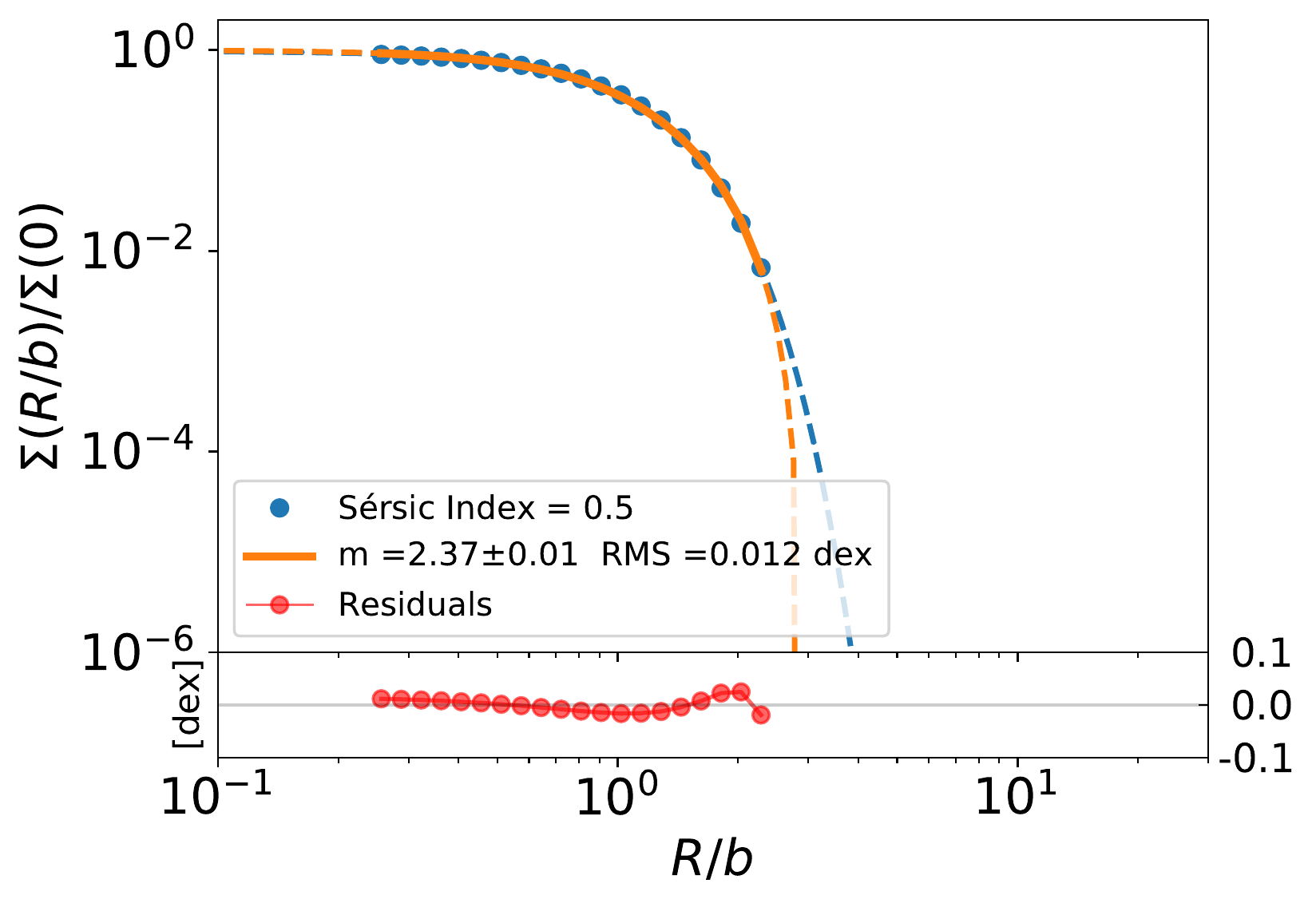}
\includegraphics[width=0.9\linewidth]{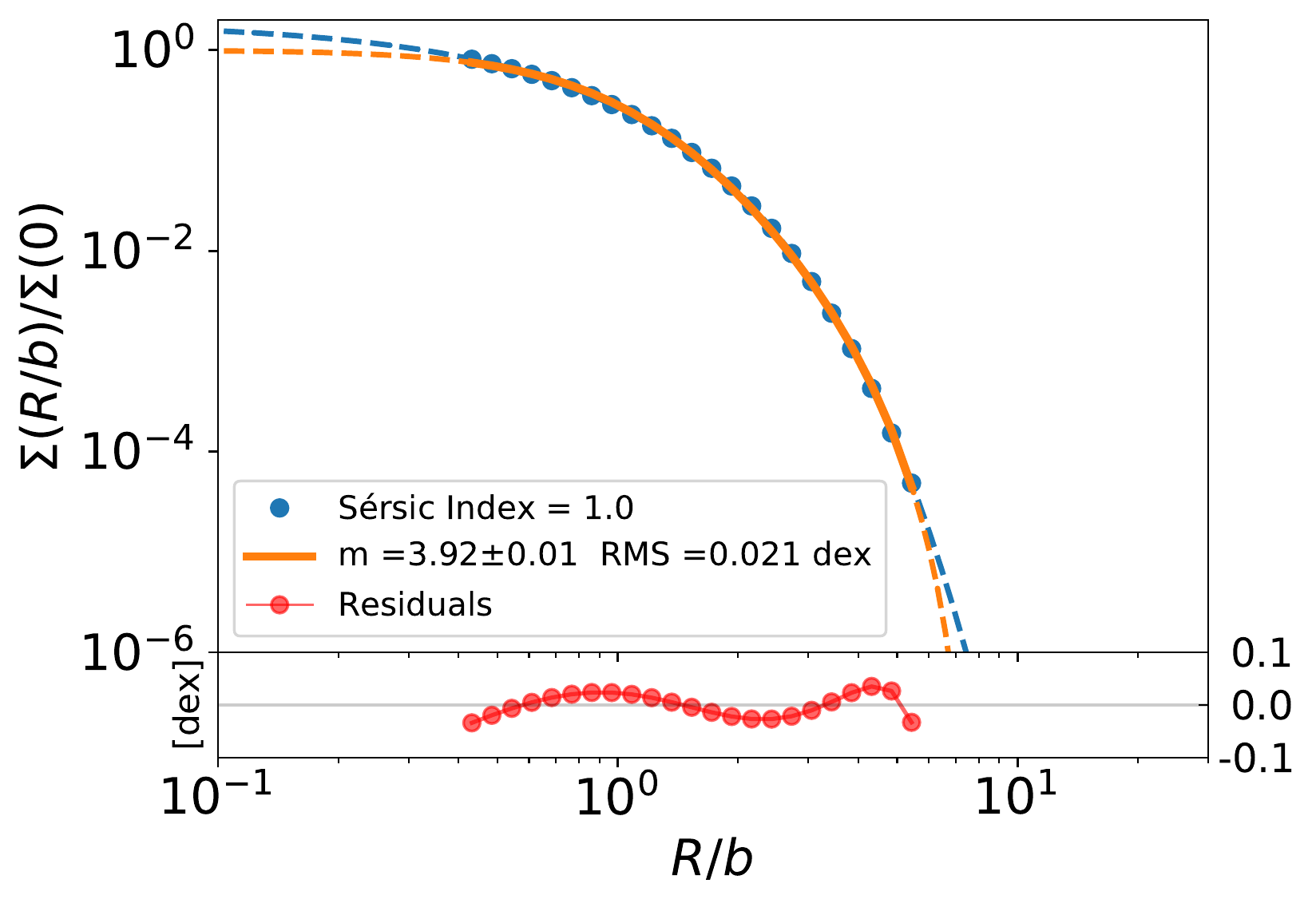}
\includegraphics[width=0.9\linewidth]{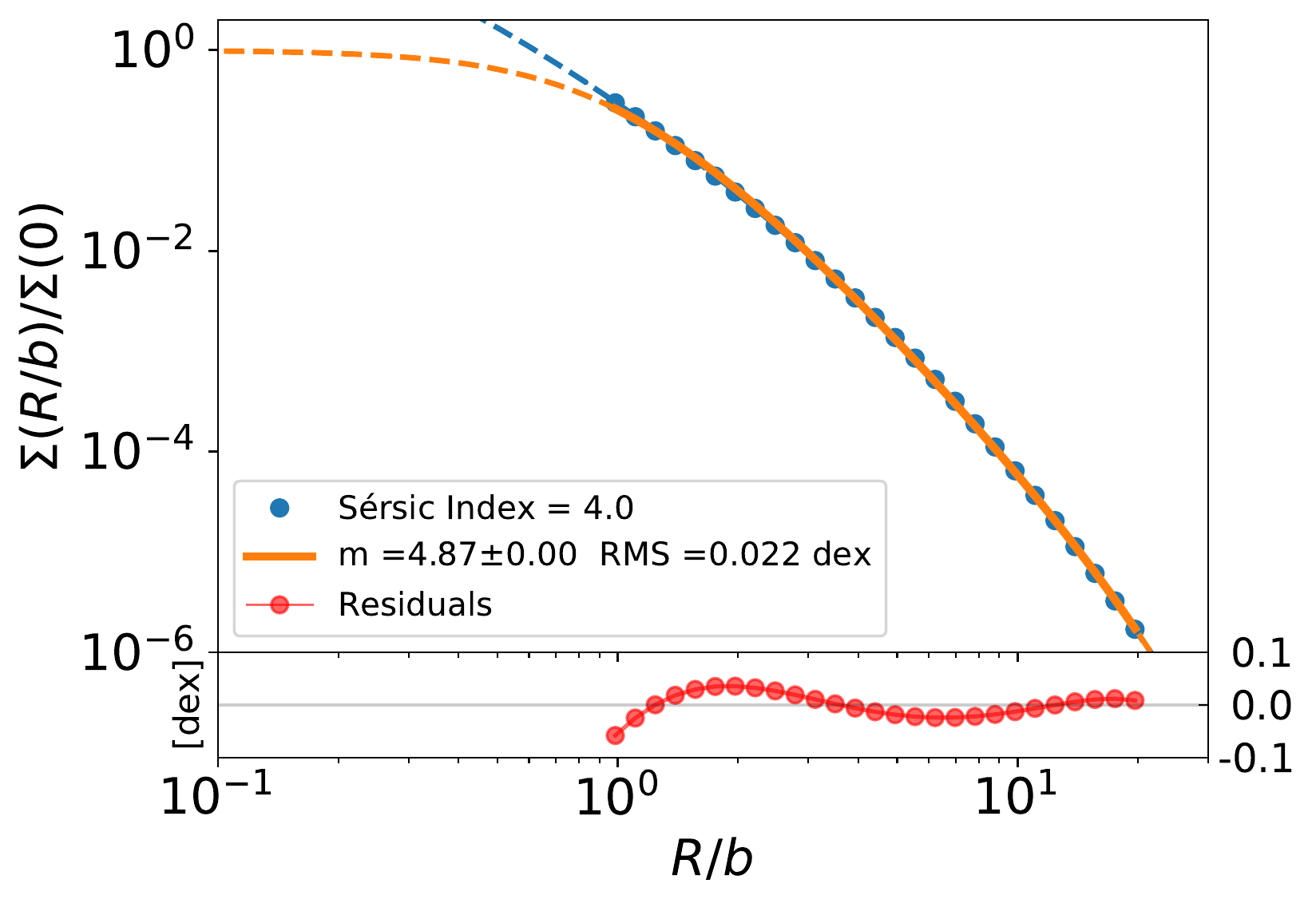}
\caption{Representative \sersic\ profiles (the blue symbols and lines) fitted with \propol s (the orange lines). The   upper, middle, and bottom panels correspond to $n=0.5$, 1, and 4, respectively.  The symbols represent the range of radii included in the fit. Except for $n=0.5$, cores have been excluded since the cores of \sersic\ profiles  and \propol s are not commensurate (see main text). Outside the core, the agreement is well within any realistic observational error (see the residuals shown in dex in the secondary panels and the RMS of the fit included in the insets). Note that the fitted range comprises a factor 20 in radii and up to a factor of $10^5$ in surface density.  The original \sersic\ index and the corresponding polytropic index are given in the insets. For display purposes, all profiles are normalized to the central surface density of the resulting \propol\ and to its radial scale-length ($b$). The equivalence between $b$ and $R_e$ is given in Fig.~\ref{fig:propol_re}.}
\label{fig:lane_emden_fitsersic}
\end{figure}

\begin{figure*}
  \centering
  \plottwo{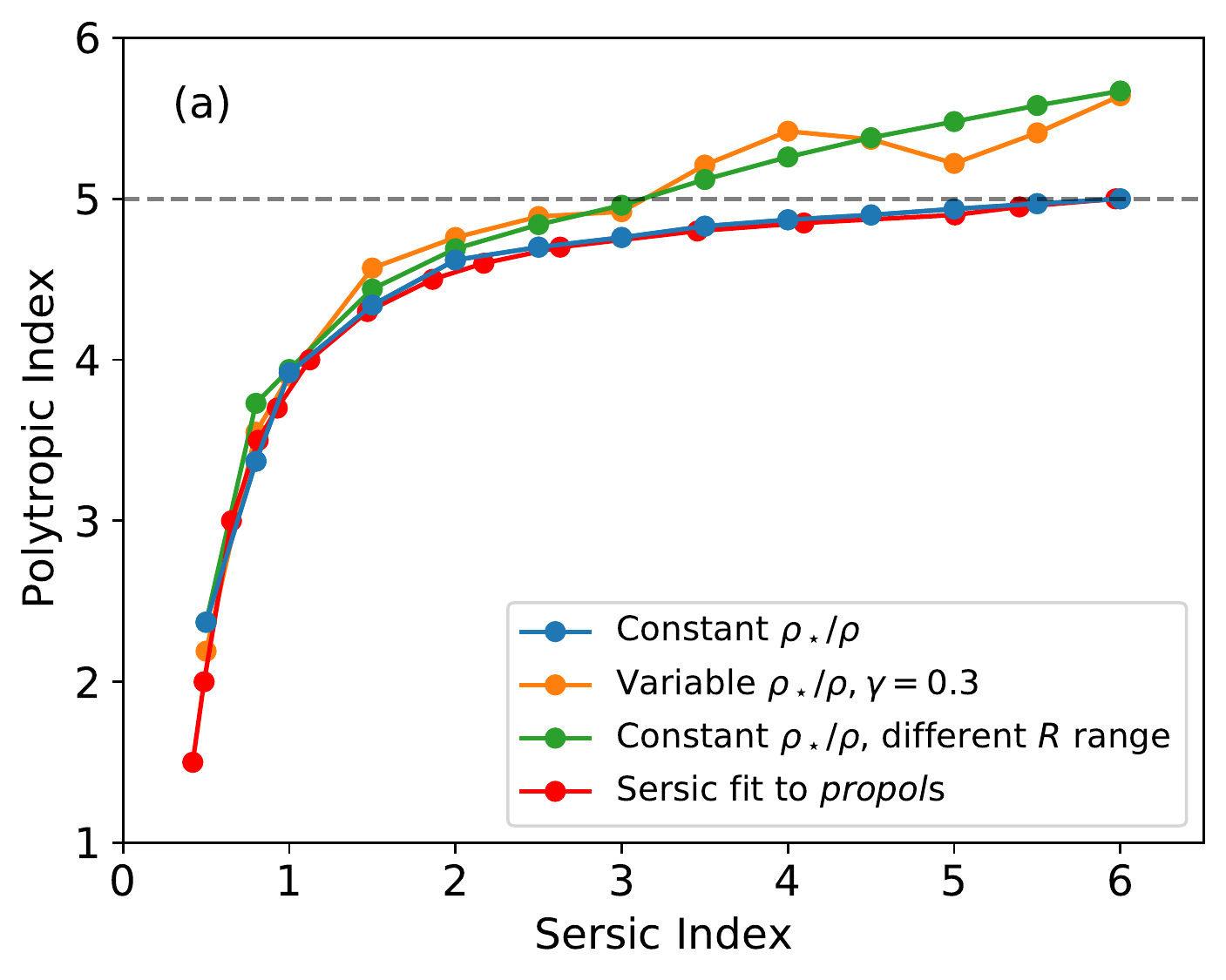}{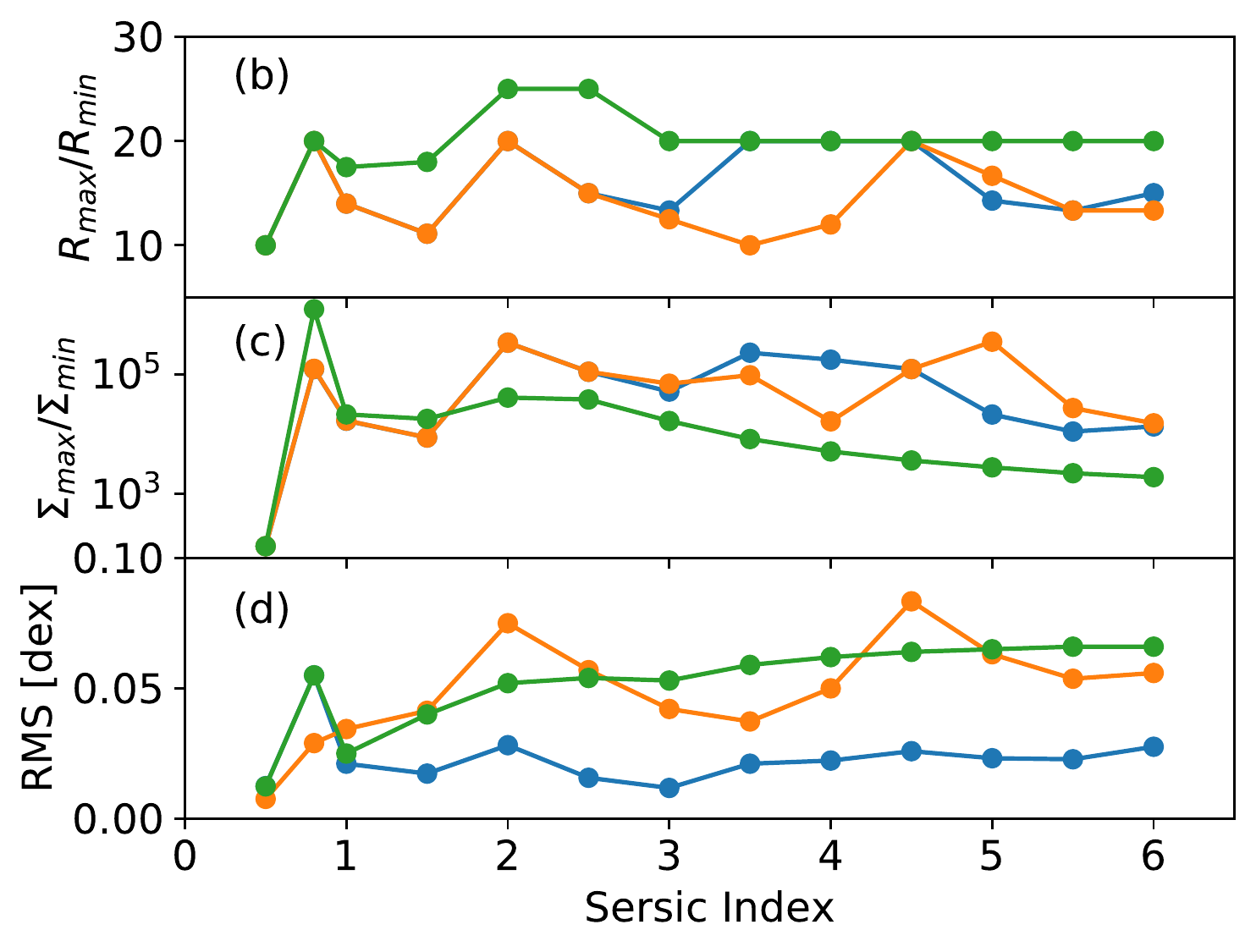}
\caption{
Fitting \sersic\ profiles using \propol s. (a) Polytropic index corresponding to each \sersic\ Index. The blue lines and symbols corresponds to fits based on {\bf grid2} in Table~\ref{tab:grids} and considering a range of radii optimized to minimize the RMS (see panel d). The orange line assumes that stellar mass and total mass do not scale with each other (fits based on  {\bf grid2g3}). The green line uses {\bf grid2} but the range of radii also includes part of the cores in the \propol s (see Fig.~\ref{fig:lane_emden_fitsersic_app} in App.~\ref{app:c}). Finally, the red line reversed the fit. Rather than fitting \propol s to \sersic\ funcions, we fit \sersic\ functions to the \propol s corresponding to the blue lines. The horizontal dashed line points out the limiting value $m=5$. (b) Ratio between the maximum and minimum radii used for fitting ($R_{\rm max}/R_{\rm min}$). (c) Ratio between the maximum and the minimum surface density of the fitting \propol\ ($\Sigma_{\rm max}/\Sigma_{\rm min}$). (d) RMS of the fits. Note that the best fits (blue lines) have an RMS around 0.02, which corresponds to a relative error of only 5\,\% over the full range of fitted radii. The color code is the same in all panels.
}
\label{fig:lane_emden6}
\end{figure*}

As it has been evidenced elsewhere \citep[e.g.,][]{2004AJ....127.1917T} and we will show in Sect.~\ref{sec:galaxies}, rather than a problem for the \propol s, this mismatch often reveals the disability  for the \sersic\ profiles to reproduce some of the central cores observed in galaxies. Thus, to study the equivalence between \sersic\ profiles and \propol s, we will fit \sersic\ profiles with \propol s using the code presented in Sect.~\ref{sec:code}, but avoiding   the central cores when $n\gtrsim 1$. 
Figure~\ref{fig:lane_emden_fitsersic} shows examples of such fits. The range of fitted radii is indicated in the figure using symbols. Even if the cores are excluded, the fits still comprise a factor 20 in radii and up to a factor of $10^5$ in surface density. Within this range, the agreement is well within any realistic observational error (see the residuals shown in dex in the secondary panels). Figure~\ref{fig:lane_emden_fitsersic} only shows 3 representative \sersic\ profiles.
The results for the whole range of observed \sersic\ indexes is summarized in Fig.~\ref{fig:lane_emden6}. The blue symbols and lines represent the fits carried out using {\bf grid2} (Table~\ref{tab:grids}) and avoiding the cores as shown in Fig.~\ref{fig:lane_emden_fitsersic}. We note that the agreement between \sersic\ profiles and \propol s is excellent,
with the root-mean-square (RMS) of the residuals well below any realistic observational error: around 0.02 dex or 5\,\% (Fig.~\ref{fig:lane_emden6}d) over 1 order of magnitude in radius (Fig.~\ref{fig:lane_emden6}b) and 5 orders of magnitude in surface density (Fig.~\ref{fig:lane_emden6}c).

The correspondence between the original \sersic\ index and the polytropic index of the fitted \propol\ is shown in Fig.~\ref{fig:lane_emden6}a. Interestingly, the range of observed \sersic\ indexes, $0.5 \lesssim n \lesssim 6.0$ \citep[e.g.,][]{2001MNRAS.321..269T,2003ApJ...594..186B,2011A&A...530A.106T,2012ApJS..203...24V}, gives rise to a range of polytropic indexes between 2 and 5  (the blue symbols and lines), right in the range where polytropes are physically sensible (Sect.~\ref{sec:poly}, Eq.~[\ref{eq:nlimits}]). In other words, the range of allowed polytropic indexes (Eq.~[\ref{eq:nlimits}]) corresponds to \sersic\ indexes in the observed range, a result whose implications are hard to judge at this point because, when $n\gtrsim  2$, the strict equivalence between \sersic\ profiles and \propol s breaks down when the cores are included.

In order to study the dependence of the above results on details of the fit, we repeat the same exercise varying some of the hypotheses and hyper-parameters that define the fits. Thus, if rather than completely excluding the cores, we partly include them, then the fits worsen and render  polytropic indexes $m >5$ for  $n \gtrsim 3$ (see the green lines and symbols in Fig.~\ref{fig:lane_emden6}a, and App.~\ref{app:c} for examples of fits and residuals). The increase of RMS is significant (from 0.02 to 0.05; cf. the green and blue lines in Fig~\ref{fig:lane_emden6}d). If rather than using pure \propol s, we assume $\rho_\star/\rho$ to vary radially ($\gamma\not = 0$ in Eq.~[\ref{eq:gamma}]), then the fits also get worse: see  Fig.~\ref{fig:lane_emden6}d, the orange line as well as the examples in App.~\ref{app:c}. In this case, {\bf grid2g3} in Table~\ref{tab:grids} was used for fitting. As it happens when the cores are partly included,  $m$ becomes  $>5$ when $n \gtrsim 3$ (Fig~\ref{fig:lane_emden6}a); examples are given in App.~\ref{app:c}. When a different more coarse \propol\ grid is used for fitting (e.g., {\bf grid3} in Table~\ref{tab:grids}) the results are indistinguishable from those shown as blue lines and symbols in Fig.~\ref{fig:lane_emden6}, which are based on  {\bf grid2}. Finally, rather than fitting \propol s to \sersic\ profiles, we also tried fitting \sersic\ profiles to  \propol s. When the range of radii are similar to the range used when fitting  \propol s to \sersic\ profiles, then the fits are equally good. The red line and symbols in Fig.~\ref{fig:lane_emden6}a correspond to this case. Summing up all these results, we conclude that the good correspondence between the outskirts of \sersic\ profiles and \propol s remains valid in quite general terms. Outside the optimal range of radii and $\gamma$, the fits worsen and the best-fitting \propol s start having $m> 5$, which corresponds to polytropes of unbound mass and energy (Sect.~\ref{sec:poly}). These trends also appear when fitting mass profiles in real galaxies (Sects.~\ref{sec:galaxies} and \ref{sec:monsters}) and in globular clusters (Trujillo \&\ S\'anchez Almeida 2021, in preparation).    
%

% 
%%%%
\section{Fitting surface density profiles of real galaxies}\label{sec:galaxies}
Differences between \sersic\ profiles  and \propol s are observed in the center ($R\lesssim R_e$) and outskirts  ($R\gg R_e$), but this mismatch with \sersic\ profiles is also shared with real galaxies. They also deviate from pure \sersic\ functions because of the presence of cores in their centers \citep[e.g.,][]{2004AJ....127.1917T,2017Galax...5...17D}, and because of the existence of truncations and stellar haloes in their outskirts \citep[e.g.,][]{2006A&A...454..759P,2008AJ....135...20E}. Thus, we try direct fits of observed mass density profiles in galaxies to see whether \propol s provide a fair description of the observation. As an exploratory work, we do not intend to reproduce the profiles in detail, rather, we want to assess the overall goodness of the fits and if they are comparable in quality to the traditional fits based on \sersic\ profiles.

%%%%%%%%%%%%
%
\begin{figure}
    \includegraphics[width=\linewidth]{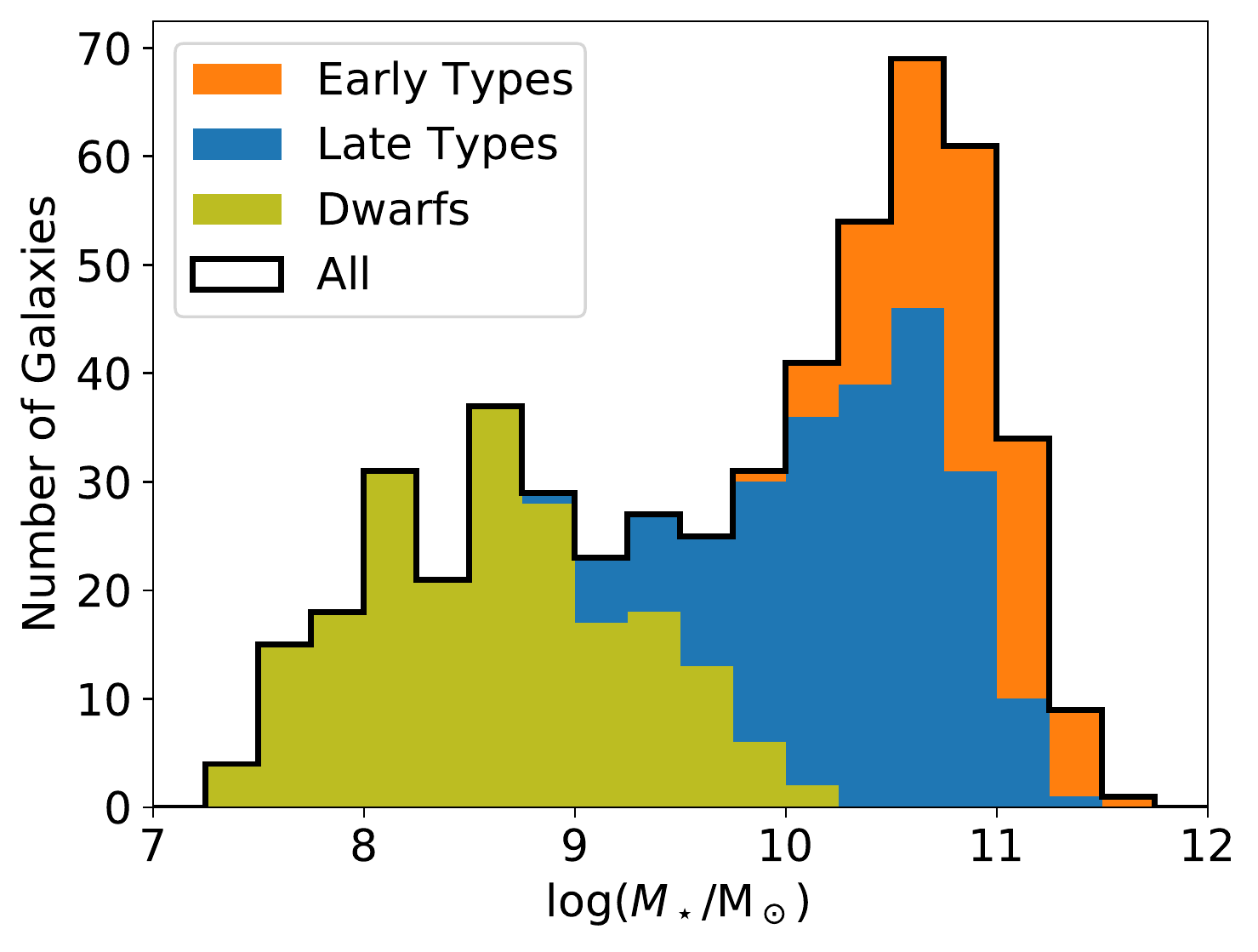}
    \caption{
      Stellar mass of the galaxies used for fitting \propol s. They have been divided into {\em dwarf} galaxies,  
        {\em early} types, and {\em late} types, as indicated in the inset. The dwarfs are low mass galaxies ($M_\star \lesssim 3\times 10^{9}\,{\rm M}_\odot$) from \citet{2013MNRAS.435.2764M} and do not have associated morphological classification, whereas the more massive  {\em early} and {\em late} types have been classified by \citet{2010ApJS..186..427N}.  Early types include S0s and are restricted to high masses ($M_\star \gtrsim 10^{10}\,{\rm M}_\odot$). The stellar masses of the galaxies, $M_\star$, are expressed in solar mass units, ${\rm M}_\odot$.
    }
  \label{fig:histmass}
 \end{figure} 
 %
%%%%%%%%%%%%%%%%%%%%%%%%%%%%
%
\begin{figure*}
  \begin{centering}
    \includegraphics[width=0.4\linewidth]{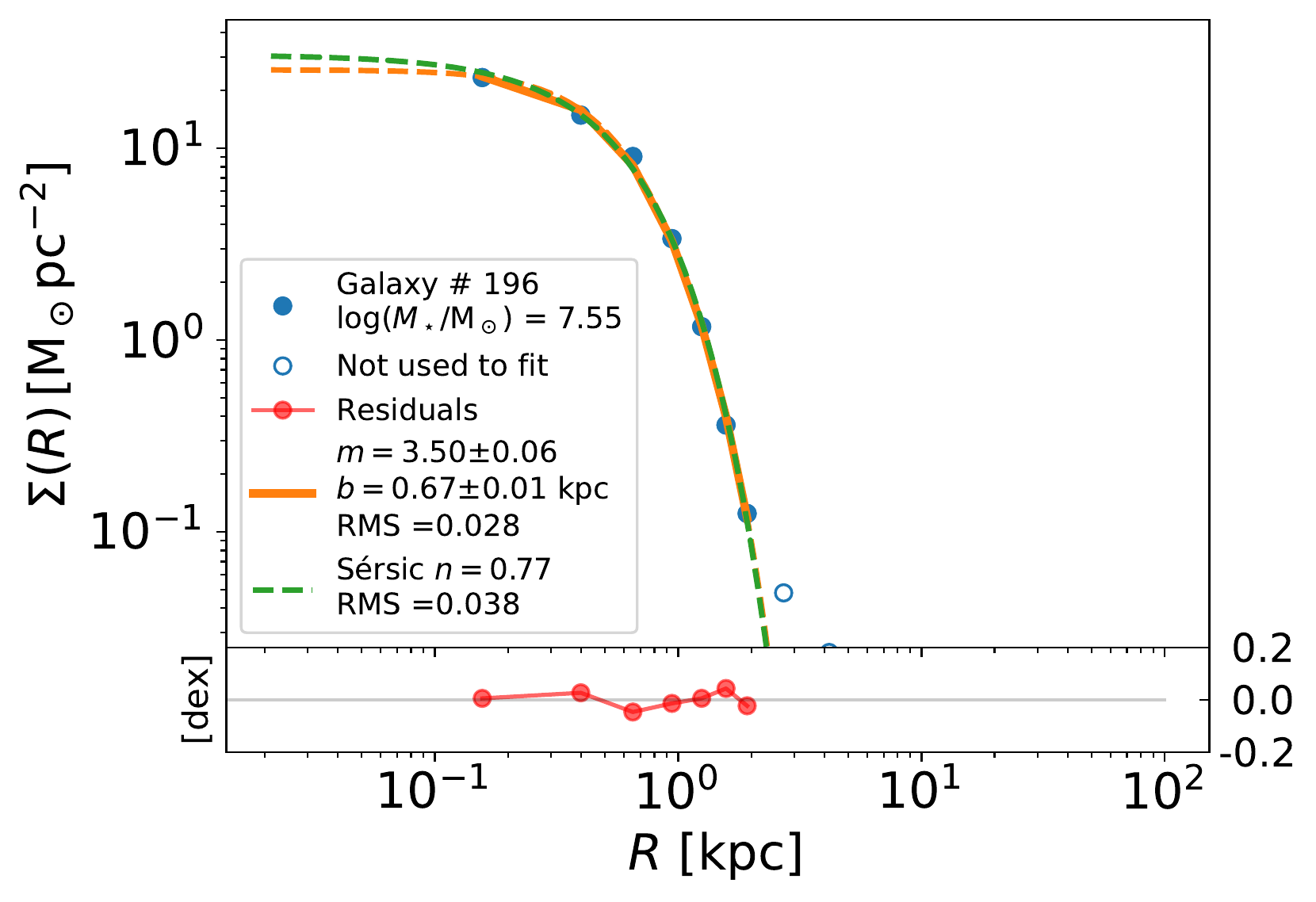}
    \includegraphics[width=0.4\linewidth]{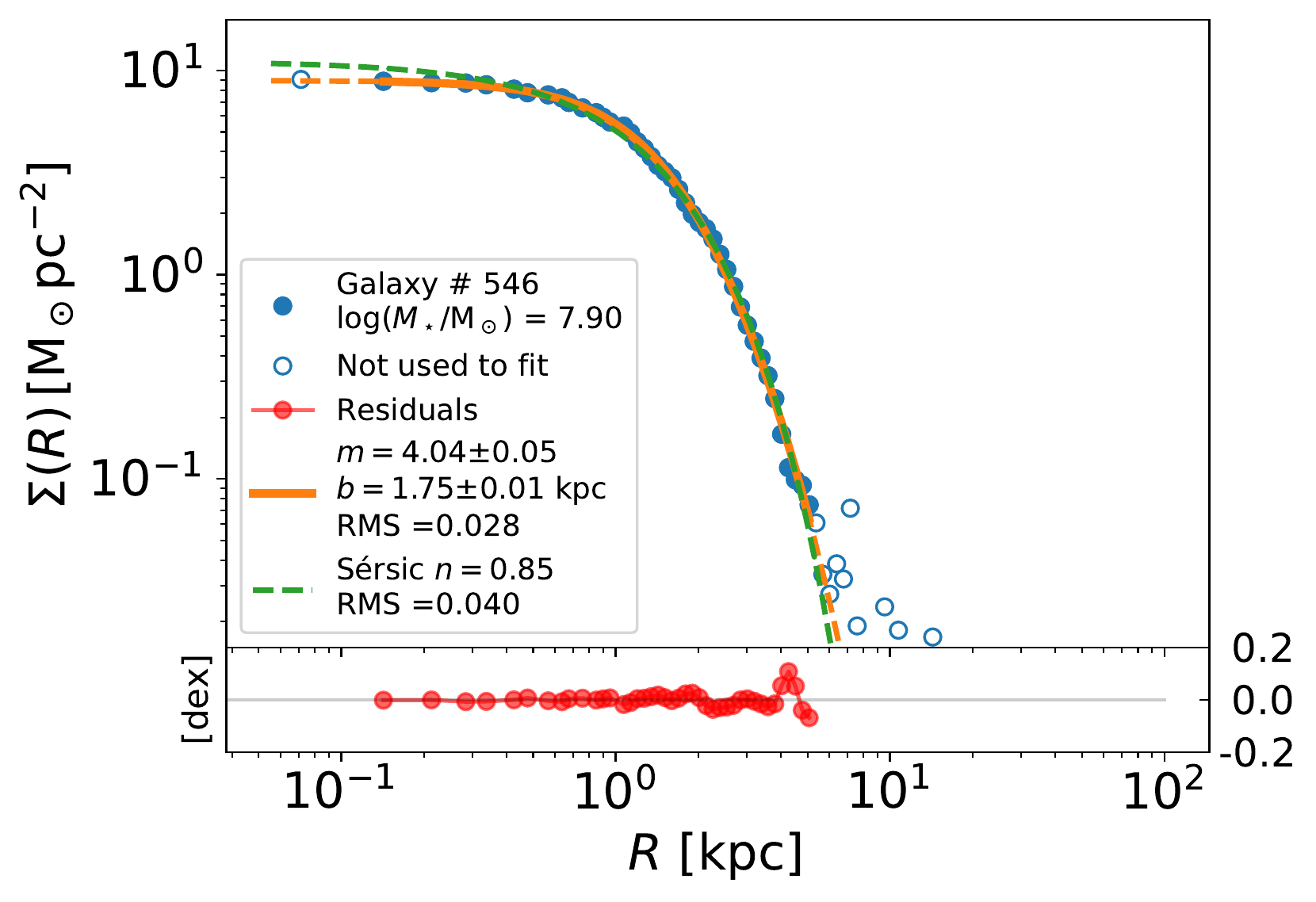}\\
    \includegraphics[width=0.4\linewidth]{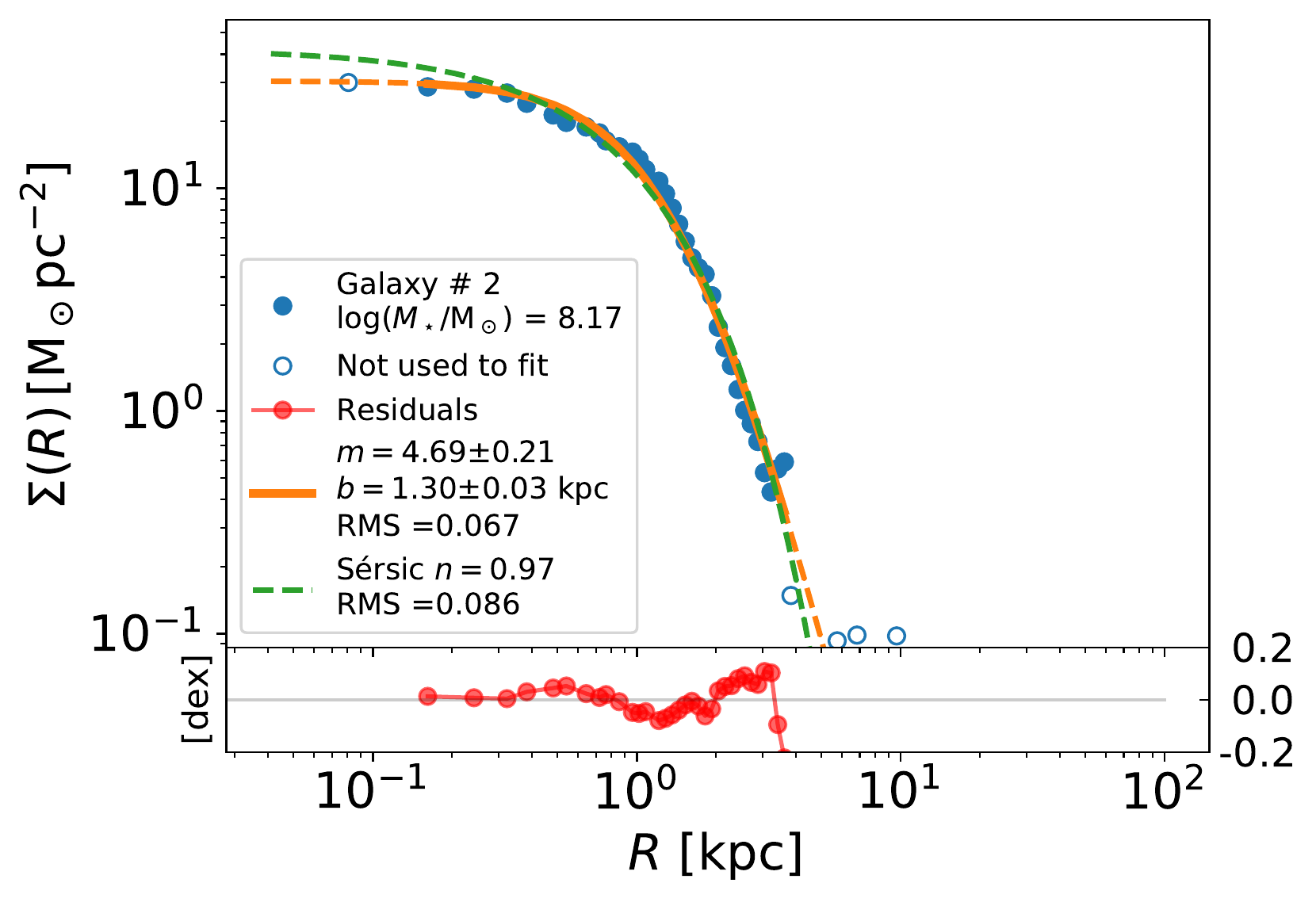}
    \includegraphics[width=0.4\linewidth]{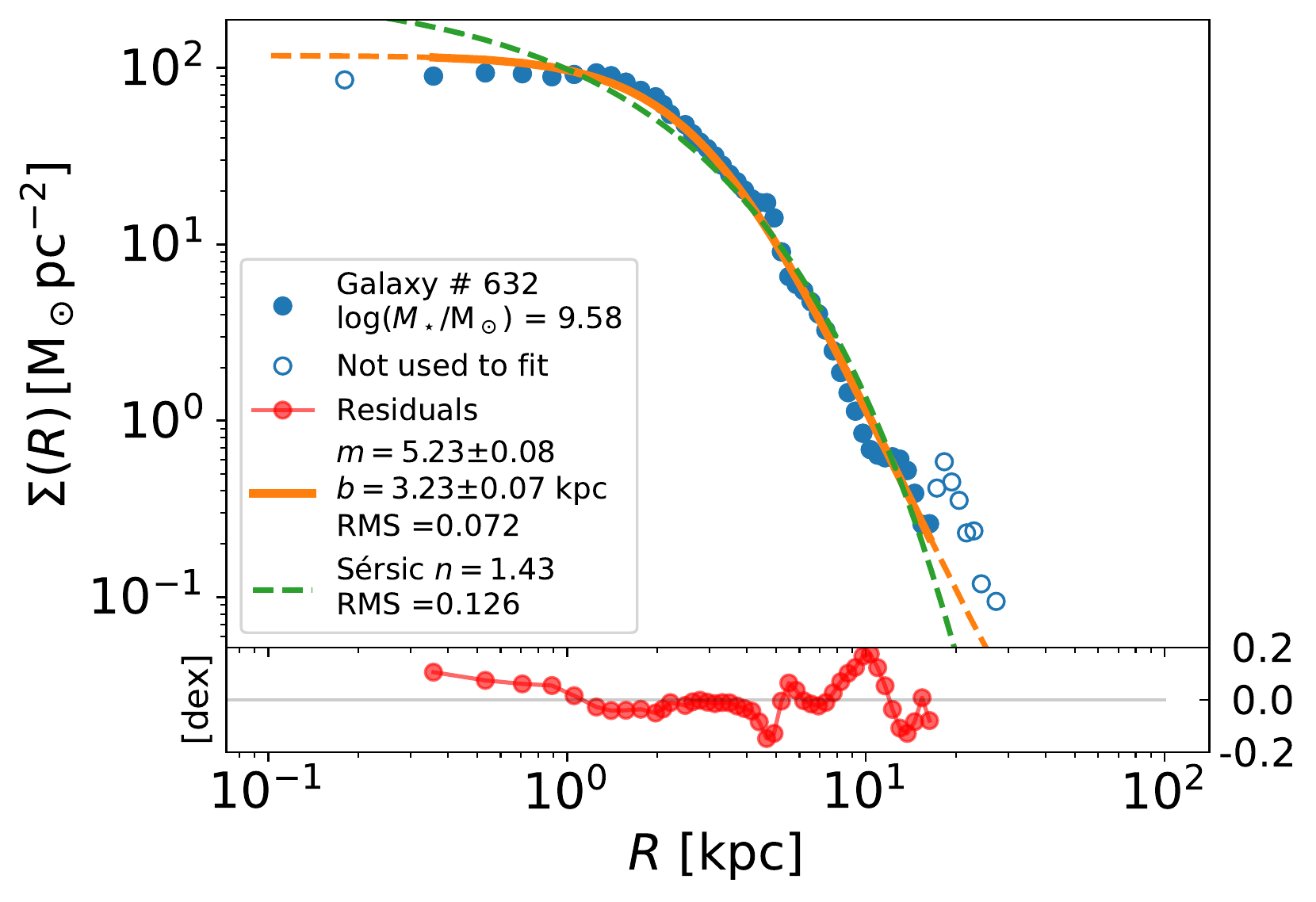}\\
    \includegraphics[width=0.4\linewidth]{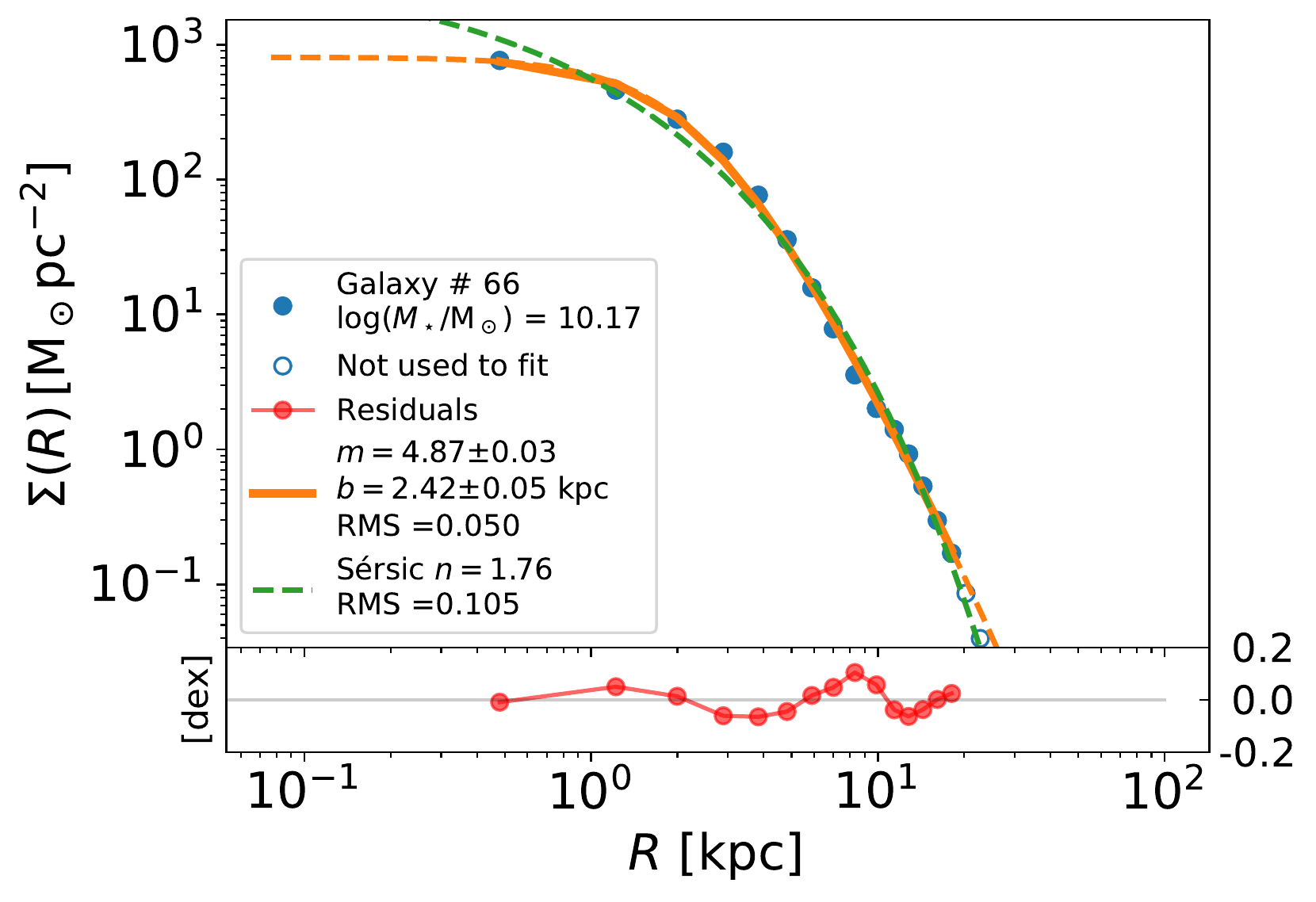}
    \includegraphics[width=0.4\linewidth]{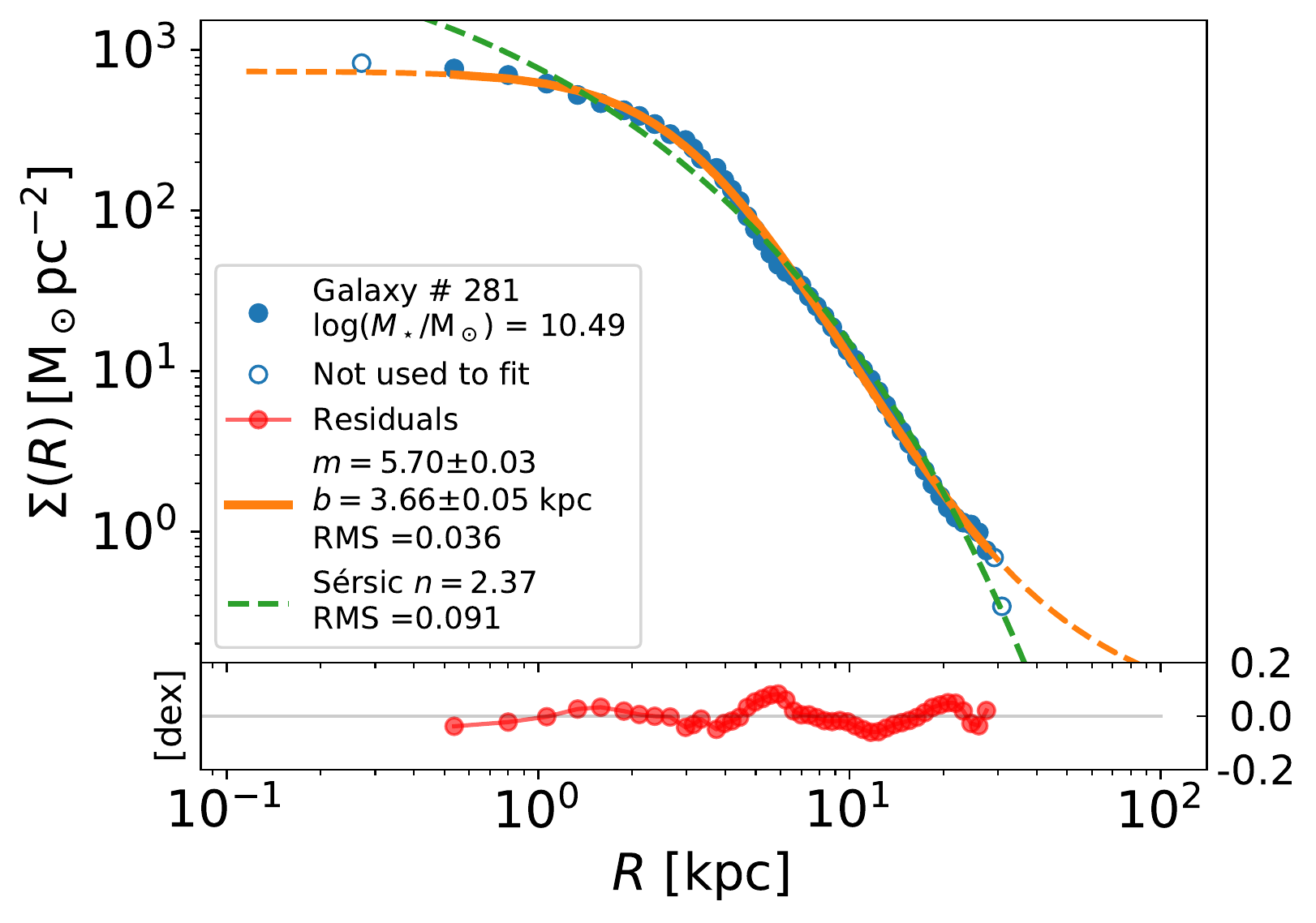}\\
    \includegraphics[width=0.4\linewidth]{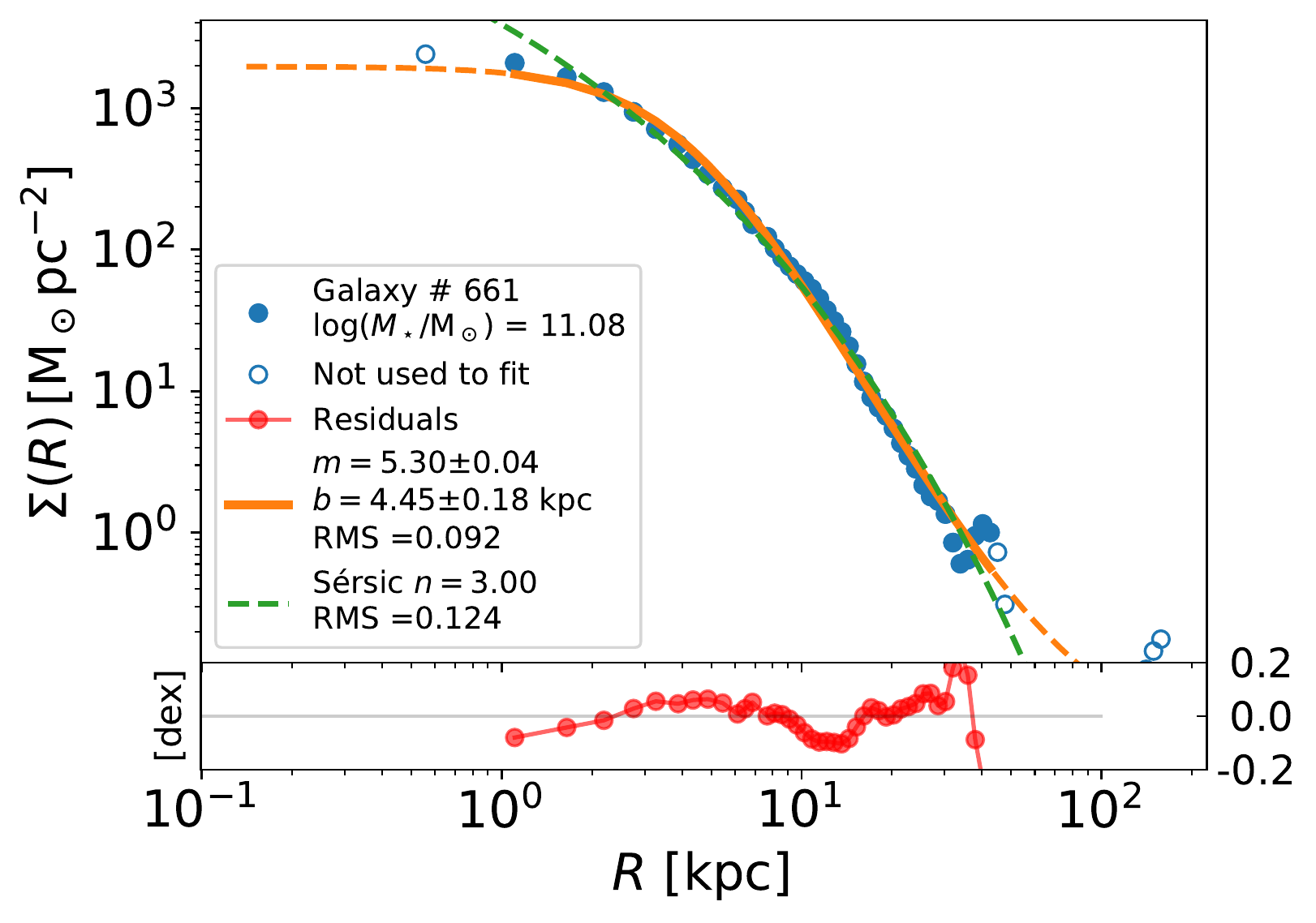}
    \includegraphics[width=0.4\linewidth]{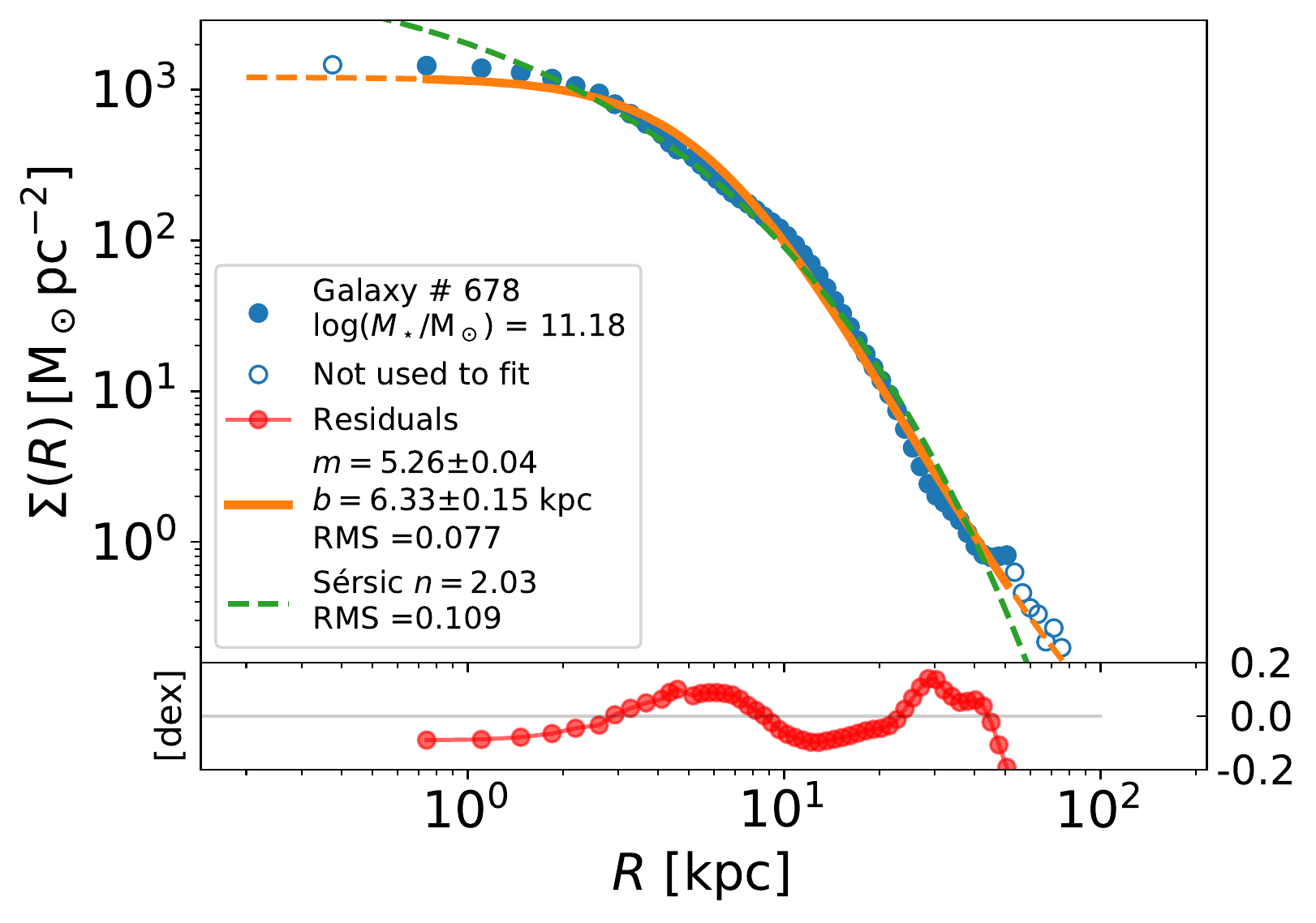}\\
    \end{centering}
    \caption{Examples of good \propol\ fits to observed mass surface density profiles from   \citet[][]{2020MNRAS.493...87T,2020MNRAS.495.3777T}. The observed profiles are represented in blue with the solid symbols marking the fitted points. (Points at the center and outskirts have been excluded to avoid contamination from PSF and noise; see main text for details.) The best fitting \propol s are shown as solid orange lines. The best fitting \sersic\ profiles are included for reference (green dashed line). All these galaxies have been chosen so that the RMS of the \propol\ fit residuals (shown as red symbols and lines) is smaller than the RMS of the sersic fit. 
The main parameters of the fits are included in the insets.
Galaxies are ordered in growing $M_\star$, from top to bottom and from left to right ($7.5 \leq \log[M_\star/{\rm M}_\odot] \leq  11.2$). The range of abscissae and ordinates is the same in all panels. 
    }
  \label{fig:lane_emden_fit_good}
 \end{figure*} 
\begin{figure*}
  \begin{centering}
    \includegraphics[width=0.4\linewidth]{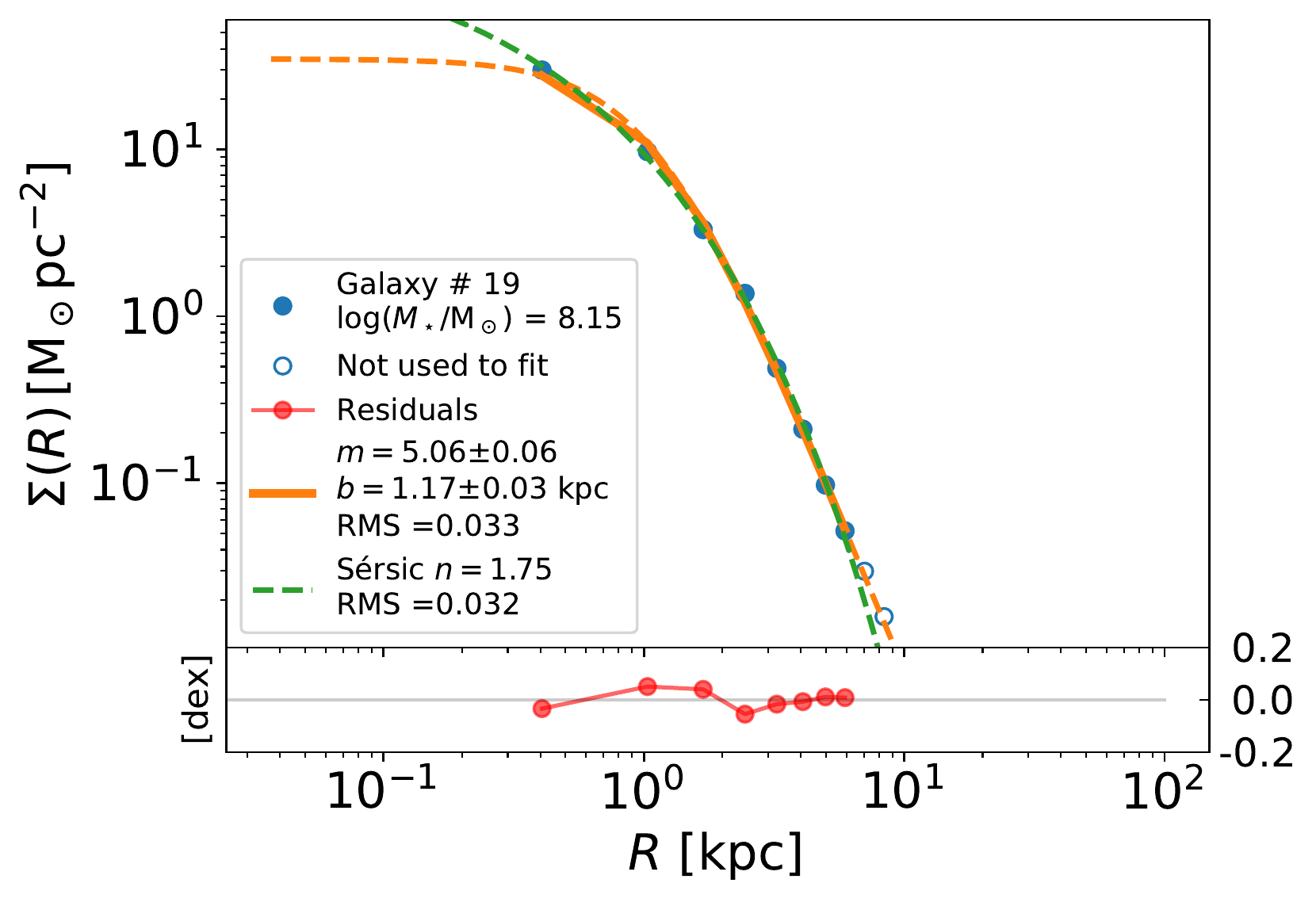}
    \includegraphics[width=0.4\linewidth]{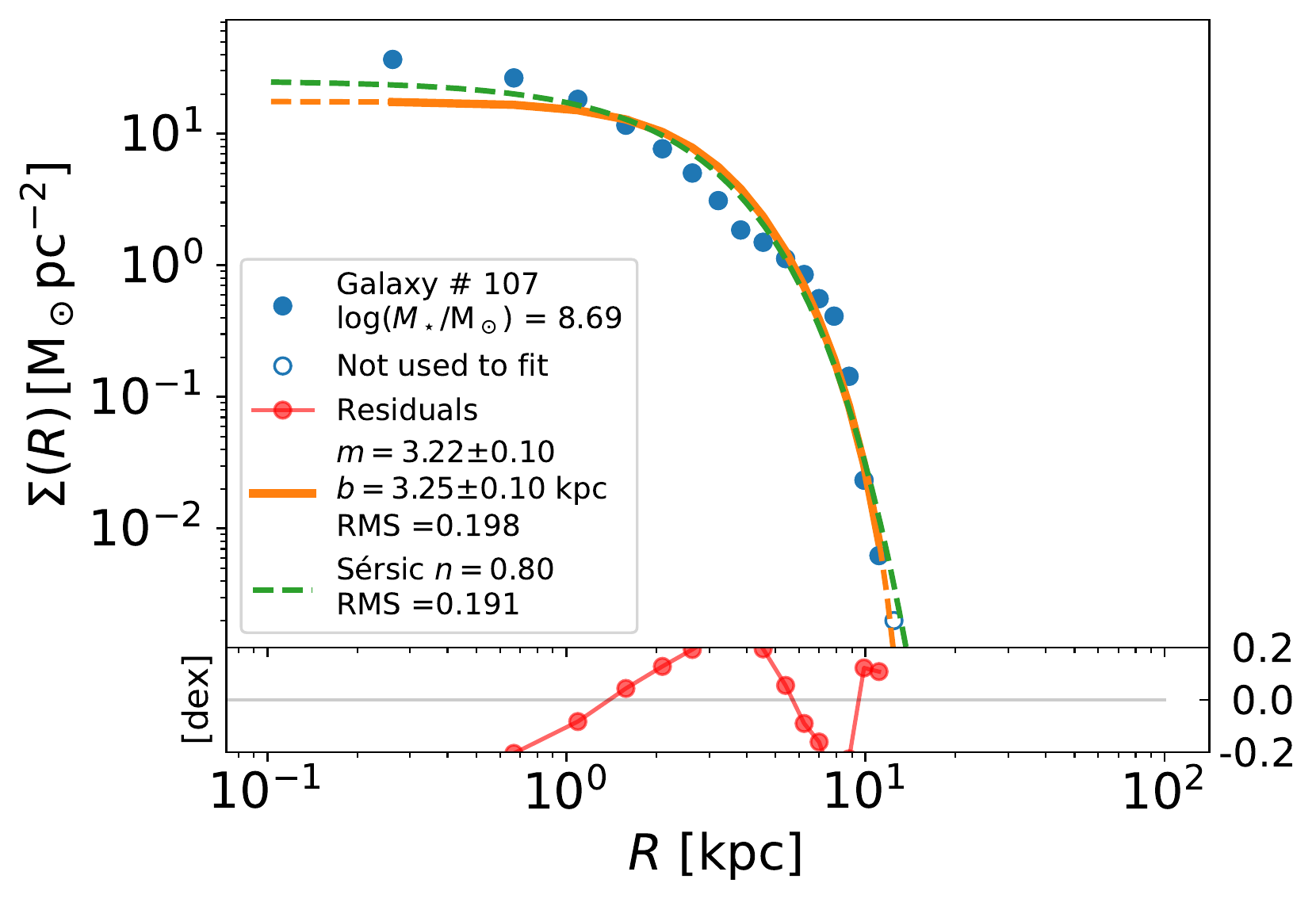}\\
    \includegraphics[width=0.4\linewidth]{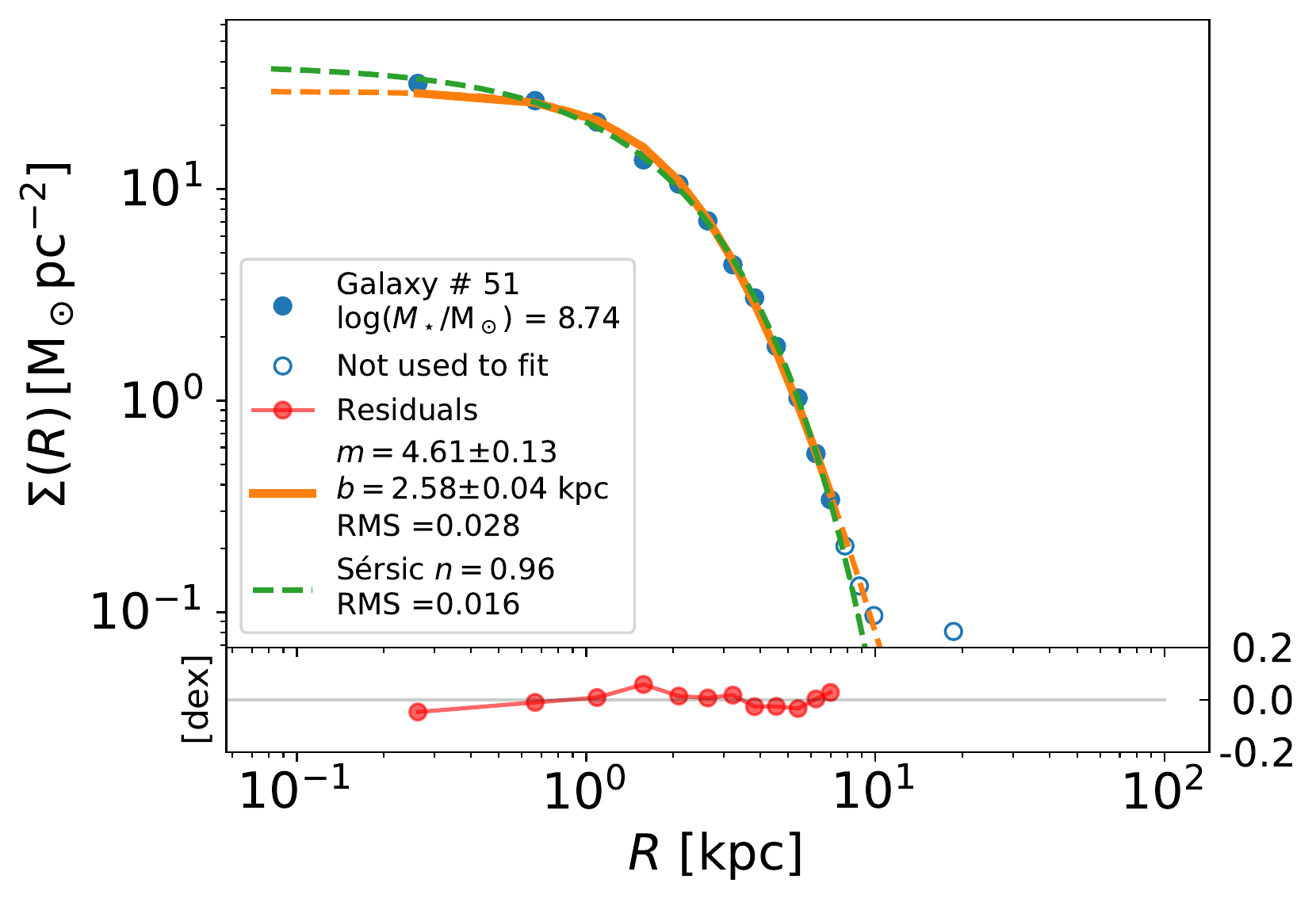}
    \includegraphics[width=0.4\linewidth]{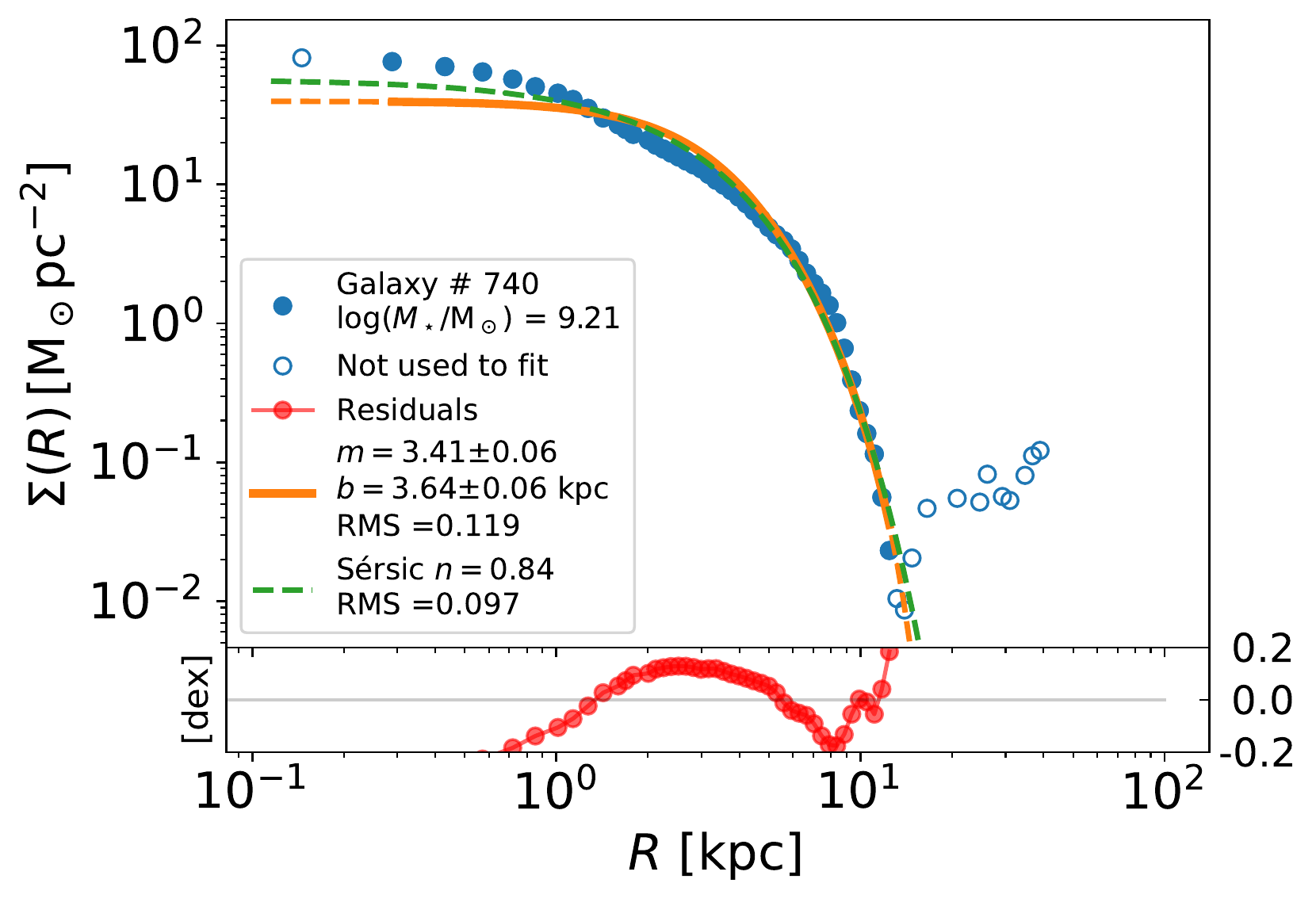}\\
    \includegraphics[width=0.4\linewidth]{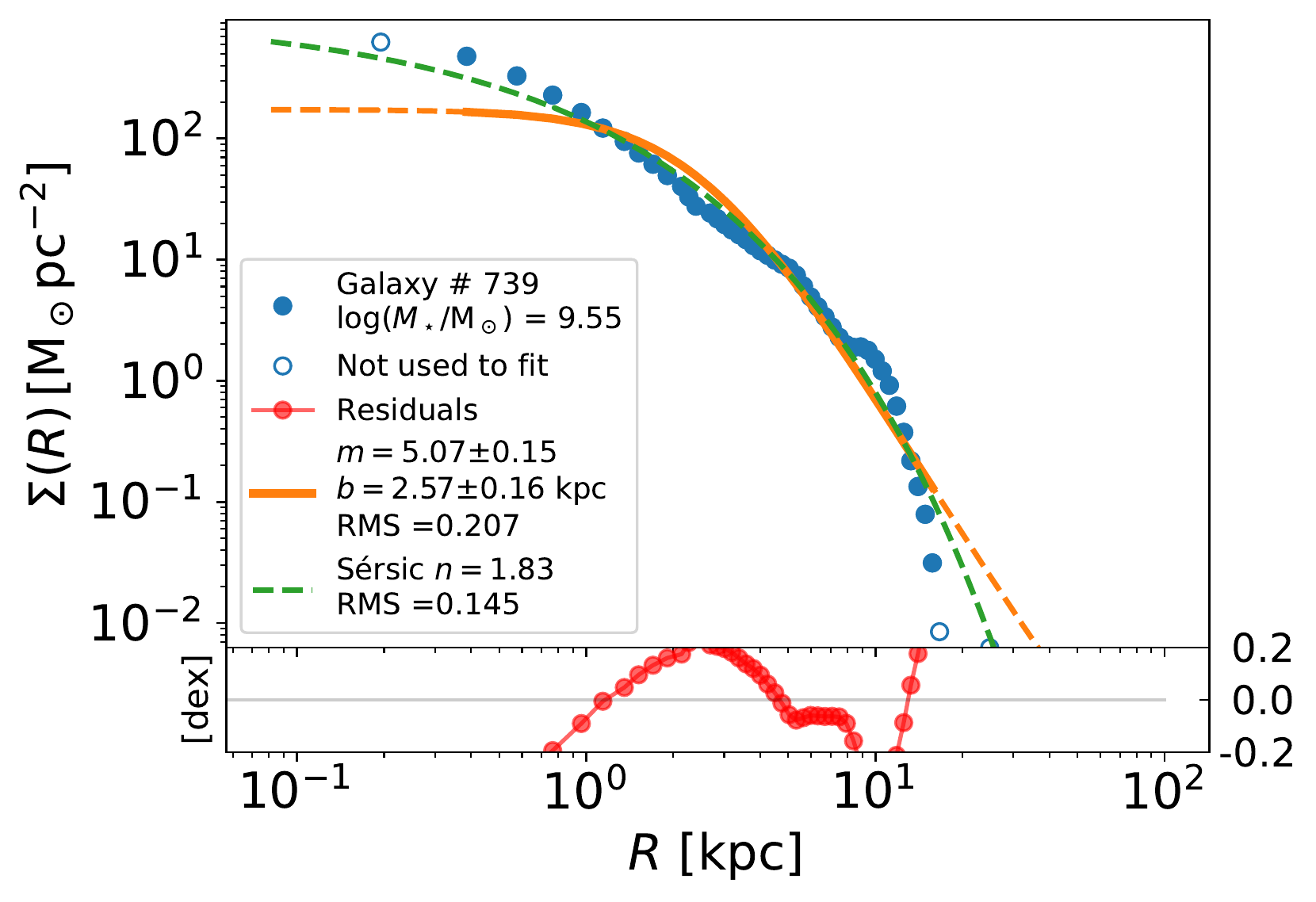}
    \includegraphics[width=0.4\linewidth]{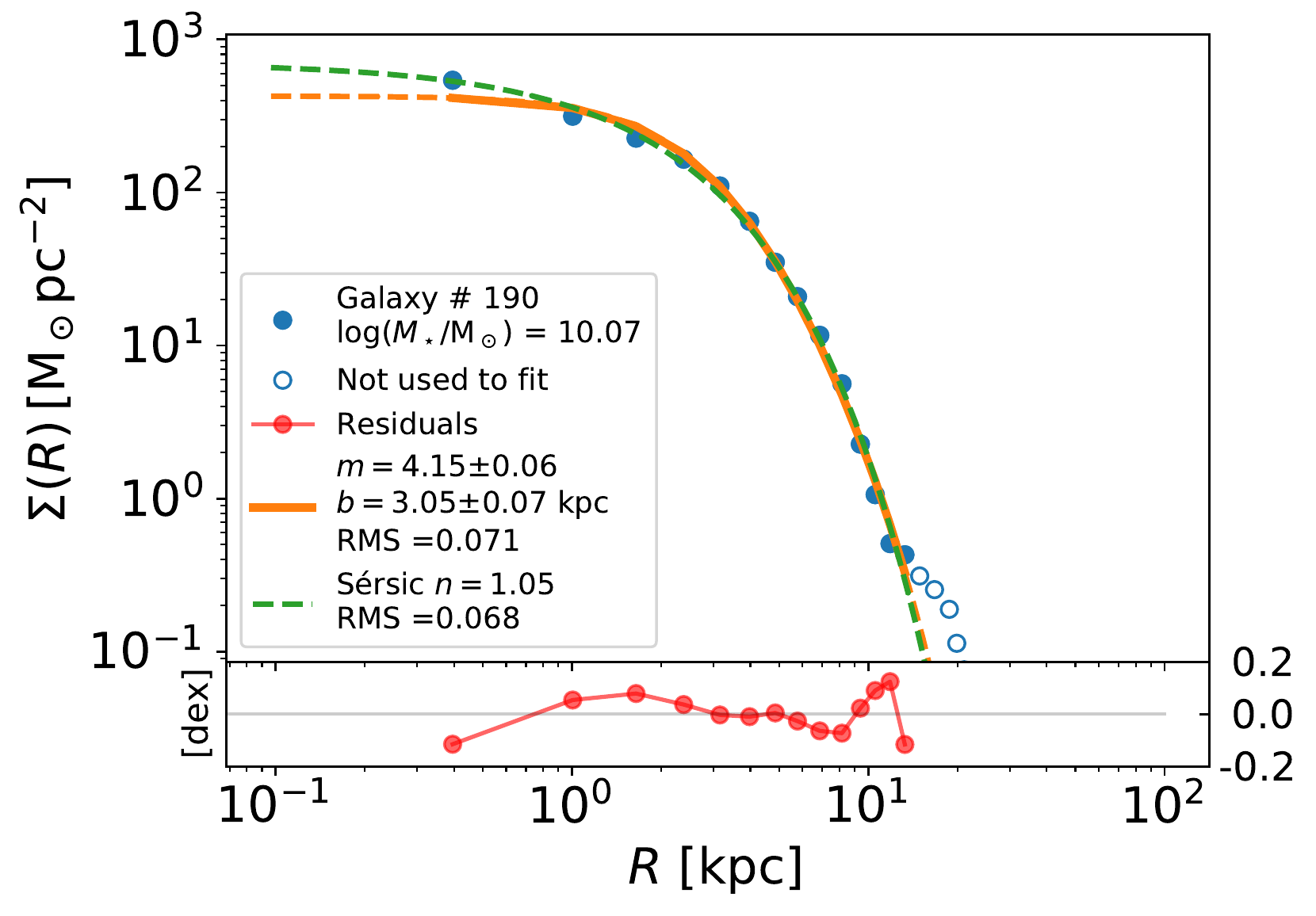}\\
    \includegraphics[width=0.4\linewidth]{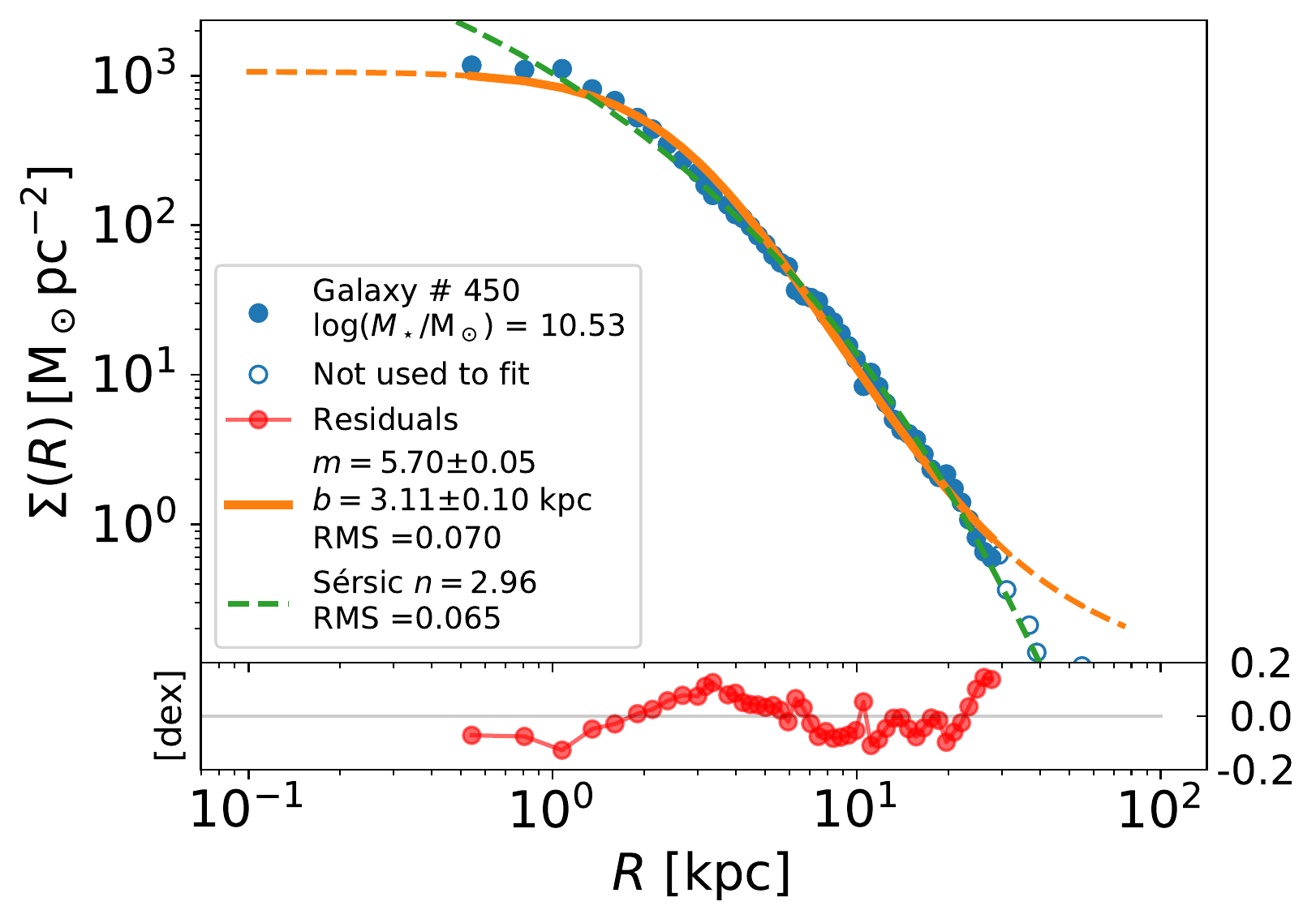}
    \includegraphics[width=0.4\linewidth]{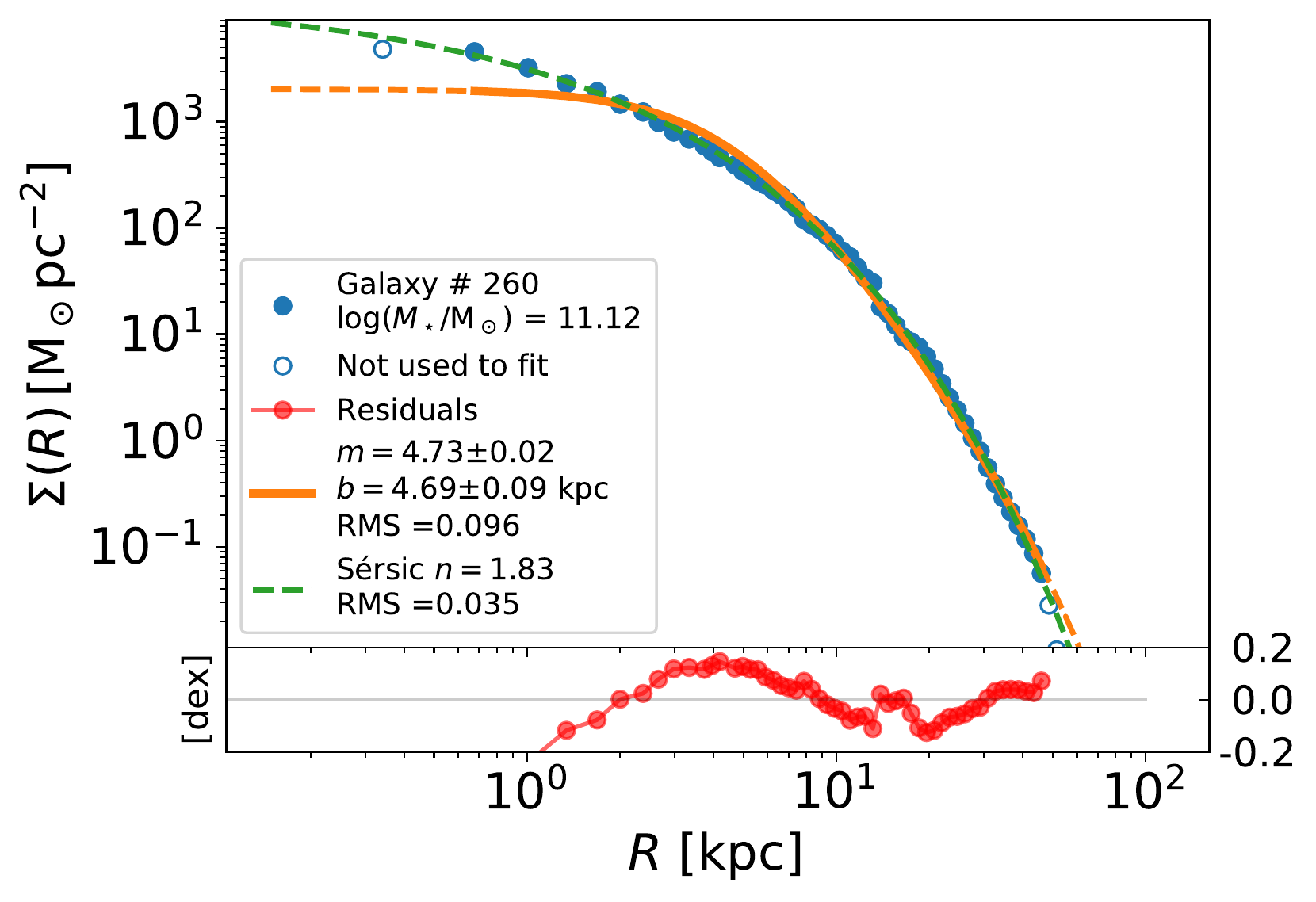}\\
    \end{centering}
    \caption{Similar to Fig.~\ref{fig:lane_emden_fit_good} showing examples of not-so-good fits to observed mass surface density profiles, chosen so that the RMS of the \propol\ fit  (the orange lines) is larger than the RMS of the sersic fit (the green dashed lines).
The layout is identical to Fig.~\ref{fig:lane_emden_fit_good}, and we refer to the caption of this other figure for details.  The range of masses in these galaxies is $8.1 \leq \log[M_\star/{\rm M}_\odot] \leq  11.1$. 
}
  \label{fig:lane_emden_fit_bad}
 \end{figure*} 
 \subsection{Description of the data set}\label{sec:dataset}
 We select the dataset described in detail by \citet[][]{2020MNRAS.493...87T,2020MNRAS.495.3777T} to study how \propol s fit real galaxies. It comprises around 750 galaxies  of the local Universe (redshift $< 0.09$), including all morphological types \citep[as classified by ][]{2010ApJS..186..427N,2013MNRAS.435.2764M}, and spanning five orders of magnitude in stellar mass ($7 < \log [M_\star/{\rm M_\odot}] < 12$; see Fig.~\ref{fig:histmass}).
The sample excludes objects with signs of strong interactions \citep[e.g., having shells or tails;][]{2020A&A...640A..38R} to keep only galaxies with relaxed outskirts.
The profiles were derived from images of the IAC Stripe 82 Legacy Project \citep{2016MNRAS.456.1359F,2018RNAAS...2..144R}, which is a re-processing of the SDSS Stripe 82 \citep{2008AJ....135..338F} with improved sky subtraction and reaching a surface brightness limit\footnote{Defined as 3-sigma fluctuations of the background of the image in averages of $10\times 10\, {\rm arcsec}^2$.}  of 29.1, 28.6, and 28.1 mag~arcsec$^2$ in the bands $g$, $r$ and $i$, respectively. The scattered light from stars in the field was removed, and the surface brightness was corrected for galaxy inclination as described by \citet{2020MNRAS.493...87T}.
The Point Spread Function (PSF), as measured from stars in the field, has a central gaussian core with FWHM~$\simeq 1.4$\,arcsec.   
Stellar mass surface densities were computed from surface brightness using mass-to-light ratios inferred from the observed colors, as described by \citet{2015MNRAS.452.3209R}.  The underlying hypotheses of this calibration include the use of \citet{2003MNRAS.344.1000B} models for the stellar populations and a \citet{2003PASP..115..763C} initial mass function.   All in all, the procedure provides mass profiles for all galaxies down to a surface density fainter than $1\,{\rm M}_\odot\,{\rm pc}^{-2}$.  Examples of profiles are shown in Figs.~\ref{fig:lane_emden_fit_good} -- \ref{fig:lane_emden_fit_bad} (the blue symbols).

 %
%\newpage 
\subsection{Fitting surface density profiles with propols}\label{sec:fit_propol_galax}
The profiles described in Sect.~\ref{sec:dataset} were fitted with the {\tt python} code introduced in Sect.~\ref{sec:code}. In order to minimize observational biases, we did not include radii smaller than 0.7\,arcsec and $g$-band surface brightness fainter than 29\,mag\,arcsec$^{-2}$. We also excluded  profiles with less than four points after thresholding.  The first constraint avoids radii that may be influenced by the PSF whereas the second discards points at or below the noise level. Apart from those discarded, all radii contribute equally to the goodness of the fit (Eq.~[\ref{eq:meritf}]). We take this approach since the residuals of the fits are mainly produced by unknown systematic errors, arising from the lack of symmetry in the galaxies, the correction of inclination, the assumed mass-to-light ratio and, obviously, the deviations of the true galaxy profiles from the \propol\ shape.
A histogram with the stellar masses of the selected galaxies is shown in Fig.~\ref{fig:histmass} -- this sample will be called {\em reference} to be distinguished from other cutouts from the original sample where some of the restriction are lifted or modified. The reference sample uses {\bf grid2} for fitting (Table~\ref{tab:grids}). 

Examples of good fits are shown in Fig.~\ref{fig:lane_emden_fit_good}.
Examples of fits that are not so good are included in Fig.~\ref{fig:lane_emden_fit_bad}.
To decide whether the fits are good or not, we use as benchmark \sersic\ profile fits to the same data. Good corresponds to \propol\ fit with RMS smaller than the \sersic\ fit, and vice versa.  Thus, together with the observed profiles (blue solid symbols) Figs.~\ref{fig:lane_emden_fit_good} and \ref{fig:lane_emden_fit_bad} include the \propol\ fit (orange solid lines) and the \sersic\  profile fit (green dashed lines). One can find good and poor fits in all the range $7.5 \leq \log(M_\star/{\rm M}_\odot) \leq  11.5$, although there is a clear trend of the \propol\ fits to worsen with increasing stellar mass, as we report on below. Note that the \propol s with $m < 5$ have finite size (see Fig.~\ref{fig:propols}). This size is always larger than the last point used for fitting, but it is often smaller than the radii of some of the excluded points (the open symbols in Figs.~\ref{fig:lane_emden_fit_good} and \ref{fig:lane_emden_fit_bad}).

\begin{figure}
    \includegraphics[width=\linewidth]{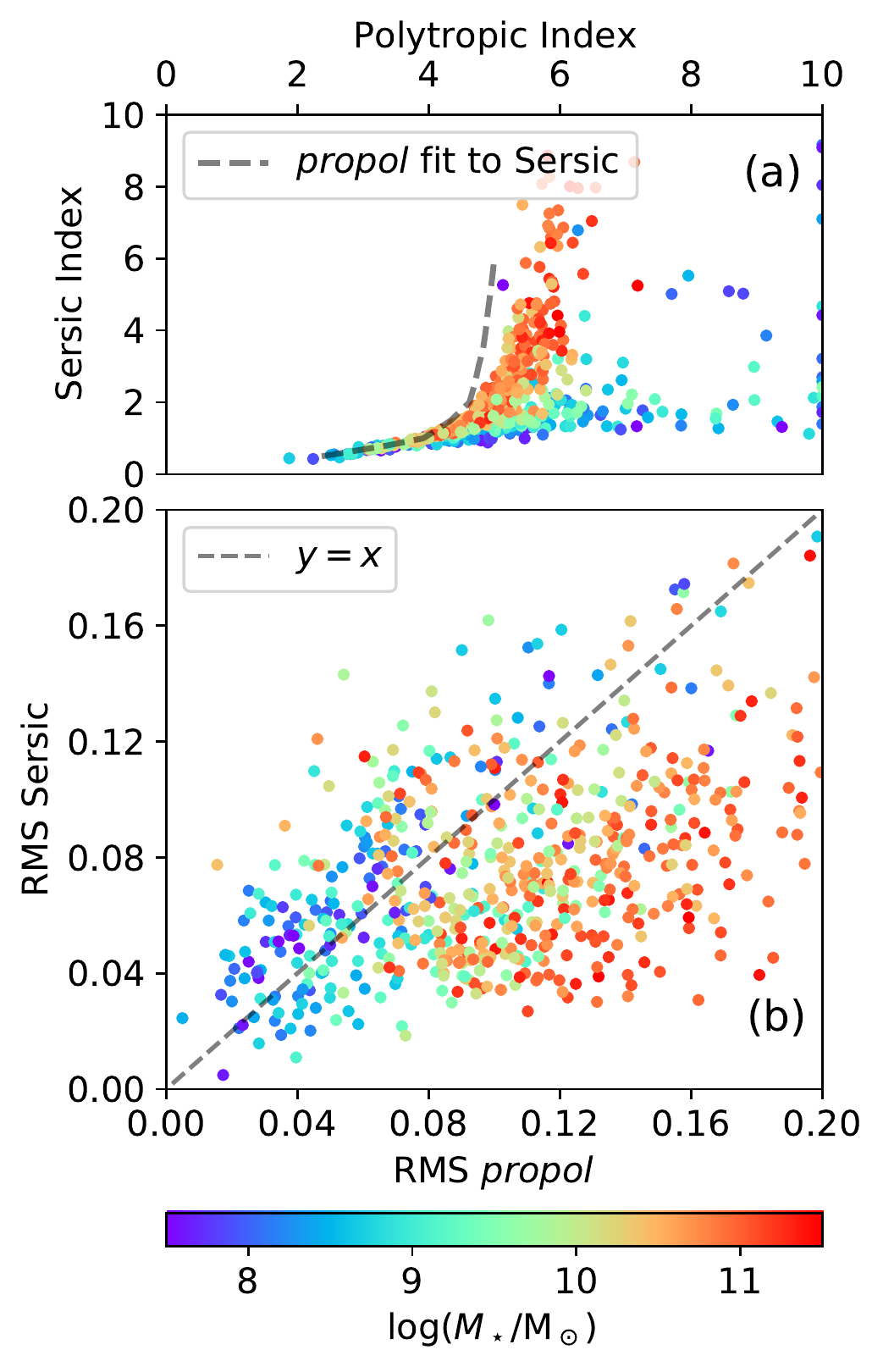}
    \caption{Diagnostic diagrams to judge the goodness of the \propol\ fits. They compare the results obtained using \sersic\ and \propol\ fits on the same mass density profiles. Each point is a galaxy from the reference sample, with the color given by its stellar mass according to the color-code in the side bar. 
      (a) \sersic\ index $n$ versus polytropic index $m$. The dashed line corresponds to the equivalence between \sersic\ index and polytropic index worked out in Sect.~\ref{sec:sersic} and represented by a blue line with symbols in Fig.~\ref{fig:lane_emden6}a.
      (b) RMS of the \sersic\ fit residuals versus RMS for the corresponding \propol\ fit. The dashed line represents $y=x$ so points above the line correspond to \propol\ fits better than the corresponding \sersic\ fit (RMS \propol\ $<$ RMS \sersic ). Note the clear dependence on stellar mass of the location of the points, so that low mass objects tend to have a better \propol\ fit, and vice-versa.
    }
  \label{fig:diagnostic_aaa}
 \end{figure} 
Diagnostic diagrams comparing the \sersic\ and \propol\ fits are included in Fig.~\ref{fig:diagnostic_aaa}, with a symbol for each galaxy color-coded according to $\log M_\star$. Figure~\ref{fig:diagnostic_aaa}a shows \sersic\ index versus polytropic index whereas Fig.~\ref{fig:diagnostic_aaa}b shows RMS of the \sersic\ fit residuals versus RMS of the \propol\ fit.  The dashed line in  Fig.~\ref{fig:diagnostic_aaa}a corresponds to the equivalence between \sersic\ index and polytropic index worked out in Sect.~\ref{sec:sersic} and represented by a blue line with symbols in Fig.~\ref{fig:lane_emden6}a. The dashed line in Fig.~\ref{fig:diagnostic_aaa}b corresponds to  $y=x$ so points above (below) the line correspond to \propol\ fits better (worse) than the corresponding \sersic\ fit.  A number of conclusions can be drawn from these figures as well as similar figures for different galaxy selections and grids included in Appendix~\ref{app:scatter} to avoid cluttering. These conclusions are:
\begin{itemize}
\item[-] The relative goodness of the \propol\  fits with respect to the \sersic\ fits depends on the galaxy stellar mass, with the goodness of the \propol\  fits increasing with decreasing stellar mass. Thus, \propol\  fits are systematically better than \sersic s for $\log(M_\star/{\rm M}_\odot)\lesssim 9$ and systematically worst for $\log(M_\star/{\rm M}_\odot)\gtrsim 10$. This effect is clear when comparing Fig.~\ref{fig:diagnostic_aaa}b with the same figure when only low-mass galaxies are plotted ($\log[M_\star/{\rm M}_\odot]< 10$; Fig.~\ref{fig:diagnostic_ccc}b).
 %Because of the strong correlation between morphological type and stellar mass in our sample (Fig~\ref{fig:histmass}), the goodness of the \propol\  fits improves substantially when only late type galaxies are considered (Fig.~\ref{fig:diagnostic_ddd}b).
%
\item [-] Even if the worsening of the fits with increasing stellar mass is clear, there are high-mass galaxies having good \propol\ fits and low-mass galaxies where \sersic\ fits do a better job (Figs.~\ref{fig:lane_emden_fit_good} and \ref{fig:diagnostic_aaa}b).
\item[-] The relation between the \sersic\ index and the Polytropic index only follows the theoretical relation worked out  in Sect.~\ref{sec:sersic} when $m\lesssim 4$ (see the dashed line in Fig.~\ref{fig:diagnostic_aaa}a). The deviations at larger indexes can be understood as the effect of including galaxy cores when fitting. The theoretical line was computed to minimize the RMS of the residuals, which forces excluding cores when $m\gtrsim 1$ (see Fig.~\ref{fig:lane_emden_fitsersic}). To show that cores are producing the deviation, we repeat the fits excluding most of the core when fitting (excluding the central $R< 2$\,arcsec). The resulting relation is shown in Fig.~\ref{fig:diagnostic_eee}a, where the points corresponding to the most massive objects have shifted toward the theoretical curve. Simultaneously, the systematic difference between the RMS of the \propol\ and \sersic\ fits at high mass tends to go away (Fig.~\ref{fig:diagnostic_eee}a).   
\item[-] Including the expected difference between the dark matter and the stellar mass density profiles does not improve the quality of the fits. Figure~\ref{fig:diagnostic_eee_other} shows another rendering of the diagnostic plot using {\bf grid2g1} rather than {\bf grid2}, where the parameter that characterizes the difference ($\gamma$ in Eq.~[\ref{eq:gamma}]) is set to 0.1 rather than 0 (Table~\ref{tab:grids}).
\item[-] Fits based on other samplings of the \propol\ grid and extending the range of indexes (e.g., {\bf grid3} in Table~\ref{tab:grids}) do not modify the above conclusions in any significant way.

\item[-] Polytropes represent spherically symmetric structures, however, deviations from this underlying hypothesis are to be expected in real galaxies. Since early type galaxies are rounder than late type galaxies, they were anticipated to have better \propol\ fits than late types. However, we do not see any obvious systematic worsening of the fits associated with late type objects (cf. Figs.~\ref{fig:diagnostic_ddd} and \ref{fig:diagnostic_ggg}). Differences can be easily ascribed to the fact that, in our sample, early types tend to be more massive (Fig.~\ref{fig:histmass}). We do not have a clear physical interpretation for this negative result. It may be due to the fact that both early and late types are massive enough for the fitting errors to hide the putative differences between the two populations.
\end{itemize}

 The discussion above is made in terms of goodness relative to \sersic\ fits, but we also quantify the fraction of good fits in absolute terms. In this context, we define as {\em good} any fit having RMS~$< 0.1$,  which is the limit for the good-fit profiles shown in Fig.~\ref{fig:lane_emden_fit_good}. As Fig.~\ref{fig:diagnostic_aaa}b evidences, this fraction very much depends on $M_\star$. Thus, we get 84\,\% of good fits when $\log(M_\star/{\rm M}_\odot)<8$, 81\,\% when $\log(M_\star/{\rm M}_\odot)$ is in between 8 and 9, 56\,\%  when in between 9 and 10, 34\,\%  when in between 10 and 11, and 23\,\%  when larger than 11. The percentage of good fits when all galaxies are included is 49\,\%.
\begin{figure}
    \includegraphics[width=\linewidth]{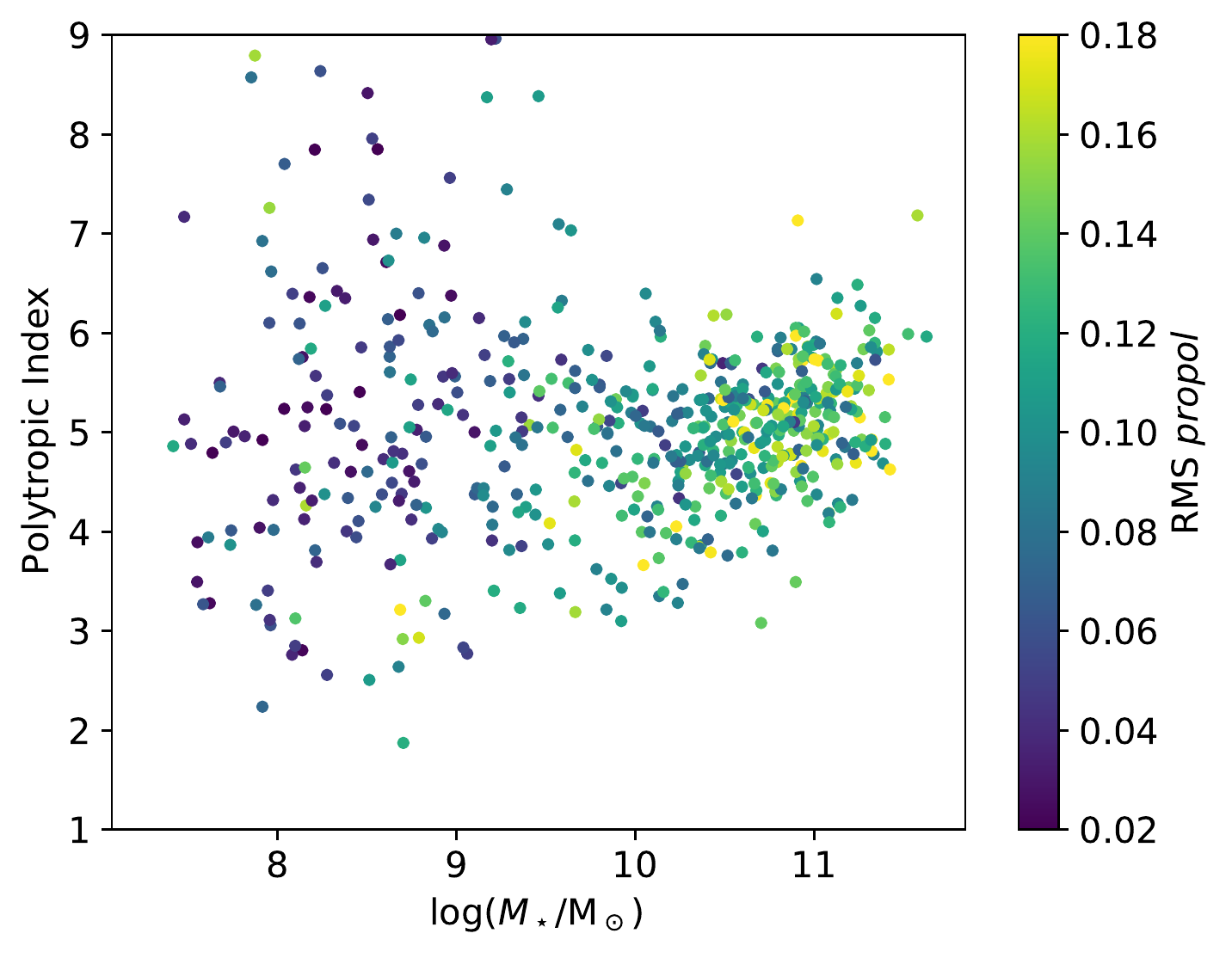}
    \caption{Polytropic index of the \propol\ fits to real galaxies versus stellar mass, color-coded with the RMS of the residuals. Indexes crowd around $m=5$, with a dispersion that increases as the stellar mass decreases. The goodness of the fit also increases with decreasing stellar mass.  
    }
  \label{fig:value_index}
 \end{figure} 
 The \propol\ fits tend to  crowd around the index $m=5$, with a dispersion that increases as the stellar mass decreases (Fig.~\ref{fig:value_index}).
 % This value, resulting from the observation of galaxies, also appears in  the context of globular cluster mass profiles (Trujillo and Sanchez Almeida 2021, in preparation).
 As we explain in Sect.~\ref{sec:poly}, the profiles having $m> 5$ have infinite mass, which may posse a problem of interpretation. The problem can be sorted out assuming that the existence of significant deviations from the \propol\ profiles outside the fitted range, which is not unexpected after all.
Breaks in the mass profiles cannot be reproduced by \propol s, although they are quite common \citep[e.g.,][]{2006A&A...454..759P}. Moreover, even when they are not present and the observed profiles vary smoothly, the outer parts of some galaxies may have not reached thermodynamic equilibrium yet (Sect.~\ref{sec:monsters}).

 %%%%%%%%%%%%%%%%%%%%%
%
 \begin{figure}
   \centering
     \includegraphics[width=0.8\linewidth]{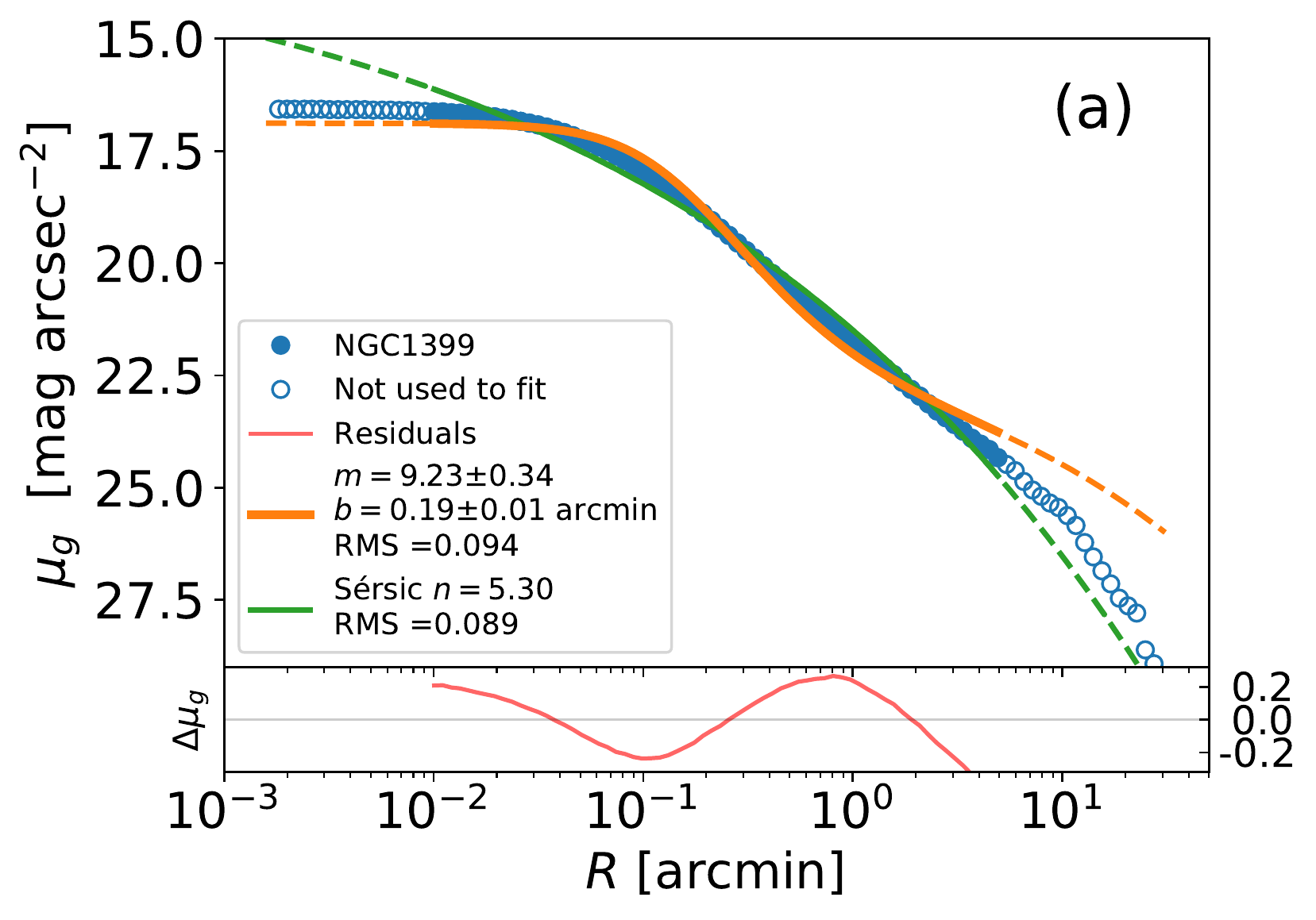}
     \includegraphics[width=0.8\linewidth]{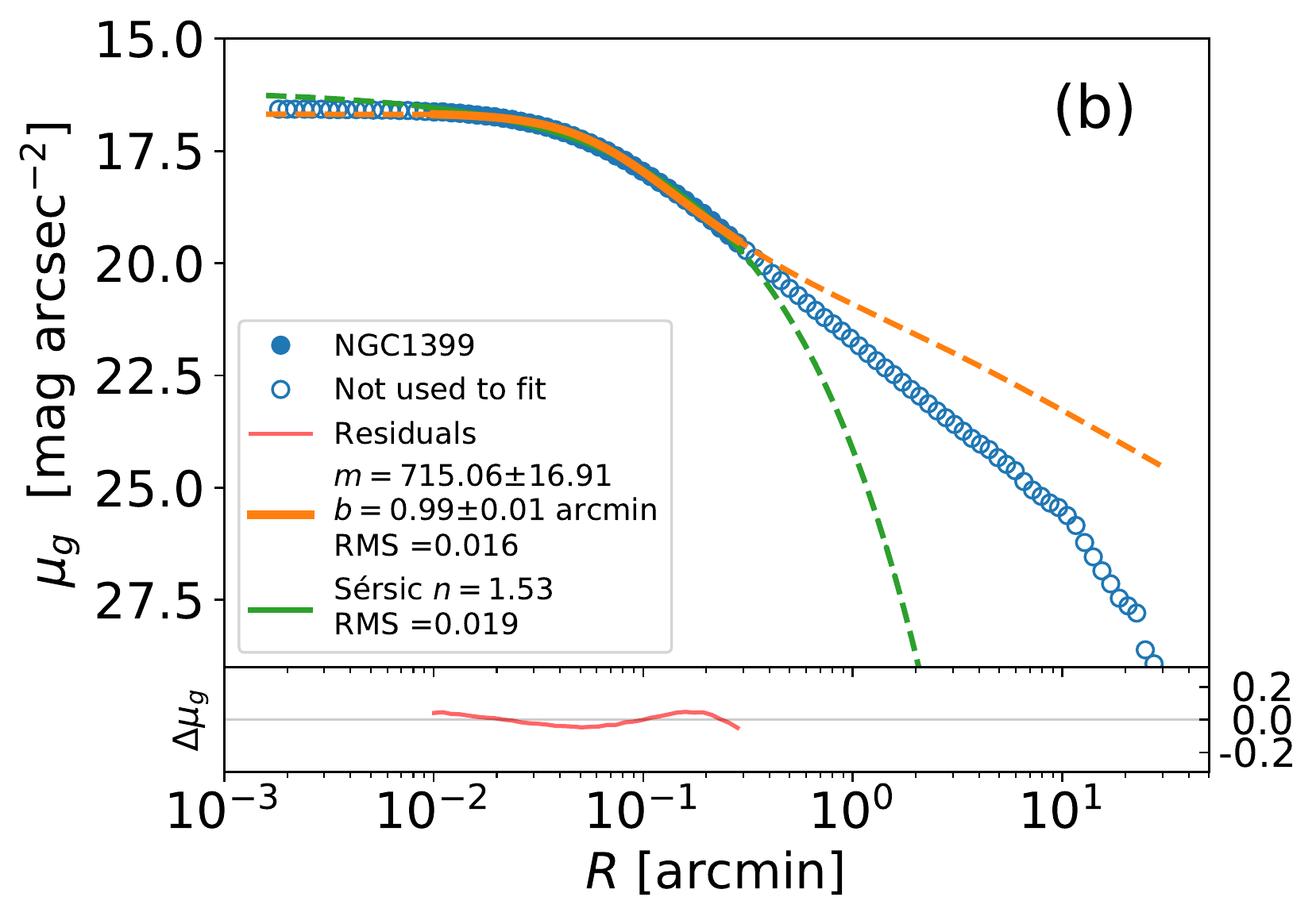}
     \includegraphics[width=0.8\linewidth]{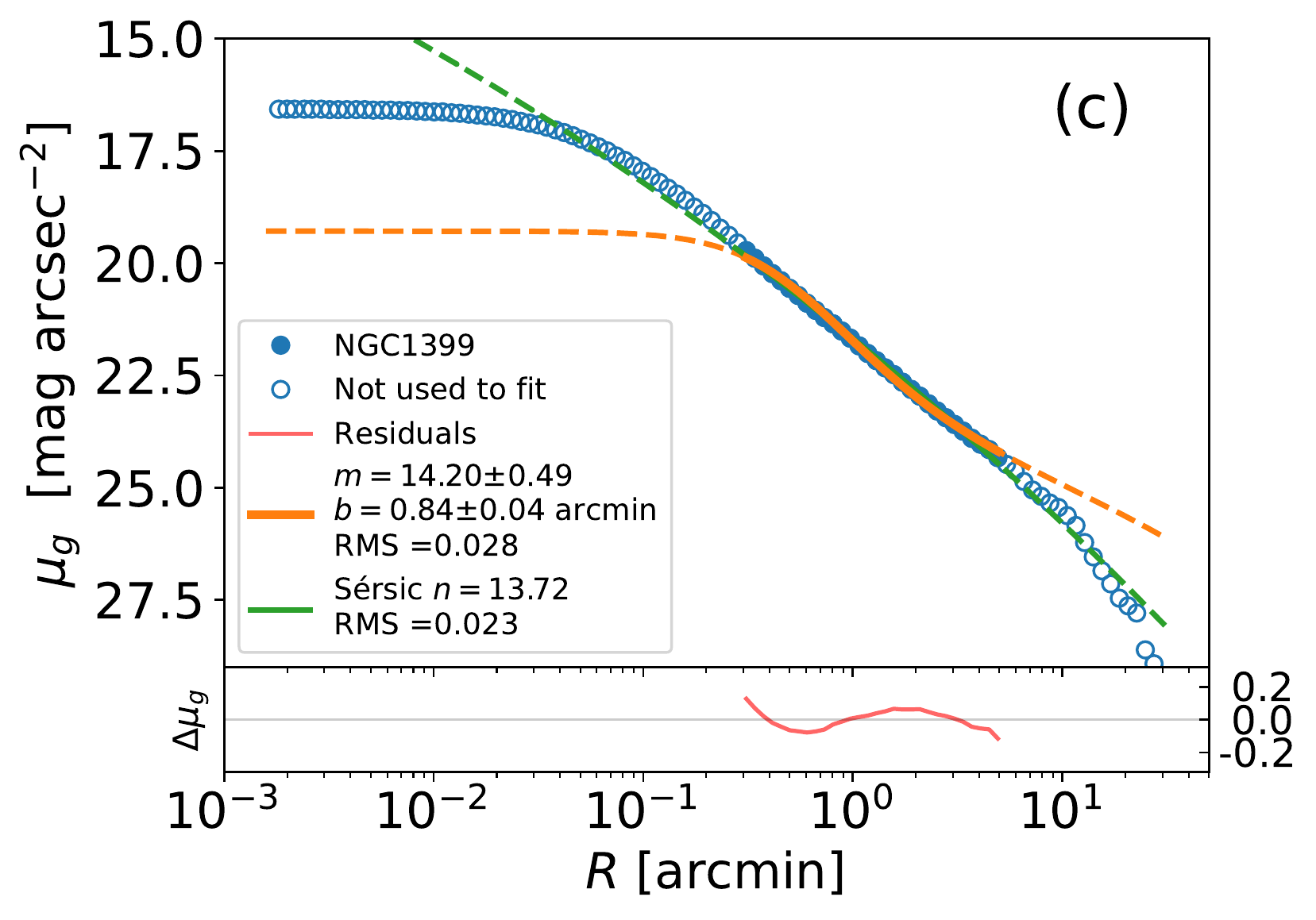}
     \includegraphics[width=0.8\linewidth]{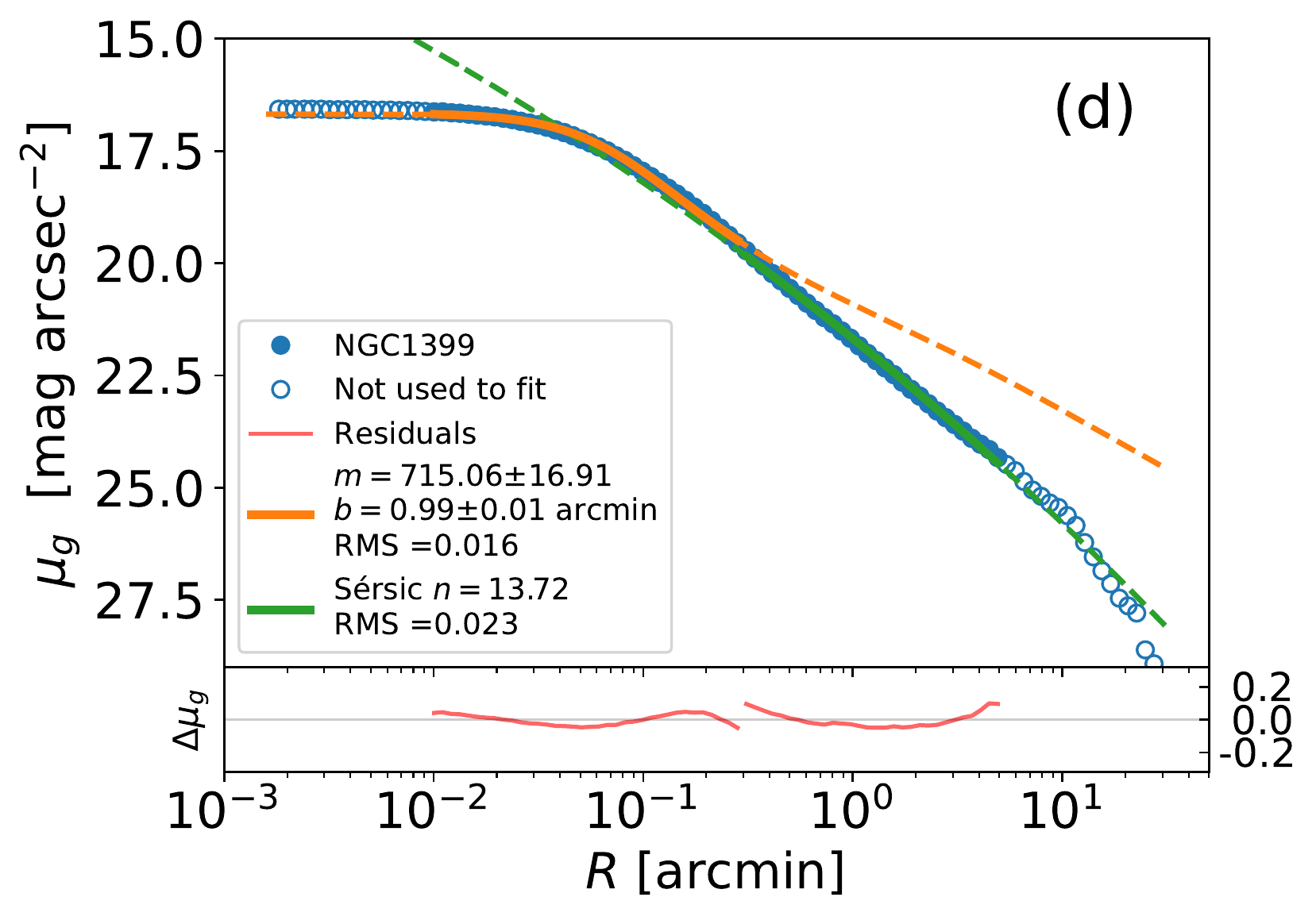}
    \caption{{\em Propol} fit to  NGC1399 ($M_\star\simeq 6\times 10^{11}\,{\rm M_\odot}$),  one of the massive elliptical galaxies in the VEGAS dataset \citep{2017A&A...603A..38S}. Example to illustrate the quality and properties of a typical fit to VEGAS galaxies, with the layout similar to Fig.~\ref{fig:lane_emden_fit_good}. The solid lines and symbols mark the range of radii used for fitting, which is different for the  different panels in the figure. The grid {\bf grid5} is used for \propol\ fitting to allow for large polytropic indexes. (a) Fit encompassing the full profile, for radii between 1\,arcsec  and 5\,arcmin. (b) Fit to the central core, which considers radii between 1 and 18\,arcsec. (c) Fit to the outskirts, for radii between 18\,arcsec and 5\,arcmin. (d) Combination of (b) and (c), where the core is represented by the \propol\ in (b) and the outskirts by the \sersic\ profile in (c). 
    }
  \label{fig:ngcs}
 \end{figure} 
 \section{(mis-)fitting massive ellipticals}\label{sec:monsters}
 Massive early type galaxies are often poorly reproduced by \propol s, as we uncover in Sect.~\ref{sec:galaxies}. Here we try to understand why.  We analyze profiles of six massive elliptical galaxies observed within the VEGAS\footnote{VST Early-type GAlaxy Survey.} project by \citet{2017A&A...603A..38S}. They were chosen because their profiles represent state-of-the-art observations of very large objects, so that the potential systematic effects arising from low noise or insufficient spatial sampling are minimized. Rather than mass profiles, the authors provide surface brightness profiles. Assuming a constant mass-to-light ratio along the galaxy, the mass and light surface density profiles are identical except for a scaling factor. Thus, we fit light profiles assuming them to present mass profiles, specifically, we fit the surface brightness in the $g$ band, $\mu_g$, provided by \citet{2017A&A...603A..38S}. (Trials with mass profiles are discussed later on, but they do not significantly improve or worsen  the goodness of the fit.) The example of one of such fits (NGC1399) is shown in Fig.~\ref{fig:ngcs}. Fits to the other massive ellipticals (NGC3923, NGC4365, NGC4472, NGC5044, and NGC5846) are  similar and will not be shown. For reference, the stellar masses of these galaxies are in the range between $4.3\times 10^{11}$ and $1.5\times 10^{12}\,{\rm M}_\odot$ \citep{2017A&A...603A..38S}. 

The profiles are well defined for radii in between 1\,arcsec and 5\,arcmin. Radii smaller than 1\,arcsec are discarded to avoid seeing smearing the profile, while radii larger than 5 arcmin are affected by breaks and truncations in the profile. The  \propol\ fit to NGC1399 is shown in Fig.~\ref{fig:ngcs}a. There are clear systematic deviations between the best fitting \propol\ and the observed profiles. The \sersic\ profile included in the same plot clearly do a better job, although the deviation at the core starts to be substantial.  The observed profile shows a central core, but the range of shapes provided by the \propol s cannot fit the core and the outskirts of the profile simultaneously. When the outskirts are not included in the fit, then \propol s make an excellent job fitting the core, as shown in Fig.~\ref{fig:ngcs}b. However, the \propol\ fit to only-the-outskirts is unrealistically  far from the observed profile (Fig.~\ref{fig:ngcs}c).       

Thus, \propol s only provide a good fit to the cores of these massive ellipticals. The full profile can be represented as an inner polytropic core plus a \sersic\ profile describing the outskirts (Fig.~\ref{fig:ngcs}d), very much in line with the piece-wise profiles describing the haloes from SIDM  (Self Interacting Dark Matter) numerical simulations \citep{2021MNRAS.501.4610R,2021MNRAS.504.2832S}. This fact suggests that the inner part of these massive ellipticals is in thermal equilibrium, which does not reach the outer part. Thus, the physical process leading to thermalization should be more effective in the high density environment corresponding to the inner part. The required scaling with density is quite natural, e.g., it appears when the equilibrium is reached via two-body gravitational interactions or by other types of collisions between dark matter particles \citep[e.g.,][]{2021MNRAS.504.2832S}.
Another pathway to thermalization may be the merger of super-massive black holes (SMBH), expected to occur at the center of massive ellipticals \citep[][]{2014ApJ...795L..31L,2016MNRAS.462.2847M}. The motion of two merging black holes produces {\em scouring} of stars. In addition, the recoil kicks the merged SMBH out of the center, forcing a final swing of the SMBH that stirs the global gravitational potential \citep[e.g.,][]{2006RPPh...69.2513M,2021MNRAS.502.4794N}. Thus, {\em scouring} plus {\em recoil} may allow the self-gravitating system to reach thermodynamical equilibrium in a timescale much sorter than the two-body collision timescale, and well within the Hubble time. 

As an additional sanity check, we also transform the surface brightness profiles $\mu_g$  into mass surface densities $\Sigma_\star$ using mass-to-light ratios pre-computed  by \citet{2015MNRAS.452.3209R} and the radial profile in the $i$-band provided \citet{2017A&A...603A..38S}. The mass profiles differ very little from the light profiles described above. In particular, the difficulty for the \propol s to provide a good overall fit remains.

%%%
\section{Conclusions}\label{sec:conclusions}

The solutions of Eqs.~[\ref{eq:lane_emden}], [\ref{eq:densityle}], and [\ref{eq:radius}]) are called polytropes. Given a total mass and energy, they describe spherically symmetric self-gravitating systems that maximize Tsallis entropy. These structures, which have been studied by astronomers for long, are nowadays gaining renewed interest because they provide valid explanation for several seemingly-disconnected properties of galaxies  (Sect.~\ref{sec:intro}). In this paper, we carry on exploring the practical interest of polytropes by studying if they account for the stellar mass distribution observed in real galaxies.

We develop a {\tt python} code (Sect.~\ref{sec:code}) to fit surface density profiles using polytropes projected in the plane of the sky (\propol s). The code is systematically applied to understand if and why \propol s have the shapes observed in the mass distribution of galaxies. \sersic\ profiles are generally regarded as good approximations to galaxy shapes. Thus, we firstly use the fitting code to address whether \sersic\ profiles and \propol s are related (Sect.\ref{sec:sersic}). We find that the cores of the \propol s and \sersic\ profiles are inconsistent when the \sersic\ index is larger than around 1. However, when the cores are excluded, \propol s and \sersic\ profiles are indistinguishable within observational errors. The RMS of the residuals is around 0.02 dex or 5\,\% (Fig.~\ref{fig:lane_emden6}d) over 1 order of magnitude in radius (Fig.~\ref{fig:lane_emden6}a) and 5 orders of magnitude in surface density (Fig.~\ref{fig:lane_emden6}c).  We find a one-to-one correspondence between the \sersic\ index and the polytropic index of the best-fitting \propol\ (Fig.~\ref{fig:lane_emden6}a, the blue line). Polytropes are physically meaningful only within a particular range of indexes (Eq.~[\ref{eq:nlimits}]) which corresponds to \sersic\ indexes between 0.4 and 6.0, in excellent agreement with the \sersic\ indexes observed in nature \citep[e.g.,][]{2003ApJ...594..186B,2012ApJS..203...24V,2001MNRAS.321..269T}.
%Thus, the ansatz that \sersic\ profiles are working approximations to the outskirts of \propol s automatically explains why galaxies show \sersic\ indexes within the observed range. To the best of our knowledge, this is the first time that such explanation is put forward in literature.  

% 
The \propol\  fitting code has been systematically applied to a large number of galaxies ($\sim 750$) with carefully measured mass density profiles (Sect.~\ref{sec:galaxies}). The galaxy sample includes all morphological types and spans five orders of magnitude in stellar mass ($7 < \log [M_\star/{\rm M_\odot}] < 12$; \citeauthor{2020MNRAS.493...87T}~\citeyear{2020MNRAS.493...87T}). The goodness of each fit is established by comparing the RMS of its residuals with those resulting from a \sersic\ fit. The relative goodness thus defined  depends on the galaxy stellar mass, with the goodness of the \propol\  fits increasing with decreasing stellar mass. The \propol\  fits are systematically better than \sersic s for $\log(M_\star/{\rm M}_\odot)\lesssim 9$ and systematically worst for $\log(M_\star/{\rm M}_\odot)\gtrsim 10$.  %Because of the strong correlation between morphological type and stellar mass in our sample (Fig~\ref{fig:histmass}), the goodness of the \propol\  fits improves substantially when only late type galaxies are considered.
Even if the worsening of the fits with increasing stellar mass is clear, there are high-mass galaxies having good \propol\ fits and low-mass galaxies where \sersic\ fits do a better job than  \propol s (Figs.~\ref{fig:lane_emden_fit_good} and \ref{fig:diagnostic_aaa}b).
The clean one-to-one relation between  \sersic\ index and Polytropic index mentioned above breaks down when the cores are included in the fits, as it usually the case when fitting real galaxies, which often have cores. The observed  polytropic indexes tend to cluster around $m=5$, although with large scatter (Fig.~\ref{fig:value_index}). We also study whether the expected difference between the total mass and the stellar mass density profiles modifies the quality of the fits, to find out that it does not.
Using a threshold in the RMS of the residuals to identify good fits in absolute terms ($< 0.1$ dex in our case), we characterize the fraction of good fits depending on $M_\star$. It goes from 84\,\% for $\log(M_\star/{\rm M}_\odot)<8$ to  23\,\%  when it is larger than 11. When all galaxies are included, the percentage reaches 49\,\%.
Section~\ref{sec:monsters} is devoted to understand why the overall good fits provided by \propol s worsen for massive galaxies. We find that \propol s are very good at reproducing the central parts, but they do not handle well cores and outskirts all together. However, the combination of a \propol\ core and a \sersic\ outskirt matches the observed profiles extremely well (Fig.~\ref{fig:ngcs}d). This fact suggests that only the inner part of a massive galaxy is already in thermal equilibrium, which is sensible considering that the physical process leading to thermalization are more effective in high density environments. The merger of SMBHs at the center of massive ellipticals also provides a pathway to thermalization (Sect.~\ref{sec:monsters}).

There is no systematic differences between the goodness of the fits in early and late type galaxies, even though the hypothesis of being spherically symmetric is better satisfied by early types. We do not have a proper interpretation of this negative result. It may be due to the fact that both early and late types are massive enough for the fitting errors to hide the putative differences between the two populations.

On the whole, \propol s do an excellent job reproducing the stellar mass distribution of low-mass galaxies, as well as the cores of the massive objects. We have also shown that the outskirts of \propol s are indistinguishable from \sersic\ profiles. In addition, \propol s should also reproduced the tight mass\,--\,size relation observed in galaxies when measuring sizes at constant surface brightness \citep[e.g.,][]{2020MNRAS.493...87T}, because the tight relation is naturally produced by \sersic\ profiles \citep{2020MNRAS.495...78S}, and so should result from \propol s as well. Finally, polytropes reproduce the cores in the total mass distribution of dwarf galaxies without being adjusted \cite[][]{2021MNRAS.504.2832S}.

Rather than being purely empirical constructs, polytropic shapes are expected when self-gravitating structures are in thermal equilibrium as defined by the Tsallis entropy \citep[][see Sect.~\ref{sec:intro}]{1993PhLA..174..384P}. Thus, our results are consistent with the ansatz that the principle of maximum Tsallis entropy dictates
the internal structure in dwarfs and in the central region of the most massive galaxies. Should this ansatz be correct, it has profound implications ranging from measuring physical properties of galaxies (should the NFW profiles\footnote{Navarro, Frenk, and White profiles, after \citet{1997ApJ...490..493N}.} be replaced with polytropes when interpreting gravitational lensing signals?) to the basic physics underlying galaxy formation (how is the thermodynamical equilibrium stablished? Is it due to some sort of non-gravitational interaction between dark matter particles?), and including galaxy modeling through numerical simulations \citep[why do cold DM numerical simulations create cusps rather than cores? See][]{2021MNRAS.504.2832S}.
%

%%
%
% references
%
%\bibliographystyle{aasjournal}
%\bibliography{../../paper129(Sersic)/ver0/bibliography}
%\bibliography{bibliography}

%%%%%%%%
%
% Appendixes 
%
\acknowledgments

We thank Nuskia Chamba for help with the handling of the profiles analyzed in Sect.~\ref{sec:galaxies}, and Marilena Spavone and Enrichetta Iodice for providing the surface brightness profiles analyzed in Sect.~\ref{sec:monsters}.  
Thanks are also due to an anonymous referee for suggestions leading to improve the presentation of results and some arguments. 
JSA acknowledges support from the Spanish Ministry of Science and Innovation, project  PID2019-107408GB-C43 (ESTALLIDOS), and from Gobierno de Canarias through EU FEDER funding, project PID2020010050.
IT acknowledges support from the Spanish Ministry of Science and Innovation,  grant PID2019-107427GB-C32, from the European Union Horizon 2020 research and innovation program under Marie Sk\l odowska-Curie grant agreement No. 721463 to the SUNDIAL ITN network, and from the European Regional Development Fund (FEDER), IAC project P/300624. The IAC project is partly financed by the Spanish Ministry of Science, Innovation and Universities (MCIU), through the State Budget, and by the Canary Islands Department of Economy, Knowledge and Employment, through the Regional Budget of the Autonomous Community.


\vspace{5mm}
\facilities{SDSS \citep{2000AJ....120.1579Y}, IAC Stripe 82 \citep{2016MNRAS.456.1359F}}


\software{{\tt astropy} \citep{2013A&A...558A..33A}  
          }



\appendix
\section{Formalism to include differences between stellar mass and gravitational mass}\label{app:a}
The Lane-Emden equation describes the distribution of total gravitational mass whereas observations refer to stellar mass. The two masses do not necessarily scale each other, in which case Eq.~(\ref{eq:needlabel}) would be useless for stellar mass. Fortunately, differences between the total density $\rho$ and the stellar density $\rho_\star$ can be easily  accommodated within our formalism. Assuming a smooth and moderate change in the stellar to total mass ratio, then,
\begin{equation}
  \ln [\frac{\rho_\star}{\rho}(r)]\simeq \ln [\frac{\rho_\star}{\rho}(0)]+\gamma\,[\ln\rho(r)-\ln\rho(0)]+\dots,
  \label{eq:tylor_exp}
\end{equation}
with
\begin{displaymath}
\gamma =  \frac{d\ln [\frac{\rho_\star}{\rho}]}{d\ln\rho}(0),
  \end{displaymath}
so that one can approximate $\rho_\star$ as
  \begin{equation}
    \rho_\star (r)\simeq\rho_\star(0)\, \Big[\frac{\rho(r)}{\rho(0)}\Big]^{1+\gamma},
    \label{eq:scalingrhos}
  \end{equation}
or, in general, 
 \begin{equation}
   \rho_\star (r)\simeq\rho_\star(r')\, \Big[\frac{\rho(r)}{\rho(r')}\Big]^{1+\gamma}.
   \label{eq:summary}
 \end{equation}
  Therefore, considering the variation of $\rho_\star/\rho$ with $r$ is equivalent to deriving $\rho$ from the Lane-Emden Eq.~(\ref{eq:lane_emden}), computing $\rho_\star$ through Eq.~(\ref{eq:scalingrhos}), which is then projected  on the plane of the sky replacing $\rho$ with $\rho_\star$ in Eq.~(\ref{eq:abeldirect0}). We want to stress that the above equations express a purely phenomenological relation between $\rho(r)$ and $\rho_\star(r)$. It does not imply any interpretation on the nature of this relation of the kind of, e.g.,  the {\em radial acceleration relation} \citep[][]{2016PhRvL.117t1101M}. The densities $\rho(r)$ and $\rho_\star(r)$ are both monotonic functions of $r$, therefore, their relative variation with $r$ can always be expressed as the Taylor series in Eq. (\ref{eq:tylor_exp}). If the ratio of densities changes moderately,  then the first two terms of the expansion suffice, leading to  Eq. (\ref{eq:scalingrhos}).

This approach requires a value for $\gamma$, which can be estimated as follows:  $\gamma > 0$  since the stars are more centrally concentrated than the DM and so both  $\rho_\star/\rho$ and $\rho$ decrease outward in a galaxy. For a significant (one order of magnitude) change from $[\rho_\star(0)/\rho(0)]$ to $[\rho_\star(r)/\rho(r)]$,  Eq.~(\ref{eq:scalingrhos}) requires
\begin{equation}
  [\rho(r)/\rho(0)]^\gamma \le \frac{1}{10}.
  \label{eq:condition}
\end{equation}
The density $\rho$ changes by many orders of magnitude in any galaxy. Thus, for a typical 
$\rho(r)/\rho(0) \sim 10^{-3}$, the condition (\ref{eq:condition}) is met for $\gamma=0.3$.
If the radial drop is larger, then a smaller $\gamma$ suffices.

%
%%%%%%%%%%%%%%%%

 \section{Computing the grid of profile shapes}\label{app:b}

\begin{figure}
 \centering 
\includegraphics[width=0.5\linewidth]{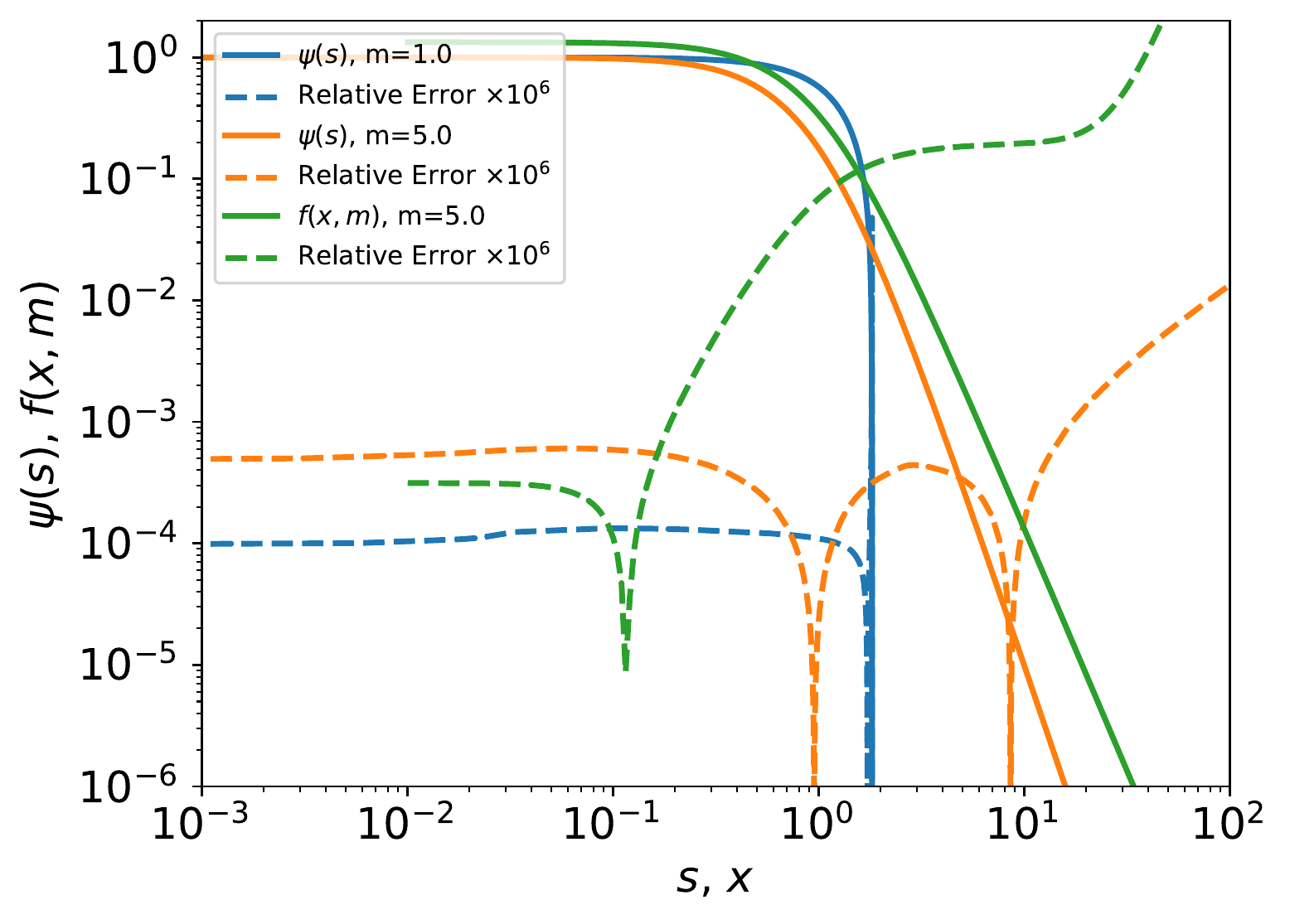}
\caption{
Relative error of the profiles in the grid used for interpolation ({\bf grid2} in this case). See text and the inset for details. Solid and dashed lines of the same color show the function and its relative error. Note that the relative errors (dashed lines) have been multiplied by a factor $10^6$. 
}
\label{fig:lane_emden_grid1_test1}
\end{figure}

 There are two critical points in the evaluation of the grid that defines the shapes of the \propol s (i.e., $f$). The first one has to do with the integration of the  Lane-Emden Eq.~(\ref{eq:lane_emden}). We split it into a system of two first order differential equations for $\psi$ and $d\psi/ds$, which were integrated from $s=0$ with the initial conditions $\psi(0)=1$ and  $d\psi/ds(0)=0$ using {\tt Lsoda} \citep{2019ascl.soft05021H} as implemented in {\tt python} ({\tt scipy.odeint}). The second one has to do with carrying out the Abel transform of the polytropes (Eq.~[\ref{eq:abeldirect}]). It was evaluated by direct integration in the 3D radial distance using the Simpson's rule  ({\tt simps} provided by {\tt scipy.integrate}). Each $x$ was integrated independently, with a mesh from $0.01$ to 200 equi-spaced  in $\log x$ steps of 0.02. The integration parameters were tuned so as to achieve a relative error in the inferred $f$ smaller than $10^{-6}$.
 The final errors in $\psi$ and $f$ were evaluated resorting to the cases where the Lane-Emden equation has analytic solutions (Eqs.~[\ref{eq:poly1}], [\ref{eq:poly5}],  and [\ref{eq:ppoly5}]).  The difference between the resulting numerical and exact solutions are shown in Fig.~\ref{fig:lane_emden_grid1_test1}.  Relative errors  in $\psi$ are of the order of $10^{-10}$ whereas they are smaller than  $10^{-6}$ for $f$. The actual figure corresponds to {\bf grid2}
 (Table~\ref{tab:grids}).

%%%%%%%%%%%%%%%%%%%%%%%%%
%
%

\section{Additional fits of S\'ersic profiles using propols}\label{app:c}

\begin{figure}
  \centering 
\includegraphics[width=0.45\linewidth]{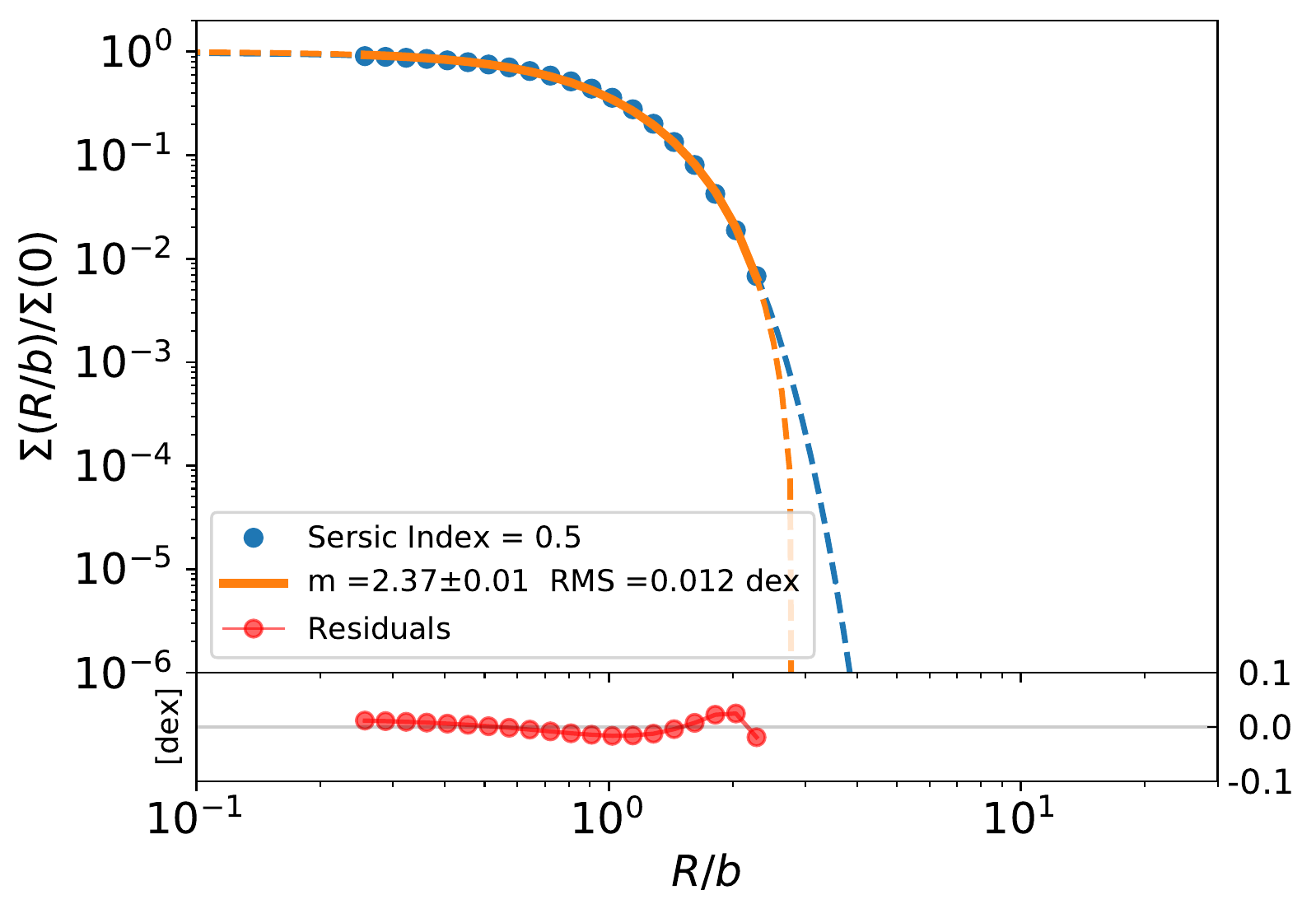}
\includegraphics[width=0.45\linewidth]{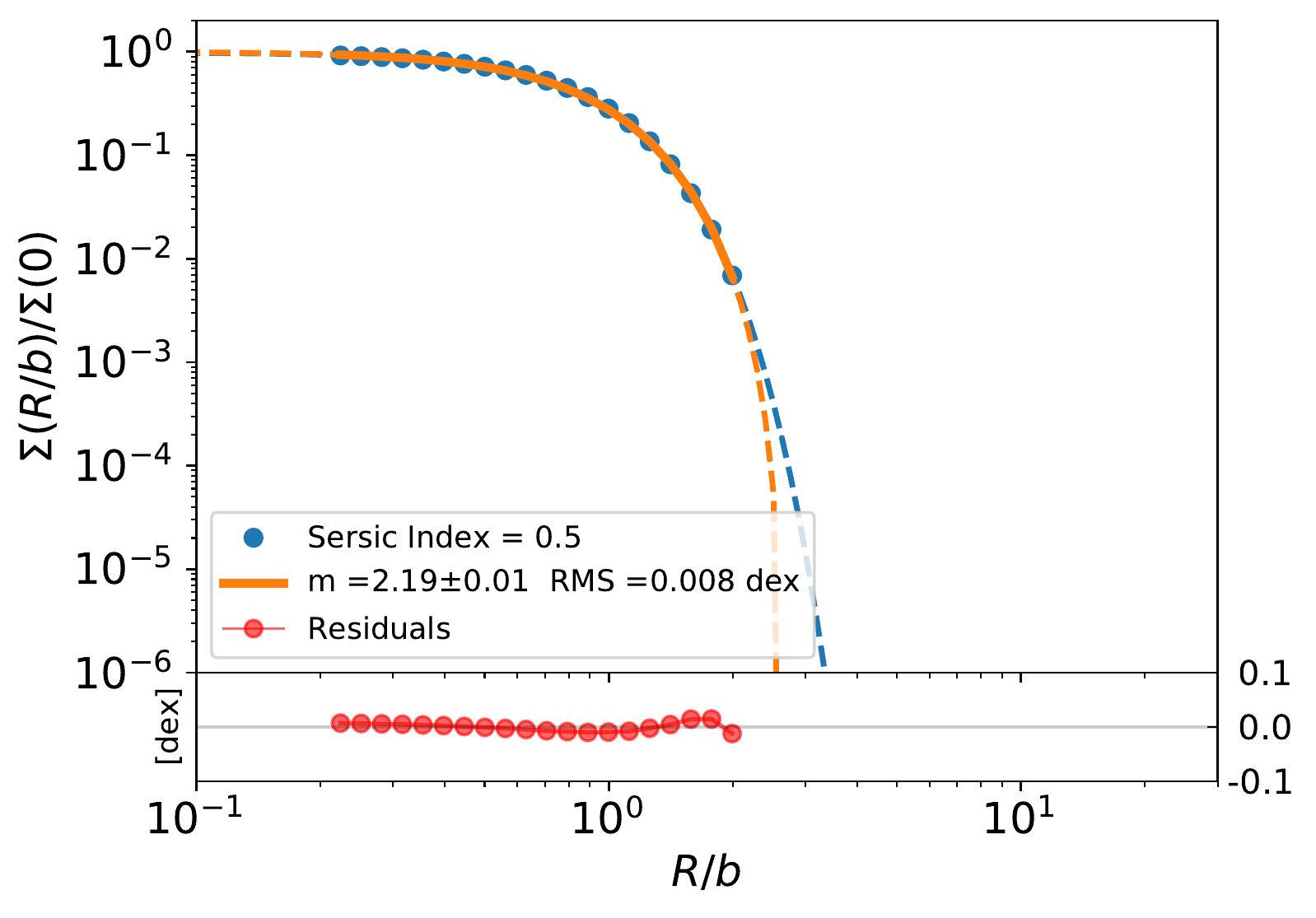}\\
\includegraphics[width=0.45\linewidth]{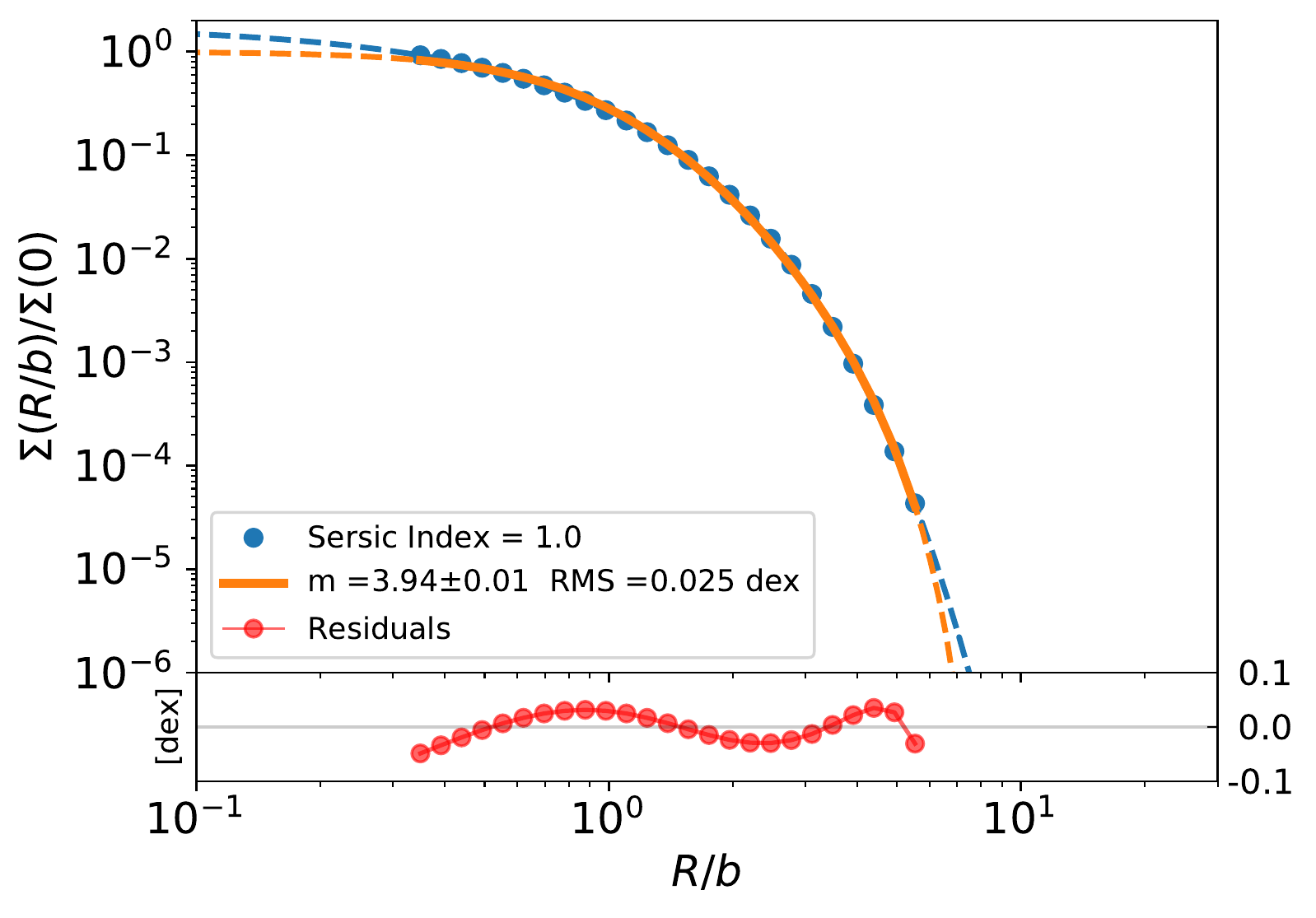}
\includegraphics[width=0.45\linewidth]{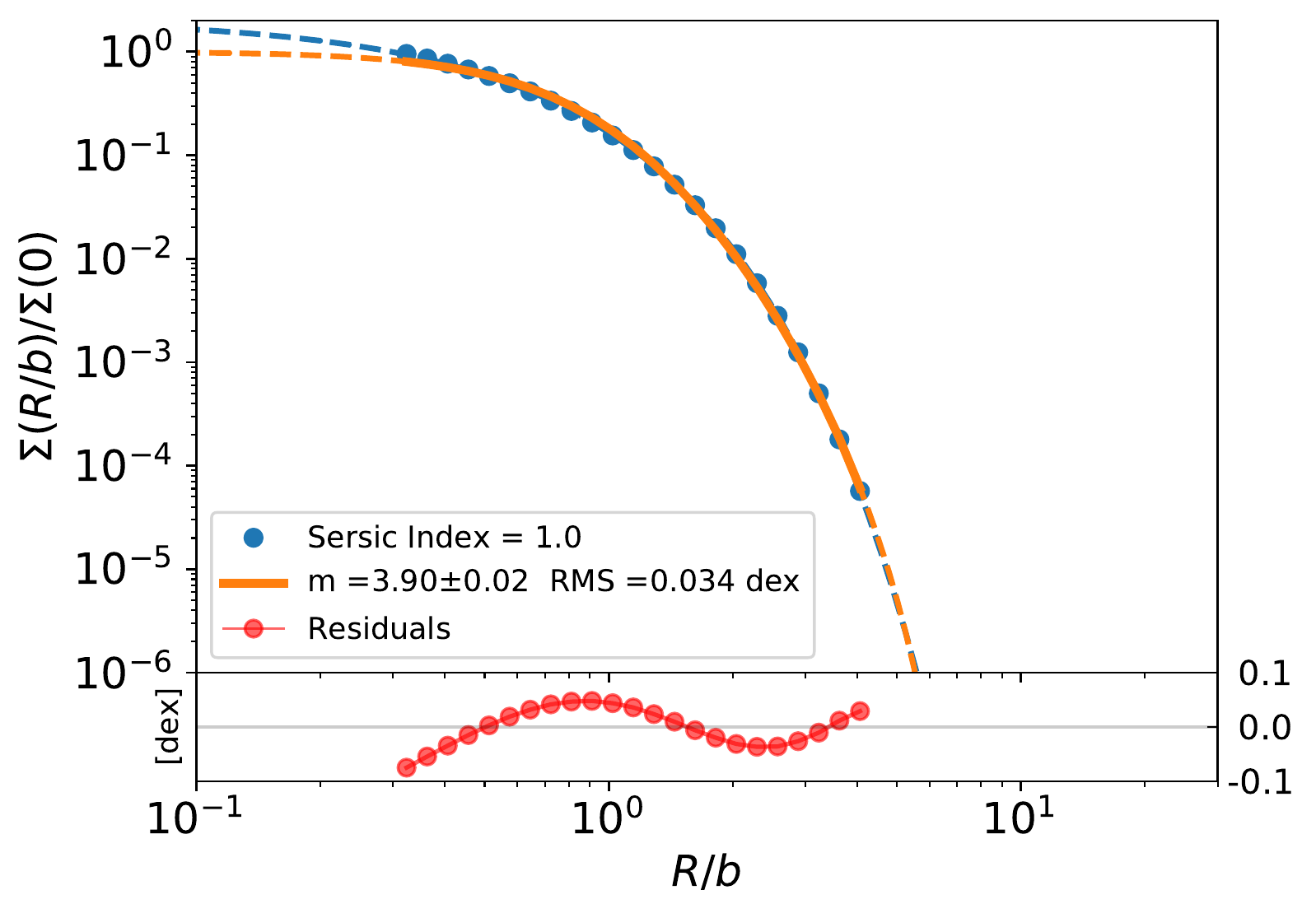}\\
\includegraphics[width=0.45\linewidth]{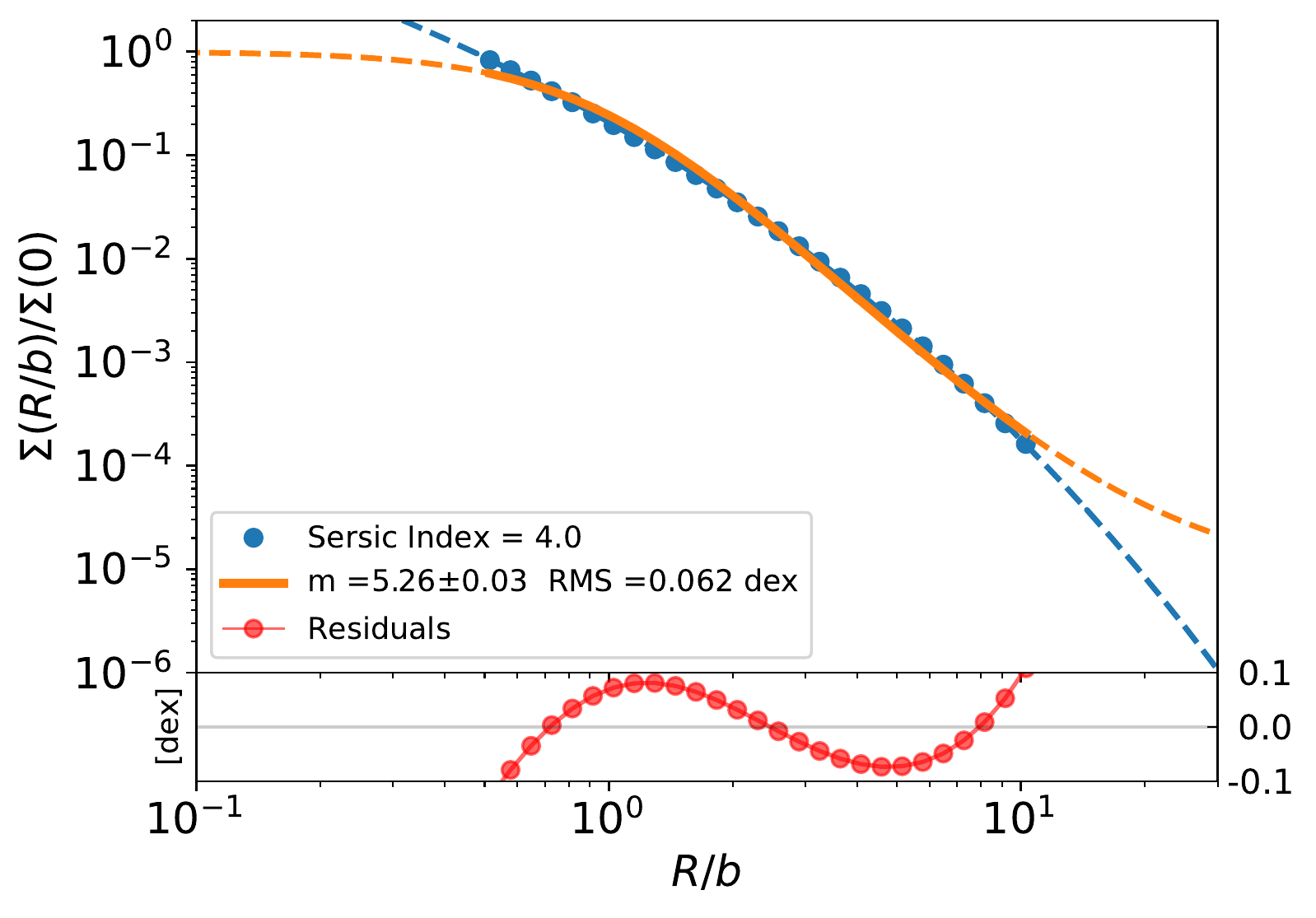}
\includegraphics[width=0.45\linewidth]{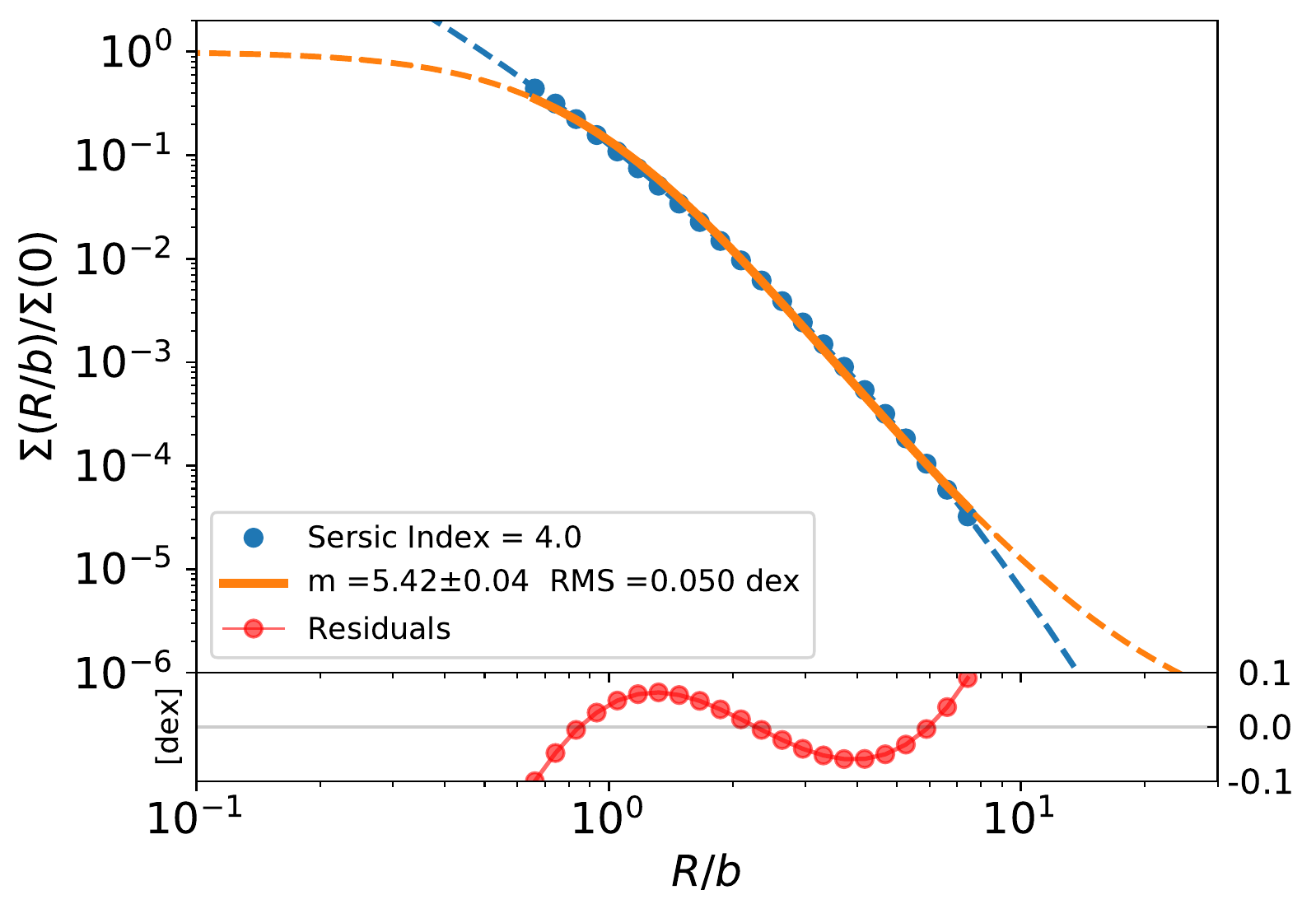}
\caption{{\em Propol} fits to \sersic\ profiles. The figure is identical to  Fig.~\ref{fig:lane_emden_fitsersic} but sampling other hyper-parameters of the fits. Left column: the central cores are partly included in the fits. They are associated with the green lines and symbols in Fig.~\ref{fig:lane_emden6}. Right column: the stellar to total mass  ratio is assumed to vary with radius with $\gamma = 0.3$ (Eq.~[\ref{eq:gamma}]). They give rise to the orange lines and symbols in Fig.~\ref{fig:lane_emden6}. 
The \sersic\ profiles to be fitted and the scaling of the axes are identical to those employed in Fig.~\ref{fig:lane_emden_fitsersic}.
}
\label{fig:lane_emden_fitsersic_app}
\end{figure}

This appendix includes examples of  \propol\  fits to \sersic\ profiles. The range of hyper-parameters characterizing the fits are varied with respect to the nominal values shown in Fig.~\ref{fig:lane_emden_fitsersic}. The left column in Fig.~\ref{fig:lane_emden_fitsersic_app} corresponds to fits where the cores are partly included, and have to be associated with the green lines and symbols in Fig.~\ref{fig:lane_emden6}. The right column in Fig.~\ref{fig:lane_emden_fitsersic_app} shows fits where the stellar to total mass density ratio is not constant with radius. They corresponds to the orange lines and symbols in Fig.~\ref{fig:lane_emden6}. 

%%%%%%%%%%%%%%%
\section{Impact of changing hyper-parameters when fitting mass surface density profiles in galaxies}\label{app:scatter}
\begin{figure}
  \centering
    \includegraphics[width=0.5\linewidth]{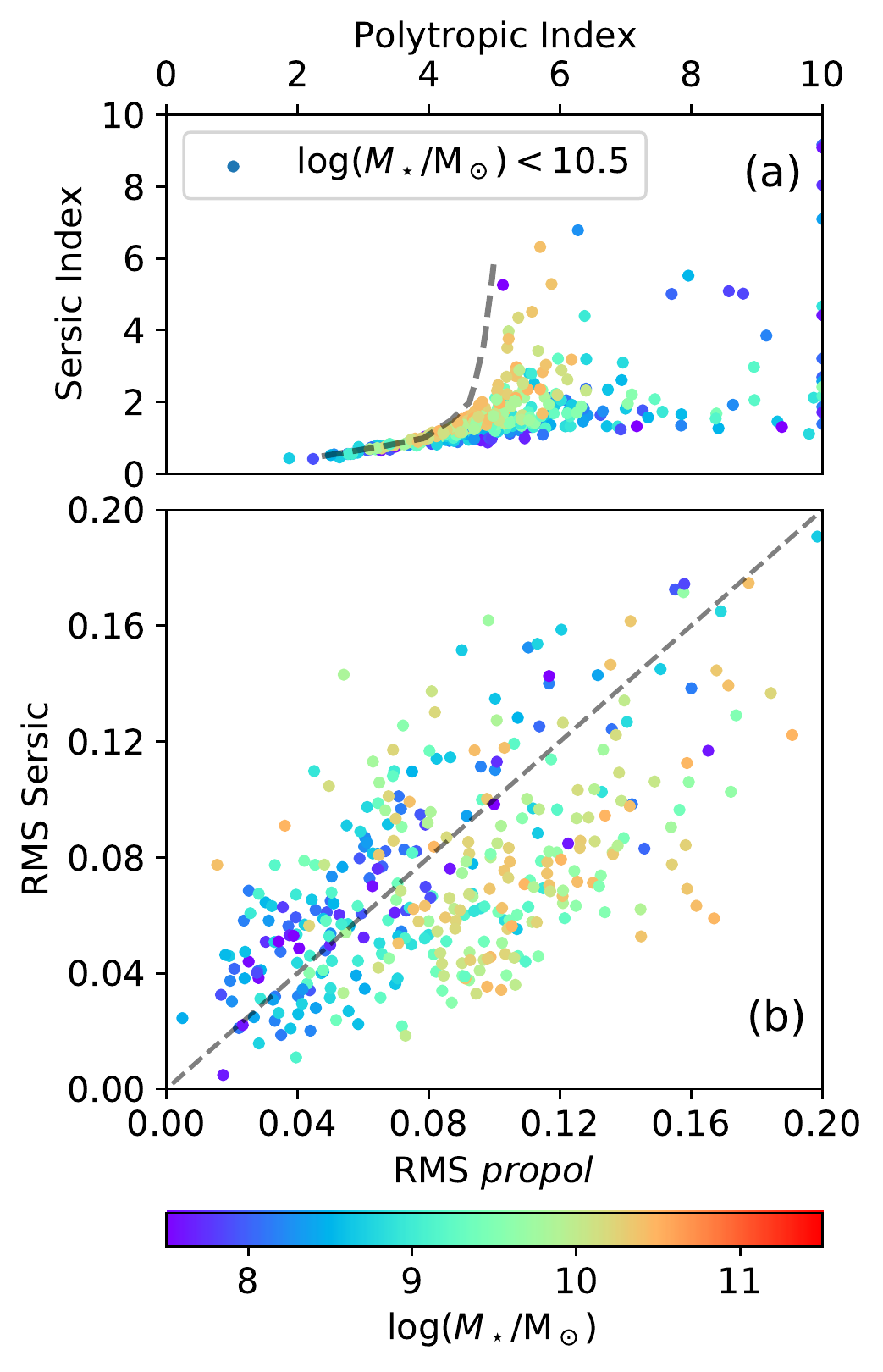}
    \caption{Figure identical to Fig.~\ref{fig:diagnostic_aaa}, except that high mass galaxies are excluded ($\log[M_\star/{\rm M}_\odot] < 10.5$ in the plot). For the layout and other details, see the caption of Fig.~\ref{fig:diagnostic_aaa}.
    }
  \label{fig:diagnostic_ccc}
 \end{figure}

\begin{figure}
   \centering
    \includegraphics[width=0.5\linewidth]{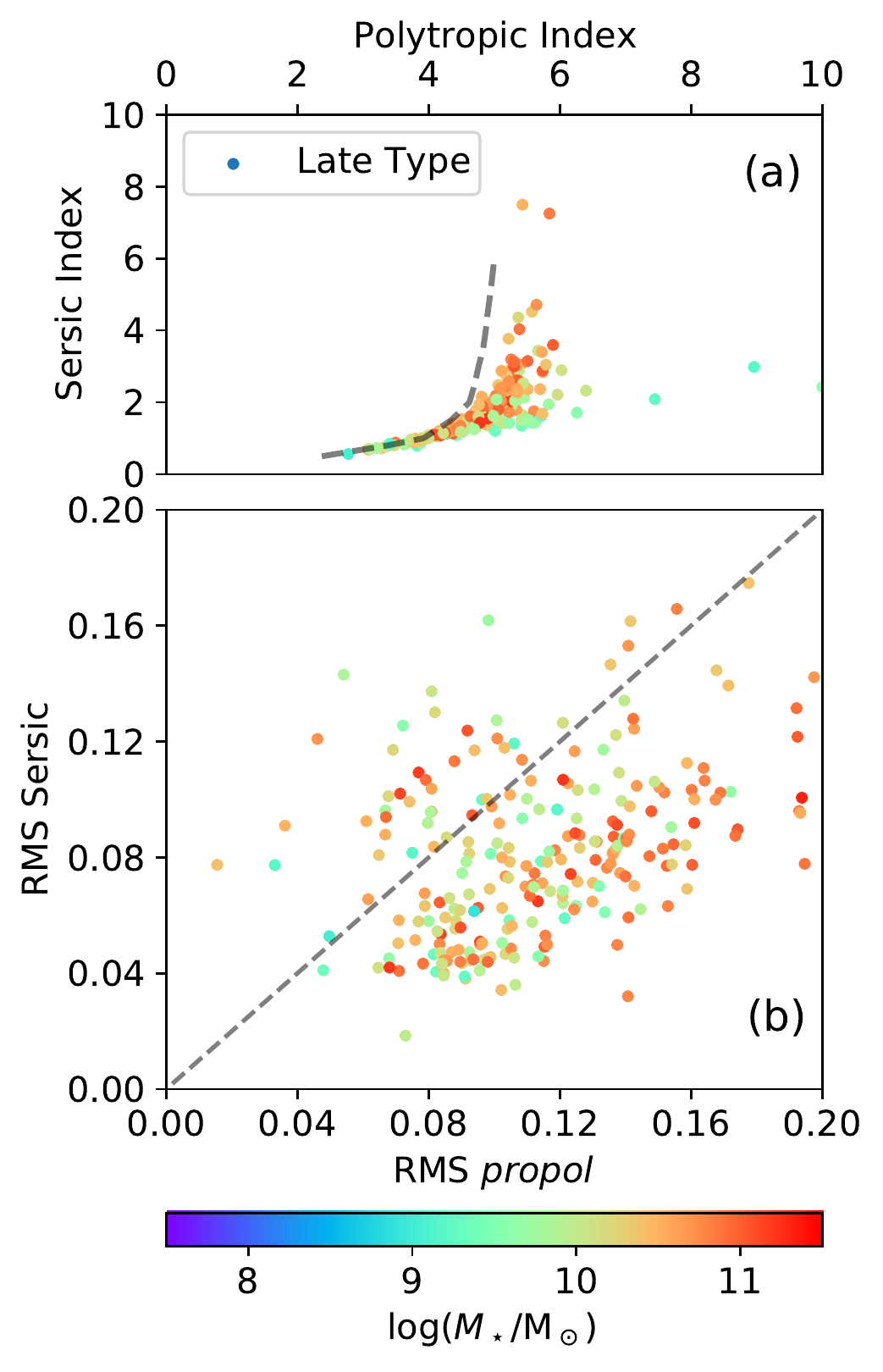}
    \caption{Figure identical to Fig.~\ref{fig:diagnostic_aaa}, except that only galaxies morphologically classified as late types are included. For the layout and other details, see the caption of Fig.~\ref{fig:diagnostic_aaa}.
    }
  \label{fig:diagnostic_ddd}
 \end{figure}

\begin{figure}
   \centering
    \includegraphics[width=0.5\linewidth]{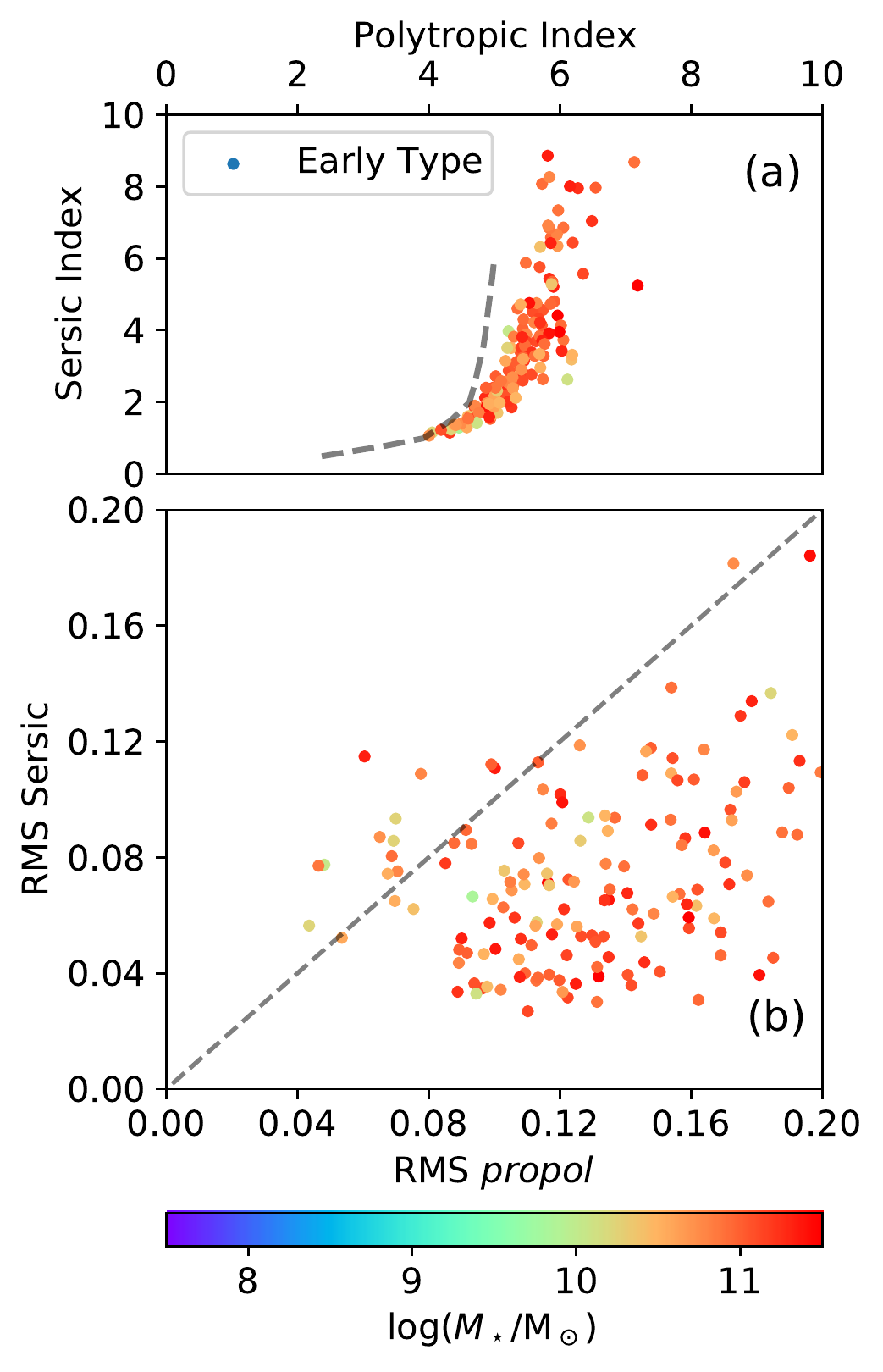}
    \caption{Figure identical to Fig.~\ref{fig:diagnostic_aaa}, except that only galaxies morphologically classified as early types are included. For the layout and other details, see the caption of Fig.~\ref{fig:diagnostic_aaa}. 
    }
  \label{fig:diagnostic_ggg}
 \end{figure}

\begin{figure}
   \centering
    \includegraphics[width=0.5\linewidth]{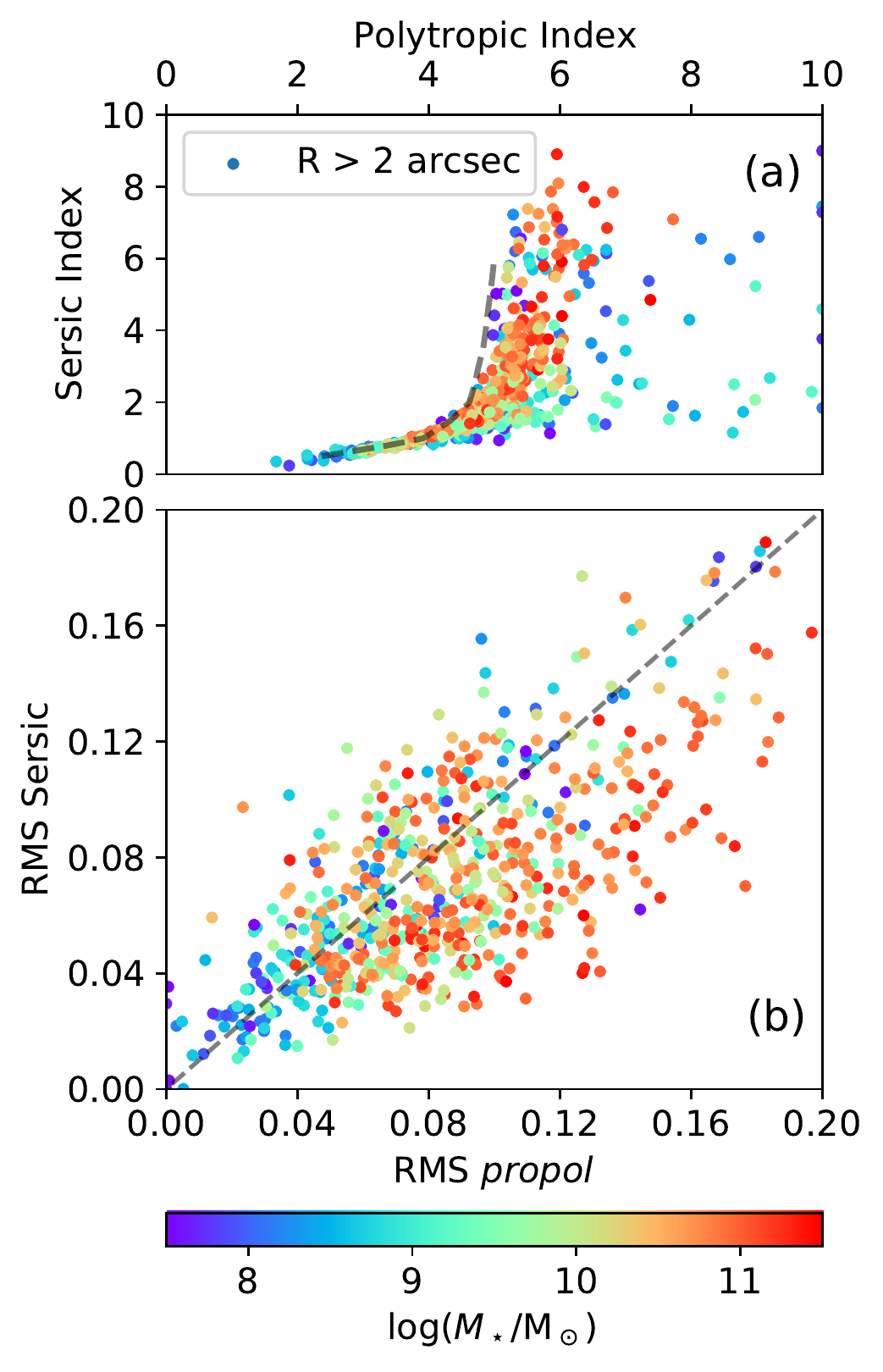}
    \caption{Figure identical to Fig.~\ref{fig:diagnostic_aaa}, except that the \propol\ fits were carried out removing most of the central cores of the galaxies ($R> 2$\,arcsec). In this case, the points corresponding to the most massive objects approach the theoretical curve worked out in Sect.~\ref{sec:sersic} (the dashed line). For the layout and other details, see the caption of Fig.~\ref{fig:diagnostic_aaa}.
    }
  \label{fig:diagnostic_eee}
 \end{figure}

\begin{figure}
   \centering
    \includegraphics[width=0.5\linewidth]{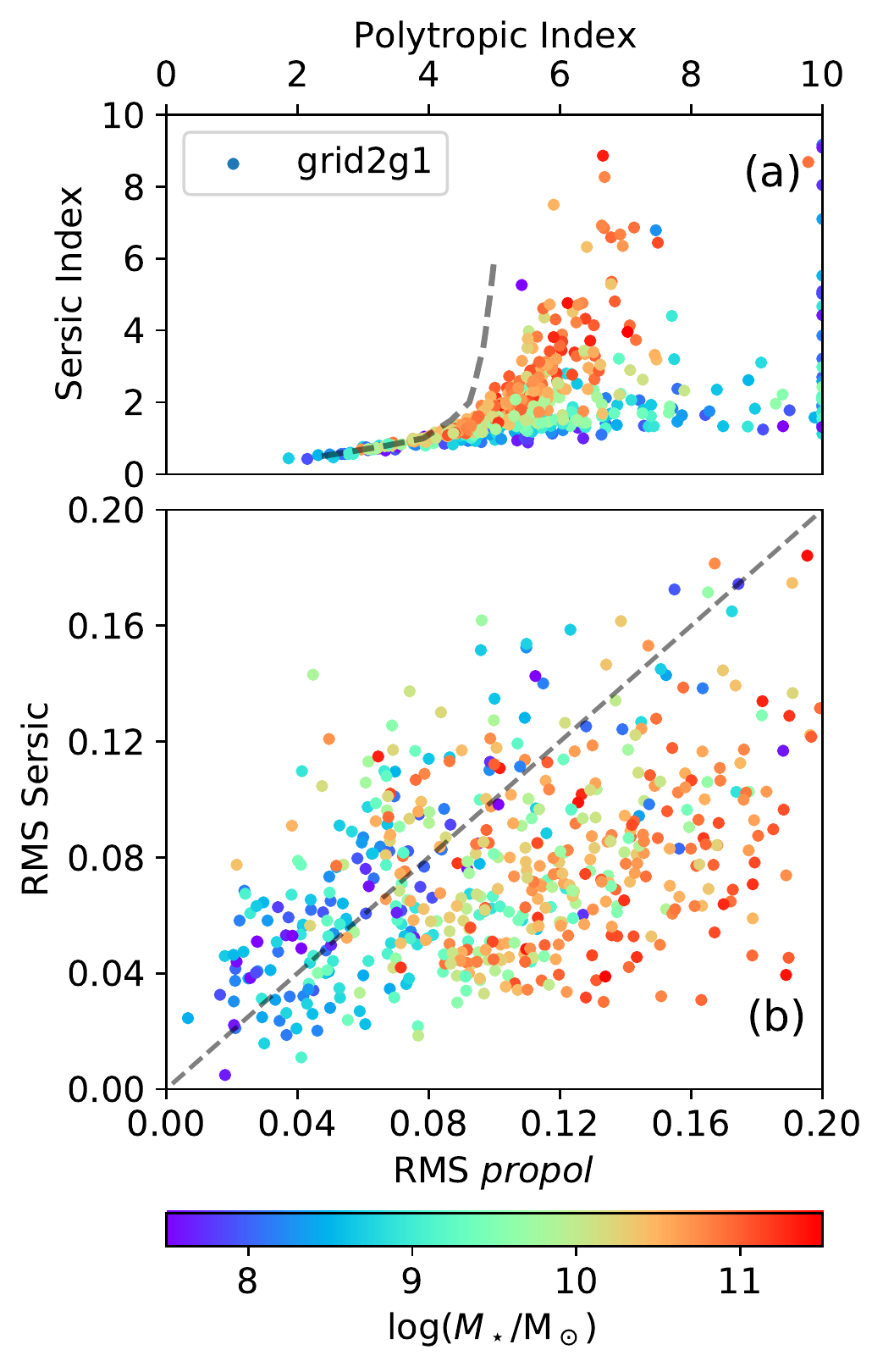}
    \caption{Figure identical to Fig.~\ref{fig:diagnostic_aaa}, except that the grid used to carry out the fits includes differences between the total mass profile and the stellar mass profile (i.e., rather than  {\bf grid2}, the fits are based on {\bf grid2g1}). For the layout and other details, see the caption of Fig.~\ref{fig:diagnostic_aaa}.}
  \label{fig:diagnostic_eee_other}
 \end{figure}

This appendix collects a number of plots that show how the goodness of the \propol\ fits to observed mass profiles changes when the hyper-parameters of the fit and the sample selection are modified. These plots are referred to and used in Sect.~\ref{sec:fit_propol_galax}. All figures are identical to  Fig~\ref{fig:diagnostic_aaa} except for: high mass galaxies are excluded in Fig.~\ref{fig:diagnostic_ccc}, only late morphological types are included in Fig.~\ref{fig:diagnostic_ddd}, only early morphological types are included in Fig.~\ref{fig:diagnostic_ggg},  central cores are not used for fitting in Fig.~\ref{fig:diagnostic_eee}, and differences between total mass and stellar mass profiles are considered in Fig.~\ref{fig:diagnostic_eee_other}.


 %



\end{document}